\documentclass{emulateapj}

\def\arcdeg{\hbox{$^\circ$}}

\def\arcsec{\hbox{$^{\prime\prime}$}}
\def\deg2{\hbox{$\rm deg^{2}$}}

\def\fs{\hbox{$.\!\!^{\rm s}$}}

\def\farcs{\hbox{$.\!\!^{\prime\prime}$}}
\def\lsim{\mathrel{\rlap{\lower4pt\hbox{\hskip1pt$\sim$}}\raise1pt\hbox{$<$}}}                
\def\gsim{\mathrel{\rlap{\lower4pt\hbox{\hskip1pt$\sim$}}\raise1pt\hbox{$>$}}}                
\newcommand\nobrkhyph{\mbox{-}}

\lefthead{Drake et~al.}
\righthead{Catalina Periodic Variables} 

\begin{document}
\title{The Catalina Surveys Periodic Variable Star Catalog}

\author{
  A.J.~Drake\altaffilmark{1}, M.J.~Graham\altaffilmark{1}, S.G.~Djorgovski\altaffilmark{1}, M.~Catelan\altaffilmark{2,3}, A.A.~Mahabal\altaffilmark{1},\\
  G.~Torrealba\altaffilmark{2}, D. Garc\'{\i}a-\'Alvarez\altaffilmark{4,5,6}, C.~Donalek\altaffilmark{1},
  J.L.~Prieto\altaffilmark{7}, R.~Williams\altaffilmark{1}, S.~Larson\altaffilmark{8},\\ E.~Christensen\altaffilmark{8},
  V.~Belokurov\altaffilmark{9}, S.E.~Koposov\altaffilmark{9}, E.~Beshore\altaffilmark{7}, A.~Boattini\altaffilmark{8}, A.~Gibbs\altaffilmark{8},\\
  R.~Hill\altaffilmark{8}, R.~Kowalski\altaffilmark{8}, J.~Johnson\altaffilmark{8}, and F.~Shelly\altaffilmark{8}
}

\altaffiltext{1}{California Institute of Technology, 1200 E. California Blvd, CA 91225, USA}
\altaffiltext{2}{Pontificia Universidad Cat\'olica de Chile, Instituto de Astrof\'{i}sica, 
Facultad de F\'{i}sica, Av. Vicu\~na Mackena 4860, 782-0436 Macul, Santiago, Chile}
\altaffiltext{3}{Millennium Institute of Astrophysics, Santiago, Chile}
\altaffiltext{4}{Instituto de Astrof\'{\i}sica de Canarias, Avenida V\'{\i}a L\'actea,
38205 La Laguna, Tenerife, Spain}
\altaffiltext{5}{Departamento de Astrof\'{\i}sica, Universidad de La Laguna,
38205 La Laguna, Tenerife, Spain}
\altaffiltext{6}{Grantecan S.\,A., Centro de Astrof\'{\i}sica de La Palma, Cuesta de San Jos\'e,
38712 Bre\~na Baja, La Palma, Spain}
\altaffiltext{7}{Department of Astronomy, Princeton University, 4 Ivy Ln, Princeton, NJ 08544}
\altaffiltext{8}{The University of Arizona, Department of Planetary Sciences,  Lunar and Planetary Laboratory, 
1629 E. University Blvd, Tucson AZ 85721, USA}
\altaffiltext{9}{Institute of Astronomy, Madingley Road, Cambridge CB3 0HA, UK}

\begin{abstract} 
  
  We present $\sim$47,000 periodic variables found during the analysis of 5.4 million variable star candidates within a
  20,000 square degree region covered by the Catalina Surveys Data Release\nobrkhyph{}1 (CSDR1).  Combining these
  variables with type-ab RR Lyrae from our previous work, we produce an on-line catalog containing periods, amplitudes, 
  and classifications for $\sim$61,000 periodic variables.
  By cross-matching these variables with those from prior surveys, we find that $> 90\%$ of
  the $\sim$8,000 known periodic variables in the survey region are recovered. For these sources we find excellent agreement between 
  our catalog and prior values of luminosity, period and amplitude, as well as classification.

  We investigate the rate of confusion between objects classified as contact binaries and type-c RR Lyrae (RRc's) based
  on periods, colours, amplitudes, metalicities, radial velocities and surface gravities.  We find that no more than few
  percent of these variables in these classes are misidentified.  By deriving distances for this clean sample of
  $\sim$5,500 RRc's, we trace the path of the Sagittarius tidal streams within the Galactic halo. Selecting 146
  outer-halo RRc's with SDSS radial velocities, we confirm the presence of a coherent halo structure that is
  inconsistent with current N-body simulations of the Sagittarius tidal stream. We also find numerous
  long-period variables that are very likely associated within the Sagittarius tidal streams system.
 
  Based on the examination of 31,000 contact binary light curves we find evidence for two subgroups exhibiting irregular
  lightcurves.  One subgroup presents significant variations in mean brightness that are likely due to chromospheric
  activity.  The other subgroup shows stable modulations over more than a thousand days and thereby provides evidence
  that the O'Connell effect is not due to stellar spots.

\end{abstract}

\section{Introduction}

The 1638 discovery of the periodic variability of Mira, by Holwarda (Hoffleit 1997), was the beginning 
of a new age of understanding in astronomy.
Periodic variables, such as Cepheids and RR Lyrae, underpin the cosmological distance scale (Freedman et al.~2001) 
while also providing insight into the processes and properties of galaxy formation (Saha 1984, 1985; Catelan 2009). 
On smaller scales, the periodic motions and interactions of binaries provide insight into stellar 
evolution and multiplicity. While under the right circumstances, such pairs can also provide estimates 
of stellar masses, sizes, densities, shapes and orbits (e.g. Southworth 2012 and refs therein).

By the early 1980s many thousands of variable stars were known (Roth 1994). 
However, it was not until the 1990s that microlensing surveys changed the scale of variable star research. Groups such
as MACHO (Alcock et al.~1993) and OGLE (Udalski et al.~1994) began to regularly deliver a treasure trove of tens of
thousands of periodic variable stars in the Magellanic Clouds and Galactic center (e.g. Alcock et al.~1996, 1997, 1998, 1999, 2000a,
2000b, 2001, 2002; Mateo et al.~1995; Udalski et al.~1996, 1998, 1999; Szymanski et al.~2001; Wozniak et al.~2002).
These surveys remain the major source of the more 200,000 periodic variables that are currently known.

Over much larger and less crowded areas of the sky, the very wide-field cameras of the ASAS survey (Pojmanski
1997) soon led to the discovery of tens of thousands of variable stars down to $V = 13$ (Pojmanski 2000, 2002, 2003; Pojmanski \&
Maciejewski 2004a,b; Pojmanski et al.~2005). Similar work was also undertaken using data from the Robotic Optical
Transient Search Experiment (ROTSE; Akerlof et al.~2000) as part of the Northern Sky Variability Survey (NSVS; Wozniak
et al.~2004; Kinemuchi et al.~2006; Hoffman, Harrison \& McNamara 2009). Survey for planetary transits 
have also discovered periodic variable stars (e.g., HATNet, Hartman et al.~2011; BOKS, Feldmeier et al.~2011).

More recently, researchers have been motivated to study the variable star populations spanning the sky to much greater
depths.  This has led to the harvest of variable stars from archival data taken by Near-Earth Object (NEO) surveys.  Data
from the Lowell Observatory Near-Earth Object Search (LONEOS; Bowell et al.~1995) has been searched by Miceli et
al.~(2008), while data from the Lincoln Near-Earth Asteroid Research (LINEAR; Stokes et al.~2000) was analysed by Sesar
et al.~(2013) and Palaversa et al.~(2013). Similarly, in our preliminary work (Drake et al.~2013a,b) we harvested 
type ab RR Lyrae variables from Catalina Sky Survey (Larson et al.~2003) photometry presented in CSDR1 (Drake et al.~2012).
In this paper, we will further investigate CSDR1 photometry in order to complete our search for periodic variable stars.

\section{Observations}

The Catalina Sky Survey\footnote{http://www.lpl.arizona.edu/css/} began in 2004 and uses three telescopes to cover the
sky between declination $\delta = -75$ and +65 degrees in order to discover Near-Earth Objects (NEOs) and Potential
Hazardous Asteroids (PHAs).  Each of the survey telescopes is run as a separate sub-surveys.  These consist of the
Catalina Schmidt Survey (CSS) and the Mount Lemmon Survey (MLS) in Tucson, Arizona, and the Siding Spring Survey (SSS) in
Siding Spring, Australia.  Each telescope has set fields that tile the sky and avoids the Galactic plane region by
10 to 15 degrees due to reduced source recovery in crowded stellar regions.  All images are taken unfiltered to maximize
throughput.  The observations analyzed in this work were taken in sets of four images separated by 10 minutes with
typical exposure times of 30 seconds. Photometry is carried out using the aperture photometry program SExtractor (Bertin
\& Arnouts 1996). In addition to asteroids, all the Catalina data is analyzed for transient sources by the Catalina
Real-time Transient Survey (CRTS\footnote{http://crts.caltech.edu/}, Drake et al.~2009; Djorgovski et al.~2011).

In this paper, we concentrate on the data taken by the CSS 0.7m telescope (CSS) between April 2005 and June 2011, that
was made publicly available via CSDR1 (Drake et al.~2012) in January 2012. Individual detections in CSDR1 are matched with
sources from a deeper ``master'' catalog derived from the median coaddition of 20 or more images.  The CSDR1 dataset has
an average of 250 observations per field. Each CSS image covers 8.2 sq. degrees on the sky. In total, the CSDR1 includes
198 million discrete sources with $12 < V < 20$ (Drake et al.~2013a). We have chosen to select only the 20,155 square
degrees covered by CSDR1 in the region $0\arcdeg < \alpha < 360\arcdeg$ and $-22\arcdeg < \delta < 65 \arcdeg$ as
sources below $\delta = -22 \arcdeg$ have better temporal coverage in SSS data that became available January 2013 as part of
Catalina Surveys Data Release 2 (CSDR2)\footnote{http://catalinadata.org} and will be anaylsed separately.

\section{Variable Source Selection}

In order to find the variable sources with CSDR1 we began by selecting variables based on values of the Stetson
variability index ($J_{WS}$, Stetson 1996) using the weighting scheme of Zhang et al.~(2003).
The current analysis method follows Drake et al.~(2013a,b). However, have revised our variable selection process to
better account for blending. This occurs in CSDR1 photometry as it is based on SExtractor aperture magnitudes. During
poor seeing or sky conditions, close pairs or groups of sources are detected as a single object, while in good seeing
conditions the objects are resolved and appear at their true brightness. Blending is important for variable star
selection since variations in seeing cause artificial changes in the apparent brightness of many catalog objects.

Candidate blended sources can be found by selecting master catalog objects that have multiple detections from a single 
observation. However, these multi-detection associations can also occur due to image artifacts and transient sources 
(such as passing asteroids). Thus, even isolated objects often have some nights where multiple detections have been 
matched from a single image.

To reduce the effect of blends we investigated the distribution of the number of detections ($N_{obs}$) for a few of 
the observation fields. Sources in the range $14 < V_{CSS} < 17$ are not affected by saturation or the natural decline in 
detection efficiency with decreasing brightness. 
Using these sources we determined the average number of detections for sources within a field ($N_{avf}$). 
By inspecting the light curves of the objects with high numbers of detections we found that objects with 
$N_{obs} > 1.07 N_{avf}$ were usually significantly affected by a neighbouring source.
We removed all variable candidates with $V_{CSS} > 14$ that had more detections than this threshold. 

In contrast to additional detections due to blending, sources with $V_{CSS} < 14$ often have $N_{obs}$ well below
$N_{avf}$. This is because of saturation since such observations were flagged removed from our period search.  This
means that the brightness sources may have few good observations. Nevertheless, we found that it was possible to detect
variability for sources brighter than $V=12$ because such sources are not saturated in poor seeing, or when they have
dimmed significantly.  Therefore, to reduce the presence of artificial variability due to saturation effects, we only
remove objects with $V < 14$ that have $N_{obs} < 0.3 \times N_{avf}$. Nevertheless it is prudent to be
cautious about the small fraction of sources near the saturation limit. The data from shallower surveys, 
such as ASAS (Pojmanski 1997) and NSVS Wozniak et al.~2004), may used to further verify the nature these sources.

In addition to the change to our treatment of blended and saturated sources, we discovered that there were systematic 
variations in the photometric noise level between fields. Therefore, instead of adopting a universal $J_{WS}$ variability 
threshold as per Drake et al.~(2013a), we derived separate thresholds for each field.

Firstly, we ordered all the sources within each field by their average magnitude. We then determined the 
average variability ($\bar{J}_{WS}$), and its standard deviation ($\sigma_J$), in groups of 200 sources. 
Since highly variable sources bias the $J_{WS}$ distribution, we perform a 3-sigma clipping on the 
values and redetermine the average and sigma values of this statistic. 
As there are thousands of sources in each field with $V_{CSS} > 16$, we combined $J_{WS}$ values in 0.25 
magnitude bins to provide more robust values. Variable candidates were selected at a $3.5 \sigma_J$ threshold 
that was interpolated to their observed average brightness. 

\begin{figure}[ht]{
\epsscale{1.0}
\plottwo{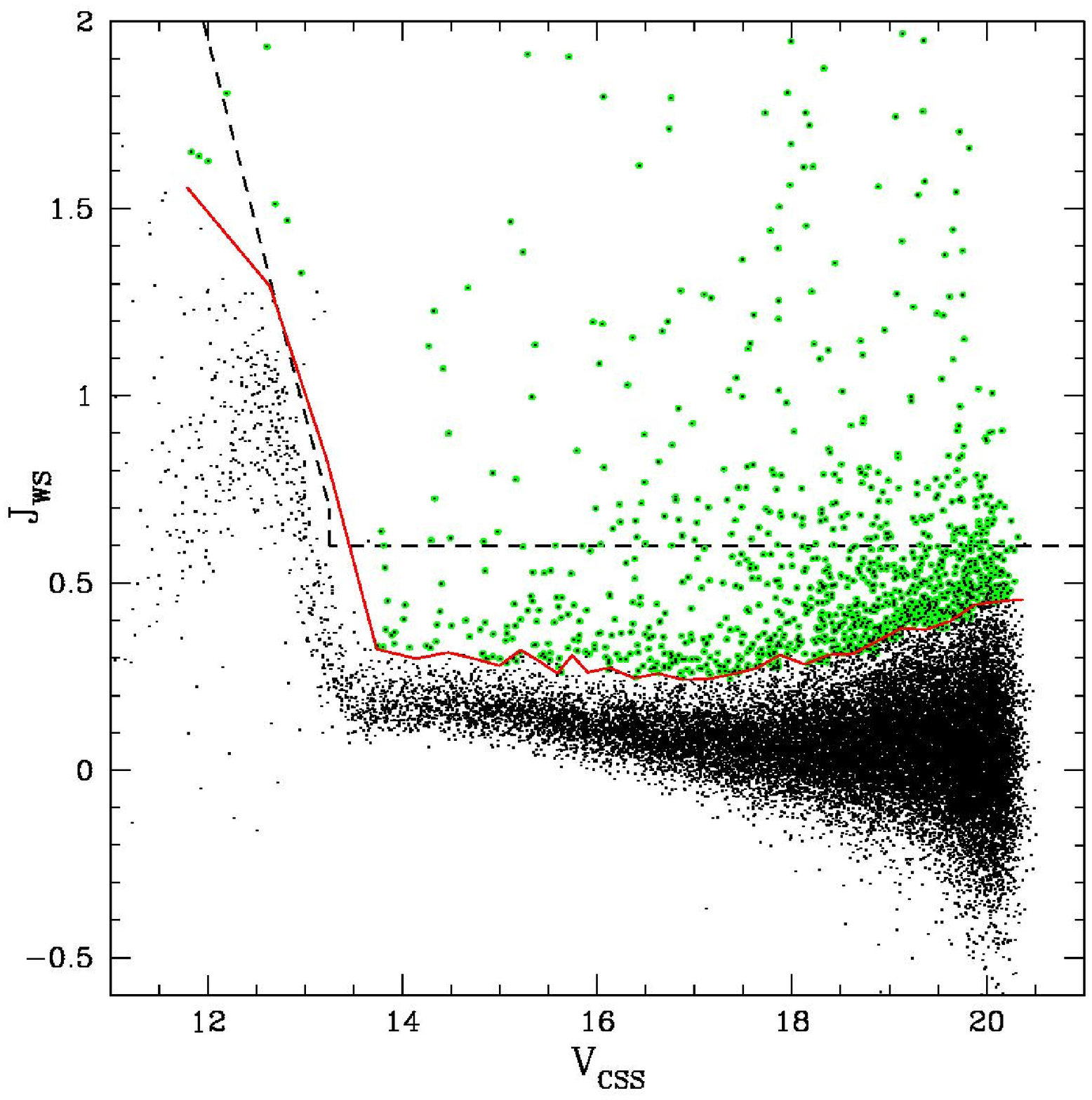}{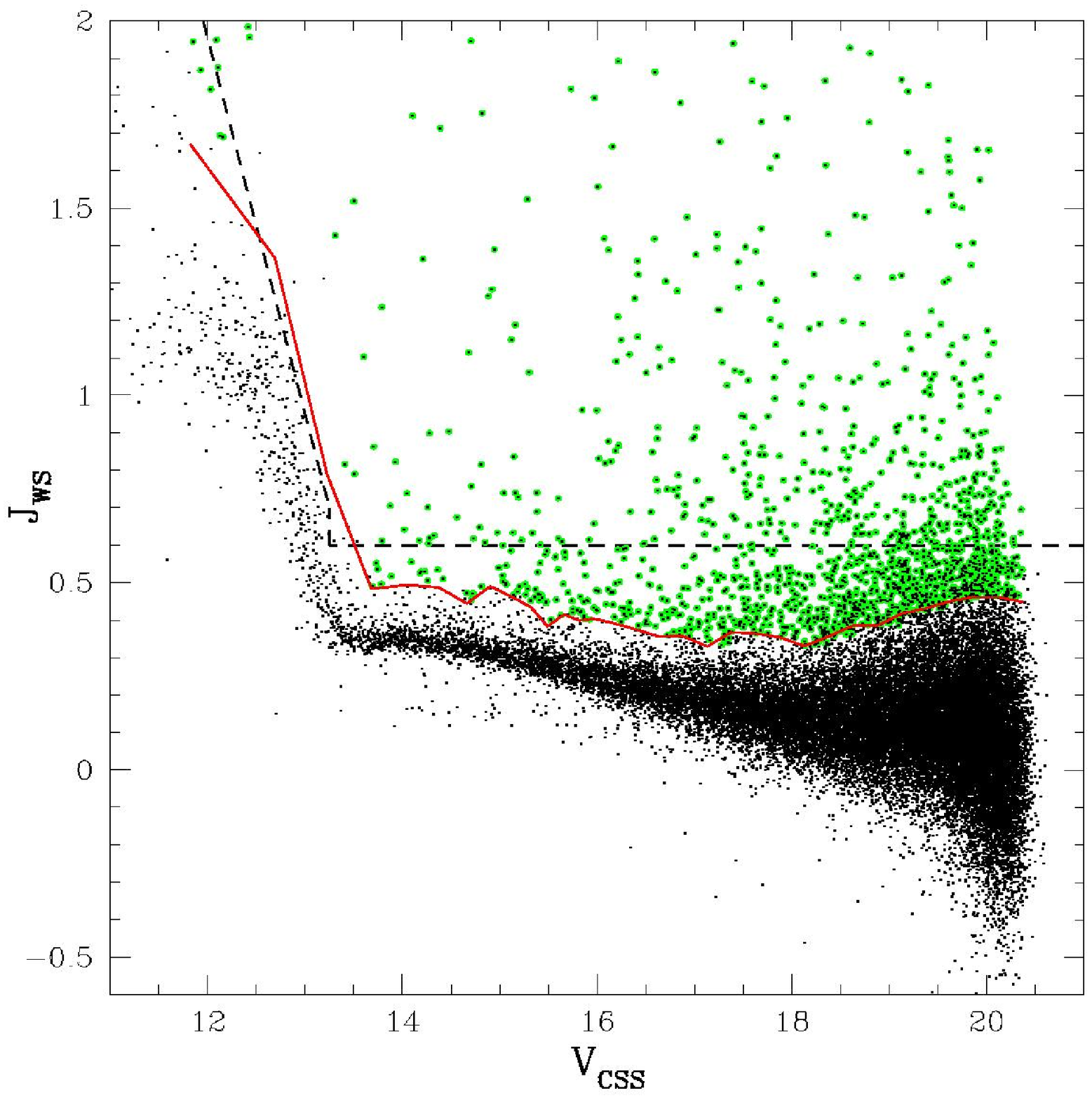}
\caption{\label{Var}
The distribution of Stetson $J_{WS}$ variability index for two fields
analysed in this work. The variable candidates are plotted as large green dots. 
The dashed line shows the selection used by Drake et al.~(2013). The solid red line 
shows the new variability threshold calculated on a brightness and field-by-field basis. 
In the left panel, we plot sources for CSS field S01004 (centered at $\alpha=00^{\rm h}39^{\rm m}22\fs 4$, 
$\delta=-01^\circ 24^\prime40\farcs 00$). In the right panel, we plot objects in CSS field N12065 
(centered at $\alpha=12^{\rm h} 17^{\rm m} 08\fs 50$, $\delta=+12^\circ 41^\prime 59\farcs 89$).
}
}
\end{figure}

In Figure \ref{Var}, we plot the magnitude distribution of $J_{WS}$ values for the sources in two fields as well as with 
the variability threshold used by Drake et al.~(2013a).  This plot demonstrates that there is a significant variation
in the distribution of $J_{WS}$ values with source brightness and field. 
In contrast to Drake et al.~(2013a), where 8 million variable candidates were selected, our new selection 
reduces this number to 5.4 million. This is 2.7\% of the total number of catalog sources.

\section{Selection of Periodic Variables}

We ran Lomb-Scargle (LS, Lomb 1976; Scargle 1982) periodogram analysis on all the variable candidates.
Candidate periodic variables were selected based on a LS significance statistic of $\eta < 10^{-5}$.  
154 thousand candidates were found in addition to the $\sim$15,000 known CSS RRab's (Drake et al.~(2013a,b).
This threshold matches that used Drake et al.~(2013b) since $\sim 15\%$ of RRab's in the CSS data 
were found to have been missed because of the higher ($\eta < 10^{-7}$) threshold used by 
Drake et al.~(2013a)

Although our variability selection process noted above removes bright blends, faint stars have fewer detections
because of lower detection efficiency. To remove blends among faint sources, we select objects with $< 5\%$ 
coincident detections from the individual images. 
This reduces the number of periodic candidates to 144 thousand sources.  We allow small numbers of multi-detections
since, apart from multiple detection due to artifacts, some faint neighbouring objects may only appear in good
seeing. Such objects only have a minor effect on the overall photometry and period determination.

As with Drake et al.~(2013a,b), the Adaptive Fourier Decomposition (AFD) method (Torrealba et al.~2014) was run on all
candidates to determine the best period for each source. As expected, there are many cases where a false signal is found
due to the sampling pattern as well as variation in lunar phase and location.  These effects lead to period estimates
that are near fractions and multiples of a sidereal day, as well as near sidereal and synodic months.  We investigated
the extent of each of these aliases and removed candidates lying within the range of each period alias. There is also
significant evidence for seasonal and annual variations.  As these features are broad they were not removed. However, as
we will show, these sources are removed during inspection. The removal of aliases reduces the numbers of candidates to 128
thousand. We expect that a small fraction (a few percent) of true variables with periods within these ranges were 
also removed by these cuts.

A clear way to reduce the number of non-periodic objects along the candidates is to select sources based on the
goodness-of-fit, or reduced-$\chi^2$ ($\chi^2_r$), of the best AFD period. As expected, we found that most of the periodic 
variable candidates had $\chi^2_r \lsim 1$.
A poor fit generally signifies that an object is either not periodic, or the best fit period is
incorrect. However, sources such as long-period variables (LPVs) exhibit significant periodicity, yet are poorly fit by a 
low-order ($\leq 6$) Fourier series due to moderately large variations in amplitude over time. 
To simultaneously retain the significantly-periodic sources, while removing poorly fit aperiodic 
sources, we removed the candidates that had both $\chi^2_r > 5$ and $\eta < 10^{-9}$. The resulting
set contains $\sim 112,\!000$ candidates.

\begin{figure}[ht]{
\epsscale{0.8}
\plotone{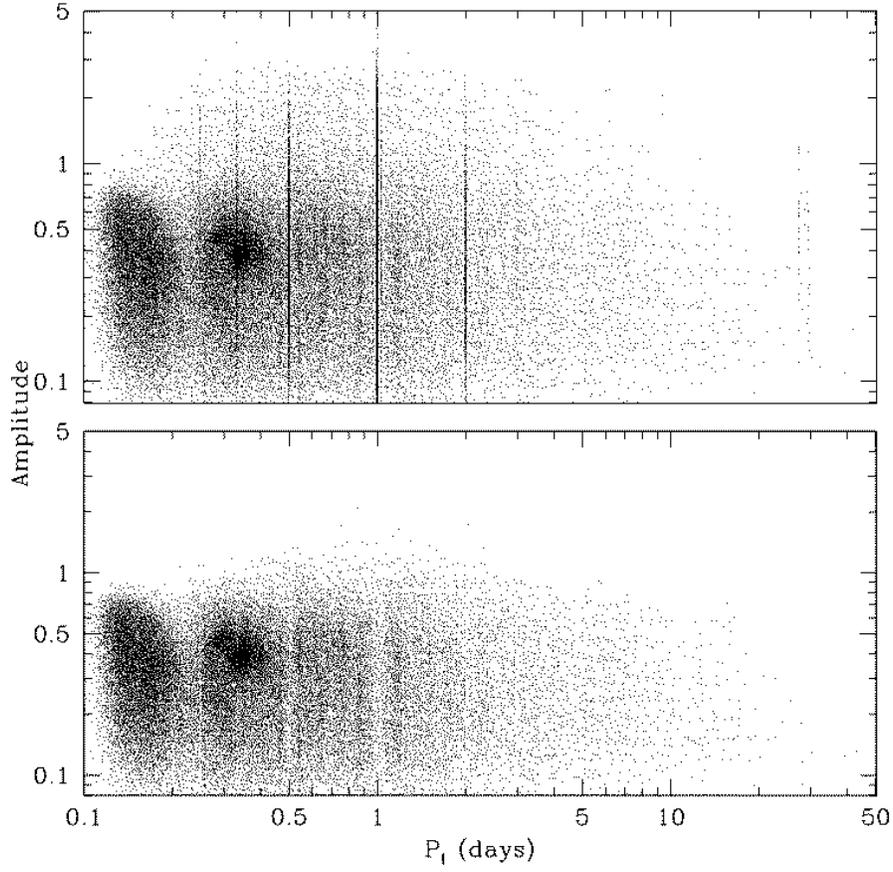}
\caption{\label{PerAmpInit}
The period-amplitude distribution of candidate variables.
In the top plot we show the period distribution before
removing the objects with Fourier fits with $\chi^2_r > 5$
or periods due to sampling aliases.
}
}
\end{figure}

In Figure \ref{PerAmpInit} we compare the period-amplitude distribution of the initial 
154,000 periodic candidates with the $\sim 112,\!000$ sources remaining after removing 
blends, period alias and sources with poor AFD fits.

All $112,\!000$ periodic candidates were inspected (by AJD) using a custom-built web service.  
This process involved classifying the variable types as well as flagging objects that had periods
which were slightly incorrect, or completely incorrect but still probably periodic variables.
The classification process itself mainly involved examination and comparison of the phased 
(an in some cases unphased) lightcurve morphology with known types of periodic variables.
To improve this process, additional consideration was always given to the range of periods, amplitudes 
and colours that different types of variables are known to have. For example, short period contact 
binaries are generally red, while RR Lyrae are blue. The values of the M-test statitistic 
(described below) were also considered. However, none of these individual parameters were 
specifically used to classify or exclude types of variable since the observational data 
itself contains some objects with unusual values. For example, unexpected colours due 
to blending that could be confirmed when examining higher resolution SDSS images.

In ambigous cases, such as may arise for faint objects, or when multiple types of variablity 
are present, the most evident and likely individual type based on the combined information.
Cases where the presence of actual periodicity was in question were vetoed.
Due to the nature of the process it is highly likely that some observational biases exist.
It is also possible that some objects have best fit periods that are aliases of their true 
period. This could lead to the type of variability being wrongly classified.

During the inspection process the initial periods of a large number of the sources were improved and corrected. In
particular, most of the contact binaries were initially found at half of their true period because of the symmetry of
their light curves. This half-period effect is a well-known problem (e.g., Richards et al.~2012). Additionally, in $\sim
2,500$ cases the AFD periods were very close to correct, but clearly not correct.  We improved the period determinations
of these sources interactively by making very slight adjustment to the period. As the actual final changes are generally
$<< 0.01\%$ the process almost always converged very quicky. However, in the case of detached eclipsing binaries, we 
often found the correct period by investigating the harmoic frequencies.

In total, $\sim 47,\!000$ light curves were classified as corresponding to periodic variables.  Of these, 396 are objects
with multiple light curves due to overlap between fields.  The duplicates include 45 RRab's previously given by Drake et
al.~(2013a,b).  The number of unique periodic variables is $46,\!668$. Another $4,\!800$ objects remain periodic
candidates and will be provided online\footnote{http://catalinadata.org}. In Table \ref{Psel}, we show the effect of
selections on the total number of candidates.

\section{General Properties of the Periodic Variables}

\begin{figure}[ht]{
\epsscale{0.8}
\plotone{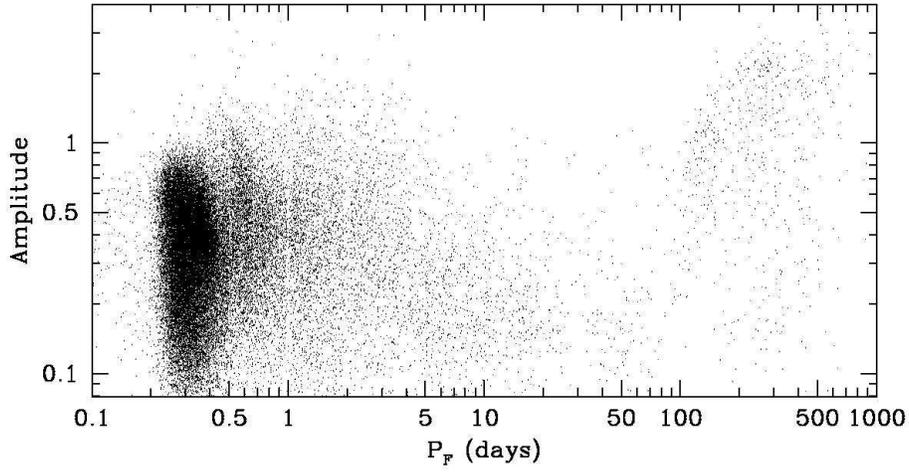}
\caption{\label{PerAmpFin}
The period-amplitude distribution of the periodic sources selected after inspection
of light curves.
}
}
\end{figure}

To see how the visually selected periodic variables differs from the initial candidates we 
investigated the period, amplitude, and magnitude properties of the sources. 
In Figure \ref{PerAmpFin}, we plot the period-amplitude distribution of the periodic variables in 
the catalog. In contrast to Figure \ref{PerAmpInit}, here we plot the final periods of the objects
rather than their initial periods.  Very few objects remain at periods less than 0.2 days. These
objects are contact binaries suffering from half-period effect, as noted above.
Additionally, very few objects with periods longer than 10 days remain. In particular, there are 
no sources with periods longer than $1,\!000$ days.
Most of the stars with long periods are LPVs.  There are very few LPVs in the 
fields we observed away from the Galactic disk.
The foreground LPVs are generally undetectable since they are bright and thus saturated within the images.
Some stars are known to have longer periods. However, like LPVs such stars are generally near our saturation 
limit. Additionally the length of our data does not provide strong constrains for objects with very long periods.
Most of the candidates that were found at very long periods had best-fit periods matching the time span of 
the data. Many of these sources were found to be quasars (QSOs) and were clearly not periodic.

A comparison of the periods for the 396 objects with multiple light curves revealed 
$\sim 10$ of the objects had different periods. Inspection of the light curves revealed
that almost all of these cases were due to contact binaries that were also well fit when 
folded at an alias of their true periods. Since contact binaries dominate the number of periodic
variables in this catalog, we expect that $\sim 10\% $ of the periods presented here are aliases.

\begin{figure}[ht]{
\epsscale{0.8}
\plotone{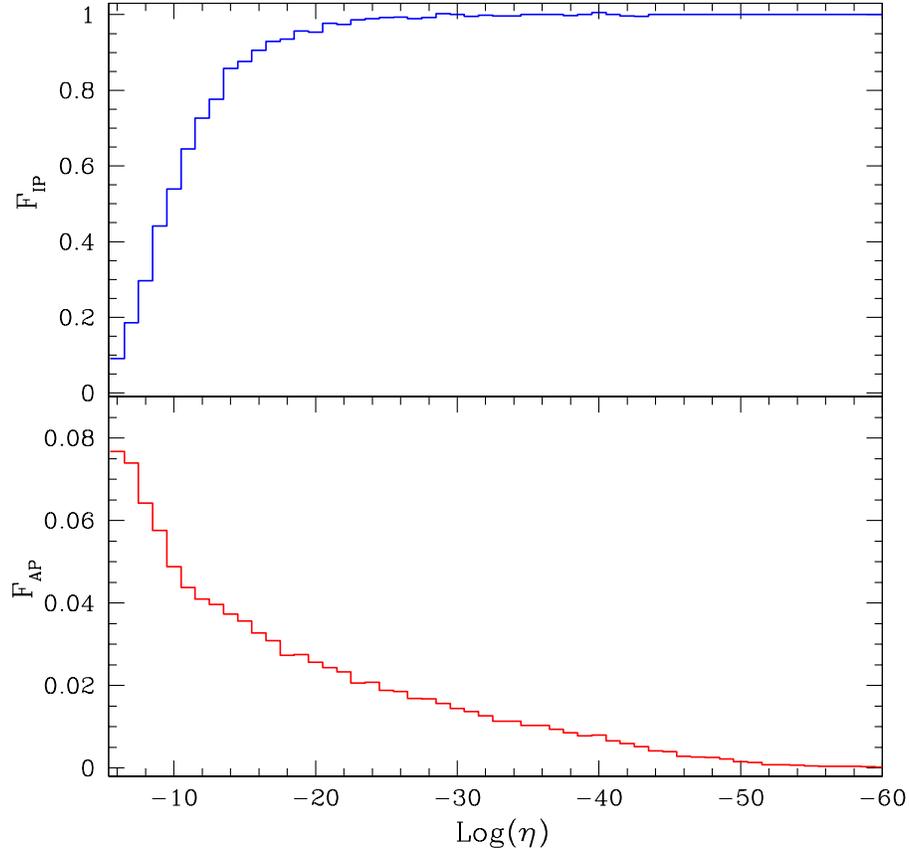}
\caption{\label{Sig}
Fraction of periodic objects compared with LS significance.
In the top panel we plot the fraction of the $112,\!000$ initial periodic
variable candidates ($F_{IP}$) that made it into our catalog. 
In the lower panel we plot the fraction of all periodic 
catalog sources ($F_{AP}$) found each LS significance level.
}
}
\end{figure}

In Figure \ref{Sig}, we present the distribution of LS periodic significance ($\eta$) for the catalog variables. 
This plot shows us that $< 10\%$ of the periodic candidates with $\eta = 1 \times 10^{-6}$ were found to be periodic. 
However, since a very large fraction of the sources found at this level, they constitute $> 7\%$ of the
total number in the catalog. This result suggests that a large number of periodic sources remain to be found 
at low levels of significance. However, since there are millions of sources at this level, their discovery 
within these data requires either new techniques, or automated methods of classification and verification.
The number of aperiodic variables within these data (such as QSOs), far outnumbers the periodic sources 
(Graham et al.~2014 in prep.).

\begin{figure}[ht]{
\epsscale{1.0}
\plottwo{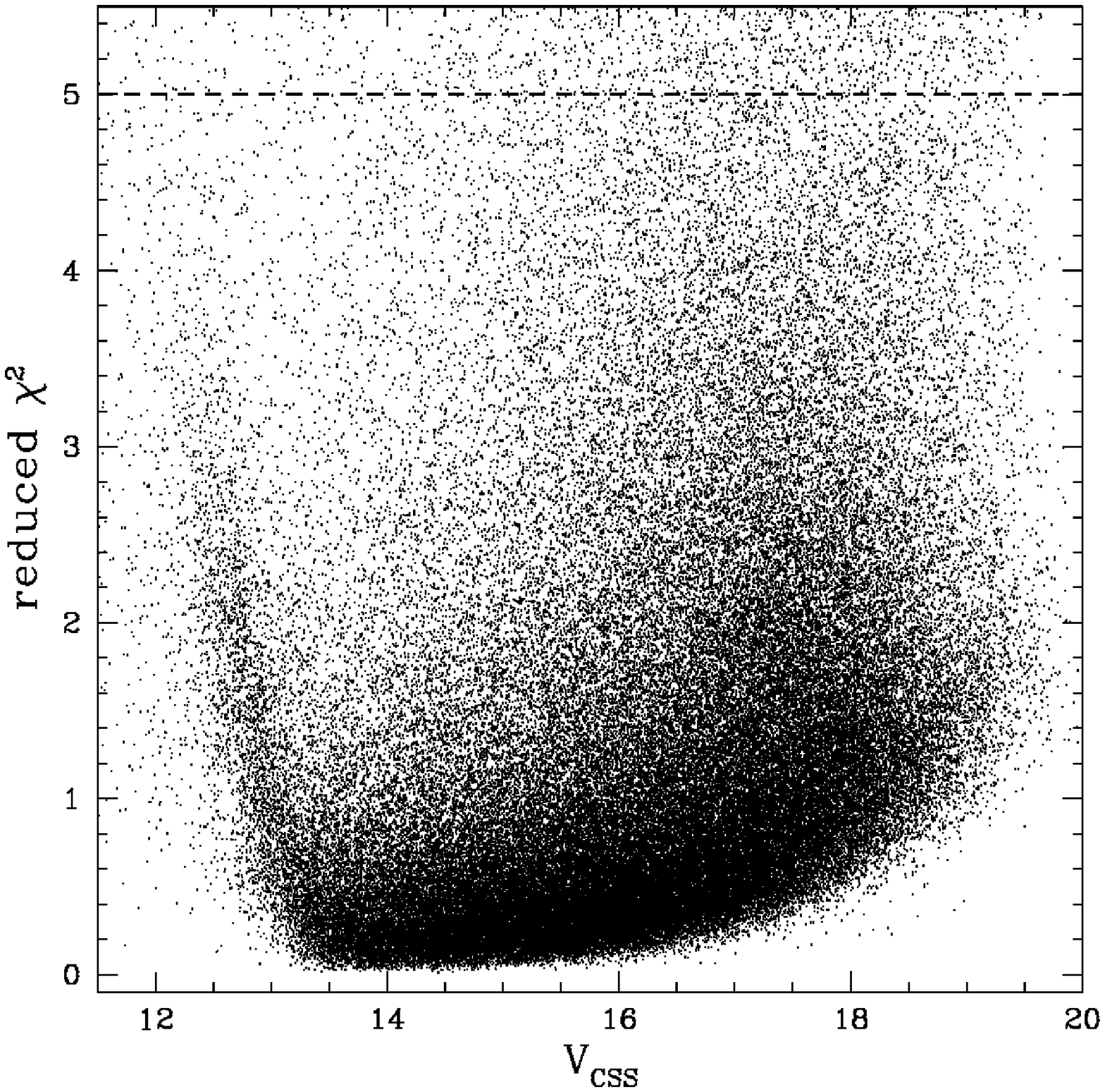}{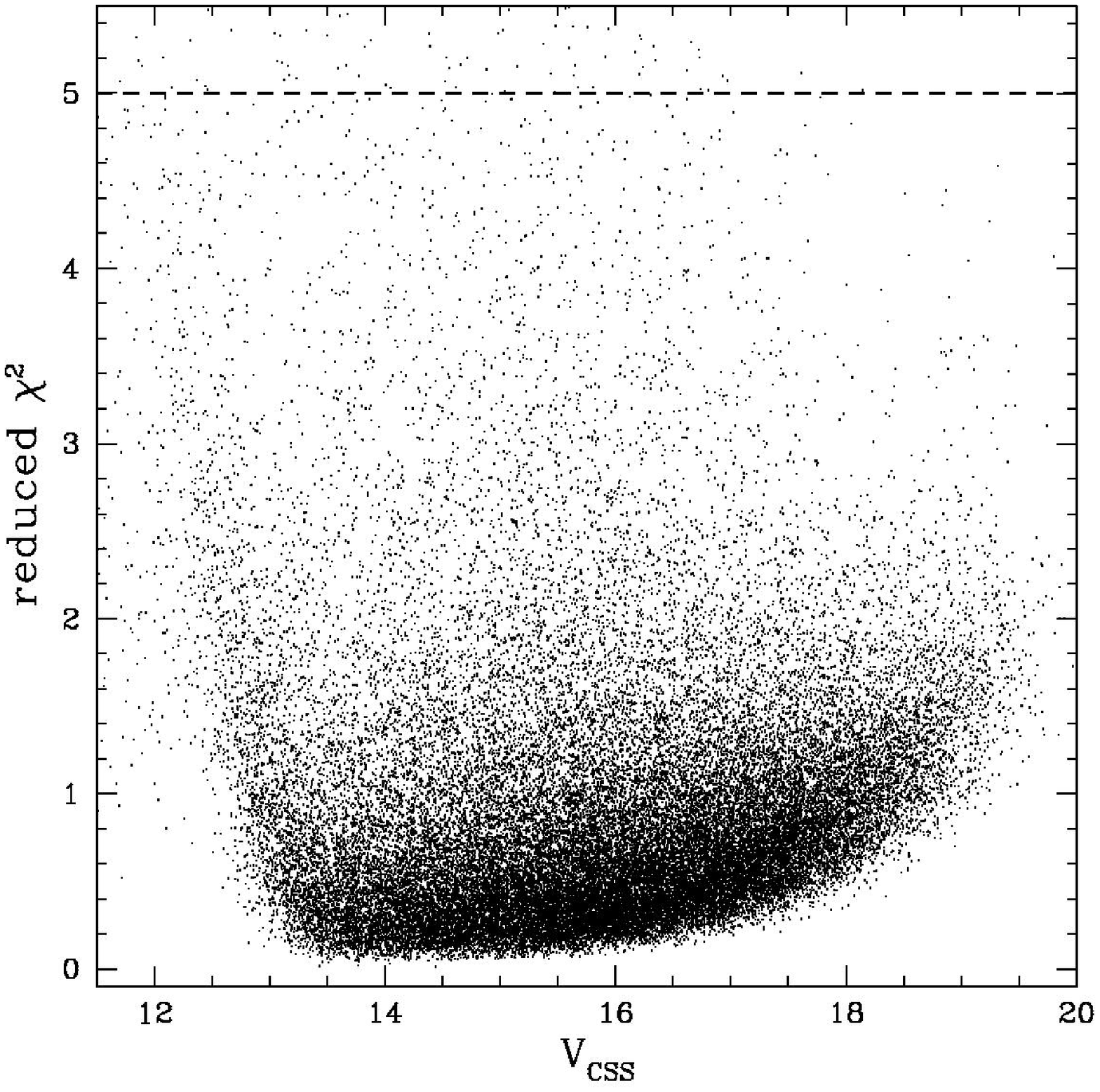}
\caption{\label{Chi}
The Fourier-fit $\chi^2_r$ values for periodic variable candidates.
In the left panel we plot the distribution for the 
128,000 periodic candidates before imposing the $\chi^2_r$
selection. In the right panel we plot the distribution
for the periodic sources in the final catalog.
}
}
\end{figure}

In Figure \ref{Chi}, we present the distribution of AFD fit $\chi^2_r$ values for the initial periodic candidates and
the final catalog sources. As expected, these figures show that most of the remaining periodic sources are moderately
good fits with $\chi^2_r < 1.5$. Most of the faint sources with large $\chi^2_r$ values have been removed.
However, many of the sources with $V_{CSS} < 13$, where saturation 
increases the value of $\chi^2_r$, were found to be periodic.
In addition, we see that most of the remaining sources in the range $14 < V_{CSS} < 18$ have $\chi^2_r$ values 
significantly less than one. Low values of $\chi^2_r$ can be caused by fitting too many parameters to the 
data, or inaccurate estimates of uncertainty. In this case, these low values are due to the systematic 
overestimation of the CSDR1 error-bars. For the brightest sources the average $\chi^2_r$ is approximately 0.3,
suggesting that the actual photometric uncertainties are $\sim 0.5$ the CSDR1 values. The fact that the CSDR1 errors
are overestimated was demonstrated by Palaversa et al.~(2013). They compared CSDR2 light curves with those from 
LINEAR. They found that the Catalina error-bars were larger, yet the objects phased with the correct periods 
generally show less scatter. This is also demonstrated by the low average values of $J_{WS}$. The overestimated
photometric uncertainties do not affect the results in this analysis.

\begin{figure}[ht]{
\epsscale{0.8}
\plotone{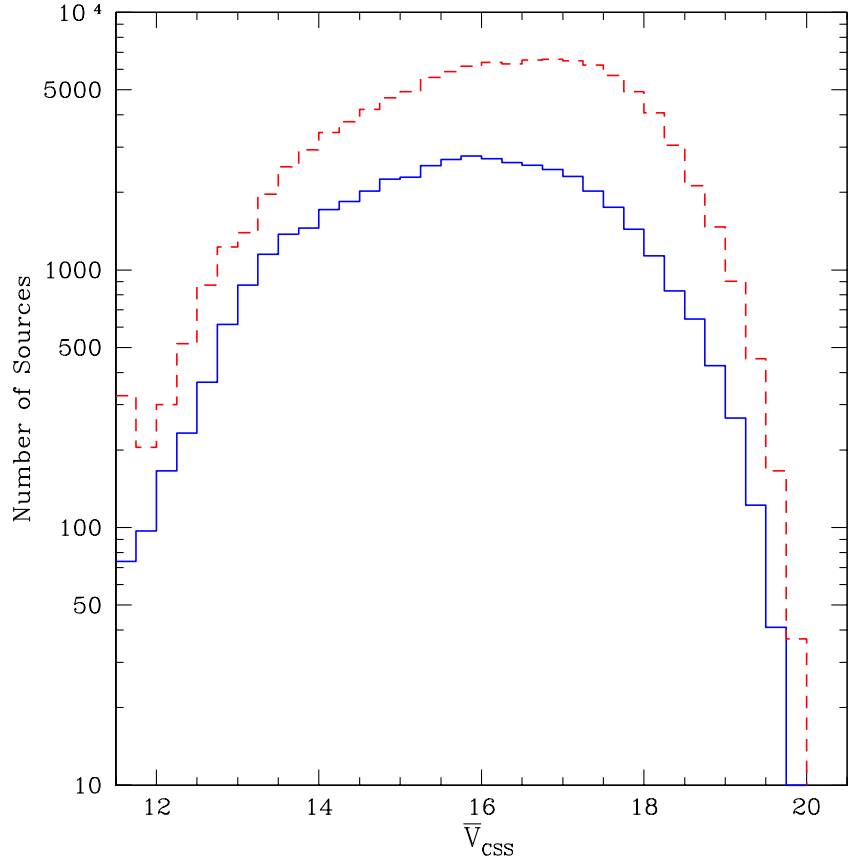}
\caption{\label{Mag}
The distribution of average CSS magnitudes. The red dashed line shows the magnitude distribution
of variables candidates before inspection. The blue solid line shows the distribution for objects 
in the periodic variable catalog.
}
}
\end{figure}

\begin{figure*}[ht]{
\epsscale{1.0}
\plottwo{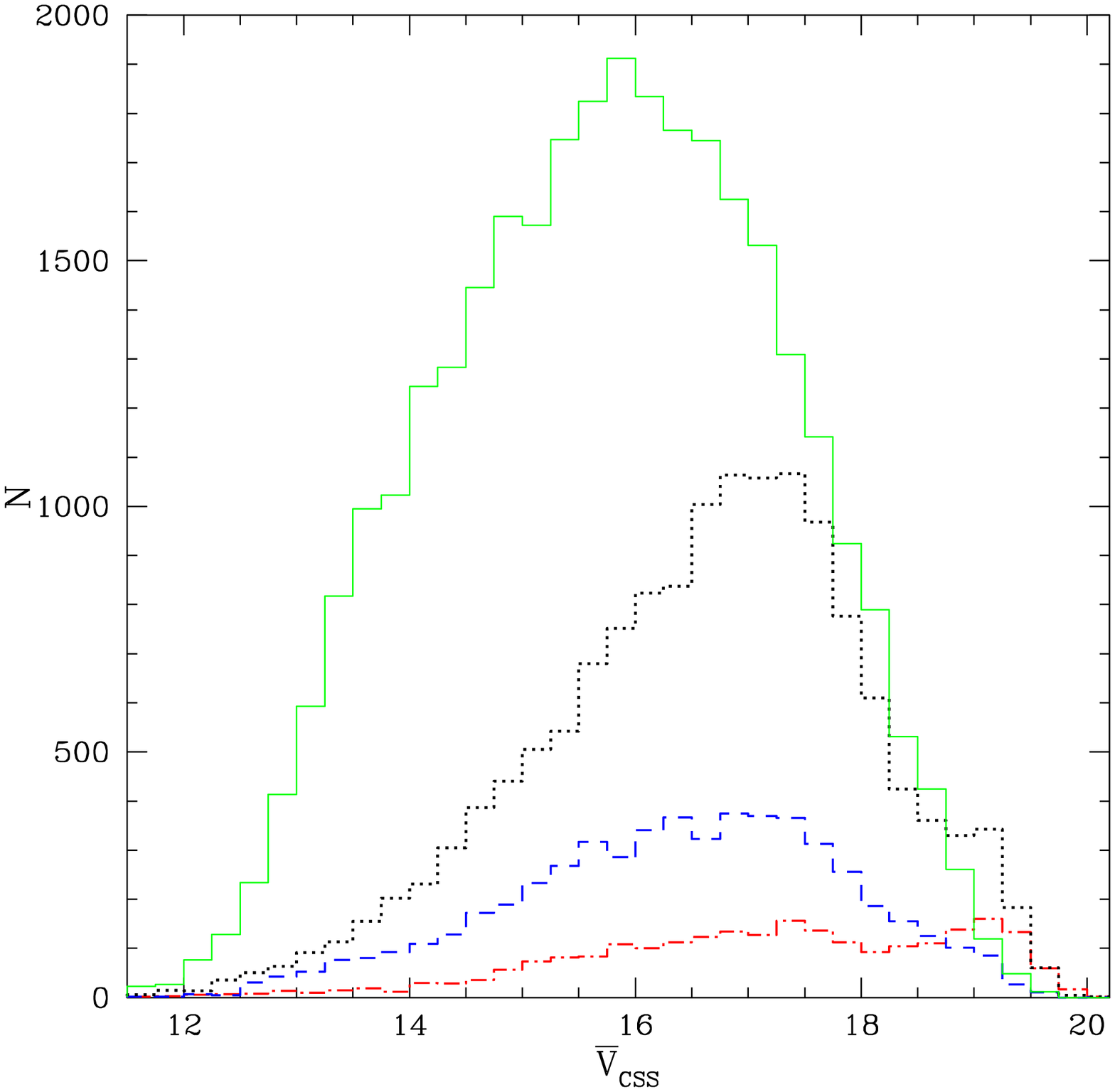}{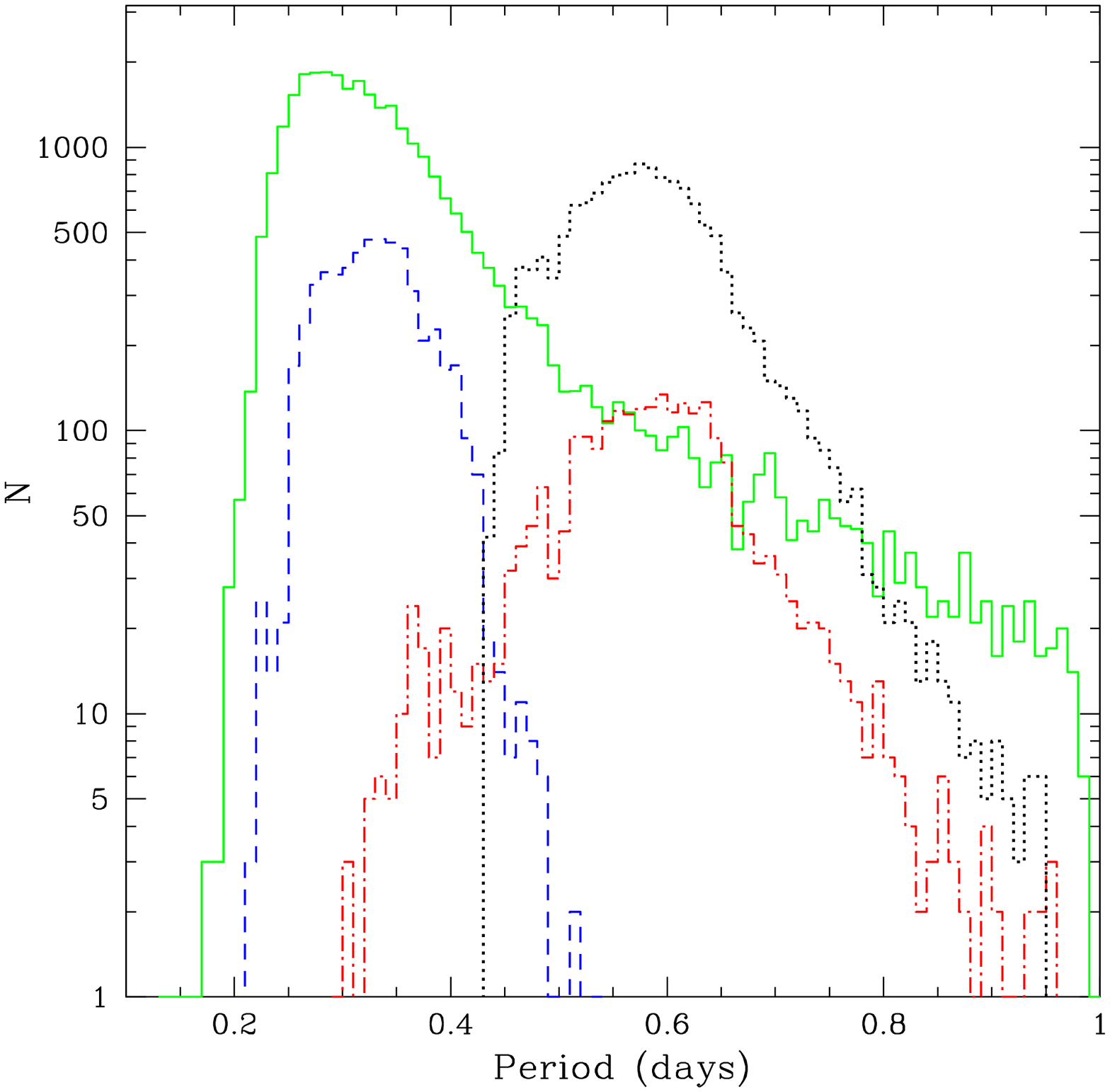}
\caption{\label{PerHist}
The distribution of the main types of periodic stars.
In the left panel we plot the magnitude distribution and in
the right we plot the period distribution.
The solid green line presents the eclipsing binaries. 
short-dashed blue line presents RRc's. The dot-dashed red
line presents RRab's from this analysis, while for comparison, 
the dotted black line gives RRab's from Drake et al.~(2013a,b).
}
}
\end{figure*}

In Figure \ref{Mag}, we compare the distribution of mean magnitudes for the initial periodic variable candidates with
the final catalog variables. The largest number of periodic sources are found near $V=16$. This result is in 
contrast to the peak of the RRab magnitude distribution from Drake et al.~(2013a), which occurred at $V=17$. 
To demonstrate the reason for this difference, we separate the main types of periodic 
variables and plot them in Figure \ref{PerHist}. Here we see that the contact eclipsing binaries dominate the number 
of periodic variables, and on average they are brighter than the RR Lyrae. Unlike the RR Lyrae, eclipsing binaries are 
mostly main-sequence stars with spectroscopic types earlier than $M$ (Norten et al.~2011). These systems generally 
lie within the Galactic disk. Hence, since CSS observations are concentrated away from the disk, only foreground 
eclipsing binaries are detected. Figure \ref{PerHist} also presents the period distributions for the main types
of periodic variables detected. It is evident that Drake et al.~(2013a) missed a small number of RRab's that  
have periods similar to RRc's and RRd's.

\section{Types of Variables}

Each of the periodic variables found among the candidates were classified into three broad classes: Eclipsing, Pulsating
and Rotational. The pulsational objects can be divided into $\delta$ Scutis, RR Lyrae, Mira and semi-regular variables,
and Cepheids. Among these we further divide the $\delta$ Scutis into those with amplitudes $V > 0.1$ (high-amplitude
$\delta$ Scutis, HADS), and those with $V < 0.1$ (low-amplitude $\delta$ Scutis, LADS) based on Alcock et al.~(2000c).
The RR Lyrae class consists of RRab's (fundamental mode), RRc's (first overtone mode), RRd's (multi-mode) and Blazkho
(long-term modulation; Blazkho 1907) types.  The Cepheid class include classical (type-I) Cepheids, type-II Cepheids (Cep-II) and Anomalous
Cepheids (ACEPs).  However, we did not find any clear classical Cepheids. We include both semi-regular variables and
Mira variables under a single, LPV classification.

The eclipsing variables in our data were generally divided into a contact and semi-detached binary (EW/EB) 
group and detached systems (EA) following Palaversa et al.~(2013). However, during a review of all the EA's 
we further divided these systems into semi-detached $\beta$ Lyrae (EB) systems and true Algol (EA) variables.
Our EB group is thus incomplete, with those systems that closely resembling contact systems being placed in 
the original EW/EB group.
We found a number of EA variables where it was not possible to determine the period due to an insufficient
number of eclipses. These objects were placed in an unknown-period ($\rm EA_{UP}$) class. We also discovered 
many eclipsing white dwarf and subdwarf binary systems. These are included in a post-common-envelope (PCEB) 
class. A small group of periodic variables exhibiting distorted light curves and varying minima and maxima
was discovered. As the nature of these sources is unclear we place them in the ``Hump'' variable class 
(see section \ref{Misc}).

The variable stars that we discovered from the rotating variable class include ellipsoid variables 
(ELL) and spotted (RS CVn) systems. Our inspection also led us to retain 4808 systems in a periodic 
candidates class (PCANDs). The light curves of these objects appeared to exhibit regular variations, yet their 
best-fit periods were clearly incorrect.  Since it is uncertain whether these objects are truly periodic 
sources, they are not included in the catalog. Based on the results of Graham et al.~(2013), it was 
expected that there would be some variables where the correct period could not be found, even 
though we applied multiple search techniques.

\begin{figure}[ht]{
\epsscale{0.9}
\plotone{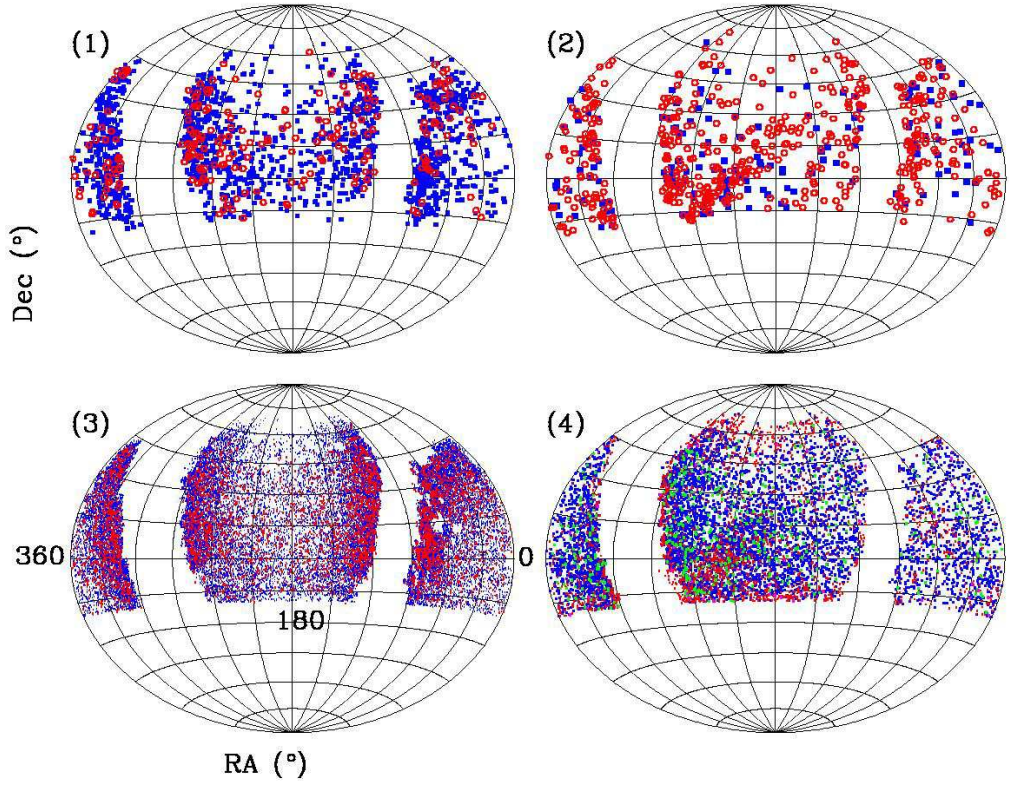}
\caption{\label{Ait}
The distribution of types of periodic variables in the CSS catalog.
In panel (1) we plot the distribution of rotational variables (blue) 
and $\delta$ Scutis (red). In panel (2) we plot LPV's (red) and Cepheid
variables (blue). In panel (3) we plot contact eclipsing binaries (blue) 
and detached binaries (red) and in panel (4) we plot RRab (red), RRc's
(blue) and RRd's (green).
}
}
\end{figure}

In Figure \ref{Ait}, we plot the sky distribution for each of the classes of periodic variables having more than 
100 members.  As expected, the eclipsing binaries are concentrated at low Galactic latitudes, while the
halo stars, such as RR Lyrae, are much more uniformly distributed.

In Table \ref{Vtype}, we present the number of periodic sources from each class.  For completeness 
and the accuracy of the fractions present in the data we have include the CSS RRab's presented by 
Drake et al.~(2013a,b). In contrast to our results, Palaversa et al.~(2013) find a smaller fraction 
of eclipsing binaries than RR Lyrae. However, as Palaversa et al.~(2013) only analysed LINEAR sources 
within the SDSS footprint region, their analysis is limited to sources at much higher Galactic 
latitudes. Considering this, the distribution of variables appear consistent in both studies.

In Table \ref{Full}, we present the parameters of all the periodic variable sources.
In the following sections we will discuss and give examples of sources from the main 
classes and subclasses and outline how the objects were separated.

\subsection{RRc's vs Contact binaries}

Determination of the correct classes of variables can be very important for studies of stellar populations. For example,
the accurate classification of RR Lyrae is necessary when they are used to trace structure within the Galactic halo.
However, the potential for misclassifying contact binaries as RRc's can limit the accuracy of results derived from
uncertain classifications (Kinman \& Brown 2010).  In Drake et al.~(2013a) we outlined how it was possible to separate
RRab's from contact eclipsing binaries based on a modified version of the so-called M-test (Kinemuchi et al.~2006, their
eq. 8) along with parameters such as the order of the Fourier fit. However, in Drake et al.~(2013a,b) we specifically
neglected RRc's and RRd's because of potential contamination.  Examples of misidentifications were illustrated by Kinman
\& Brown (2010) based on RR Lyrae presented by Akerlof et al.~(2000).  Here we will investigate the extent of the
potential eclipsing binary-RRc misclassification problem within our periodic variable catalog.

\begin{figure}[ht]{
\epsscale{0.7}
\plotone{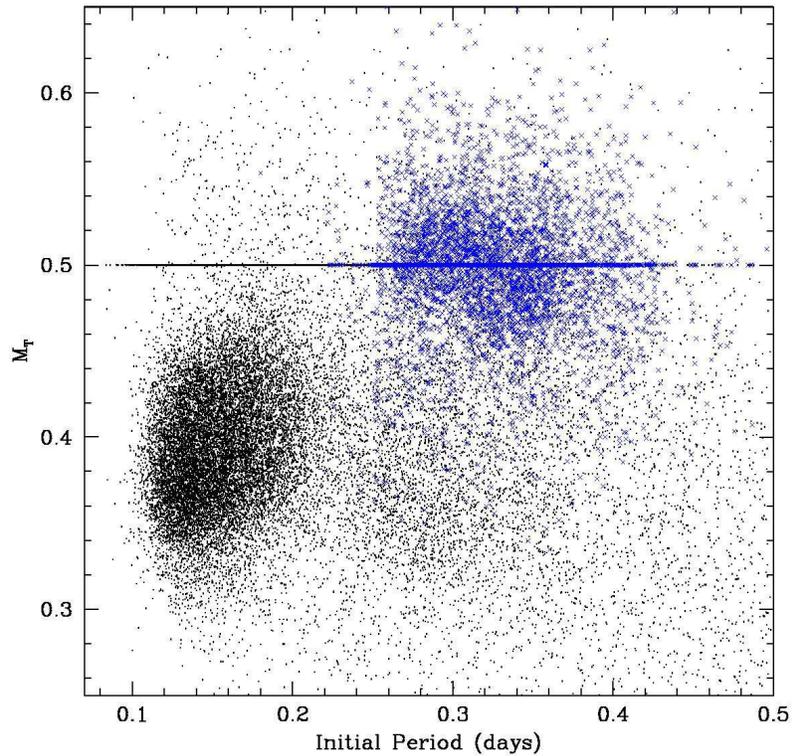}
\caption{\label{Mtest}
M-test statistic values for contact eclipsing binaries and RRc's as a function
of period. The RRc's are marked as blue crosses at their observed periods while 
the eclisping binaries are given as black dots at half their observed periods.
}
}
\end{figure}

As with Drake et al.~(2013a) we start with investigating how M-test statistic values ($M_t$) 
vary between contact binaries and RRc's. In Figure \ref{Mtest}, we plot the $M_t$
for these sources based on our catalog. It is immediately clear that there are two separate
groups. Here we have plotted the eclipsing binaries at half their true periods since this 
is the period which most were initally discovered with. 
The objects we classify as eclipsing binaries are strongly grouped at half-periods of $0.1 < P_F/2(d) < 0.22$ 
and M-test values $0.34 < M_t < 0.46$, while RRc's are concentrated at $0.24 < P_F(d) < 0.42$ 
and $0.45 < M_t < 0.55$. 

As with the Drake et al.~(2013a) RRab's, the $M_t$ values of RRc's vary with period 
due to changes in their light curve morphology. The difference between the $M_t$ values 
of the eclipsing binaries and RRc's is indicative of real morphological differences 
in the light curves. The contact binaries spend less time below their average brightness
than RRc's that are closer to sinusoidal. However, apart from the two main clumps of 
variables, there are still many binaries with $M_t$ values and periods that overlap 
the RRc's.

\begin{figure}[ht]{
\epsscale{0.7}
\plotone{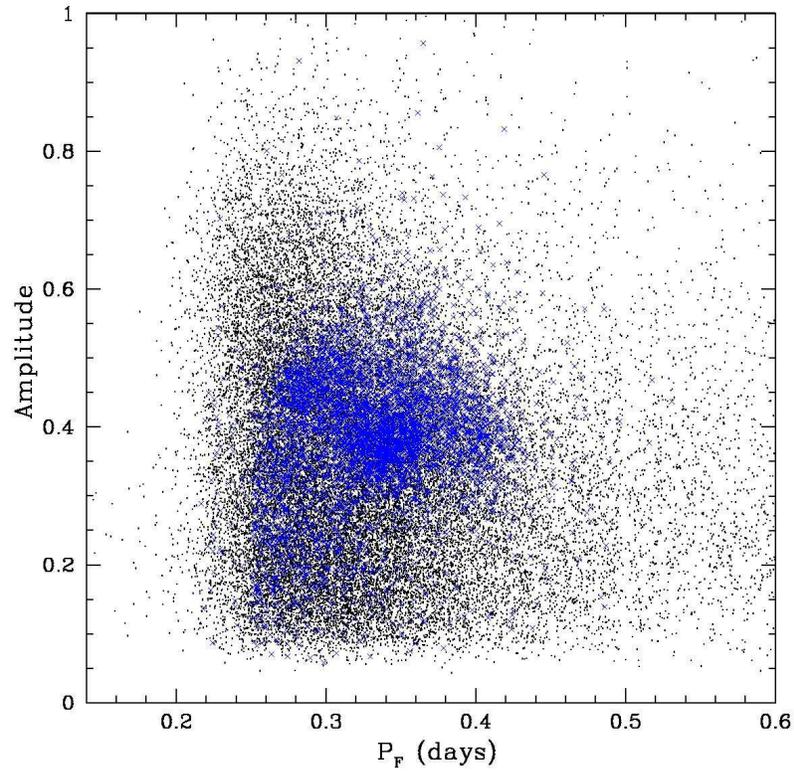}
\caption{\label{PerAmp}
The period-amplitude distribution for eclipsing binaries and RRc's. 
The symbols follow those in Figure \ref{Mtest}.
}
}
\end{figure}

Another way of distinguishing RRc's and binaries is by their amplitudes.  In Figure \ref{PerAmp}, we plot the
distribution of contact binaries and RRc's at their final periods. As noted earlier, most of the contact binaries were
initially found at half-periods. Also, the contact binaries have amplitudes $A_V < 0.8$, while the RRc's are generally
concentrated near $A_V \sim 0.4$. When RRc's and contact binaries are plotted with their true periods (Figure
\ref{PerAmp}) they completely overlap.  However, in actuality the difference between these two classes of
sources is clear from the light curves since the eclipsing binaries have two cycles per orbital period, while the RRc's
have only one.
The misidentification of contact binaries as RRc's should only occur for objects with amplitudes in 
range $0.3 < A_V < 0.5$, and periods $0.44 < P(d) < 0.82$. From Figure \ref{PerHist} it is clear that only a 
small fraction of contact binaries and RRc's have such periods. Nevertheless, since there are large very 
numbers of contact binaries, it is worth considering additional information.

\subsubsection{Additional Information from SDSS and WISE}

The Wide-field Infrared Survey Explorer (WISE, Wright et al.~2010) provides mid-IR data for sources across the entire
sky.  To further investigate the level of misclassification among the eclipsing binaries and RRc's, we matched all of
our periodic sources with the WISE catalog.  Of our $112,\!000$ initial candidates, $103,\!000$ had WISE matches with
$w1$-band ($3.4\mu m$) data within $3\arcsec$, and $43,\!000$ for sources from our final catalog ($\sim 95\%$).  For
each of the contact binaries and RRc's with WISE matches we determine their $V_{CSS} - w1$ colours.

\begin{figure}[ht]{
\epsscale{0.7}
\plotone{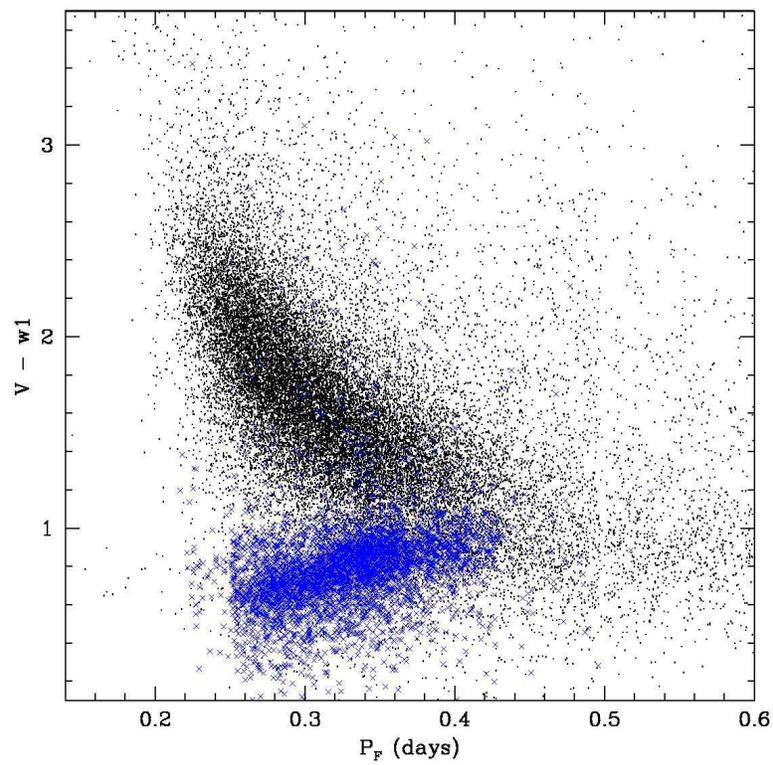}
\caption{\label{WISE}
The period-colour distribution of RRc's and contact 
binaries from WISE and CSS photometry.
The symbols follow those in Figure \ref{Mtest}.
}
}
\end{figure}

In Figure \ref{WISE}, we plot the colours and periods of the contact binaries and RRc's. The results demonstrate
the significant colour variation between contact binaries and RRc's in addition to the clear colour evolution 
with period.  The combination of $w1 - V_{CSS}$ colours and period thus provides an excellent means of separating 
relatively red contact binaries from bluer RRc's. However, at long periods, contact binaries still have similar 
colours to RRc's.

Another means of separating eclipsing binaries from RRc's is via multi-band optical photometry. We matched the initial
periodic candidates with photometry from SDSS Data Release 10 (SDSS-DR10, Ahn et al.~2013). In this case we found 
only $68,\!000$ matches since SDSS images cover much less area than WISE or CSS. However, the SDSS data has significantly 
better depth and resolution than CSS.

\begin{figure}[ht]{
\epsscale{1.0}
\plottwo{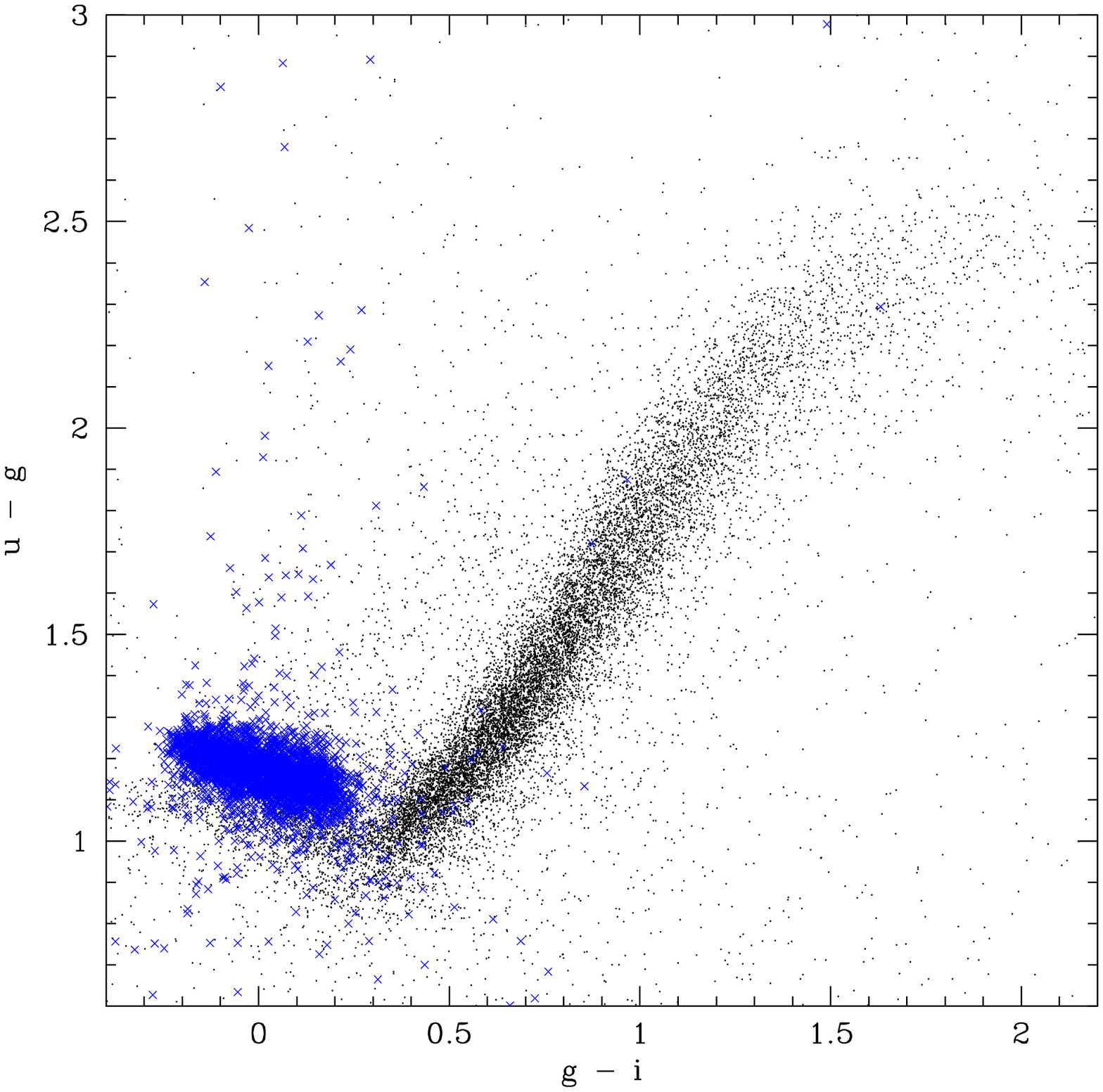}{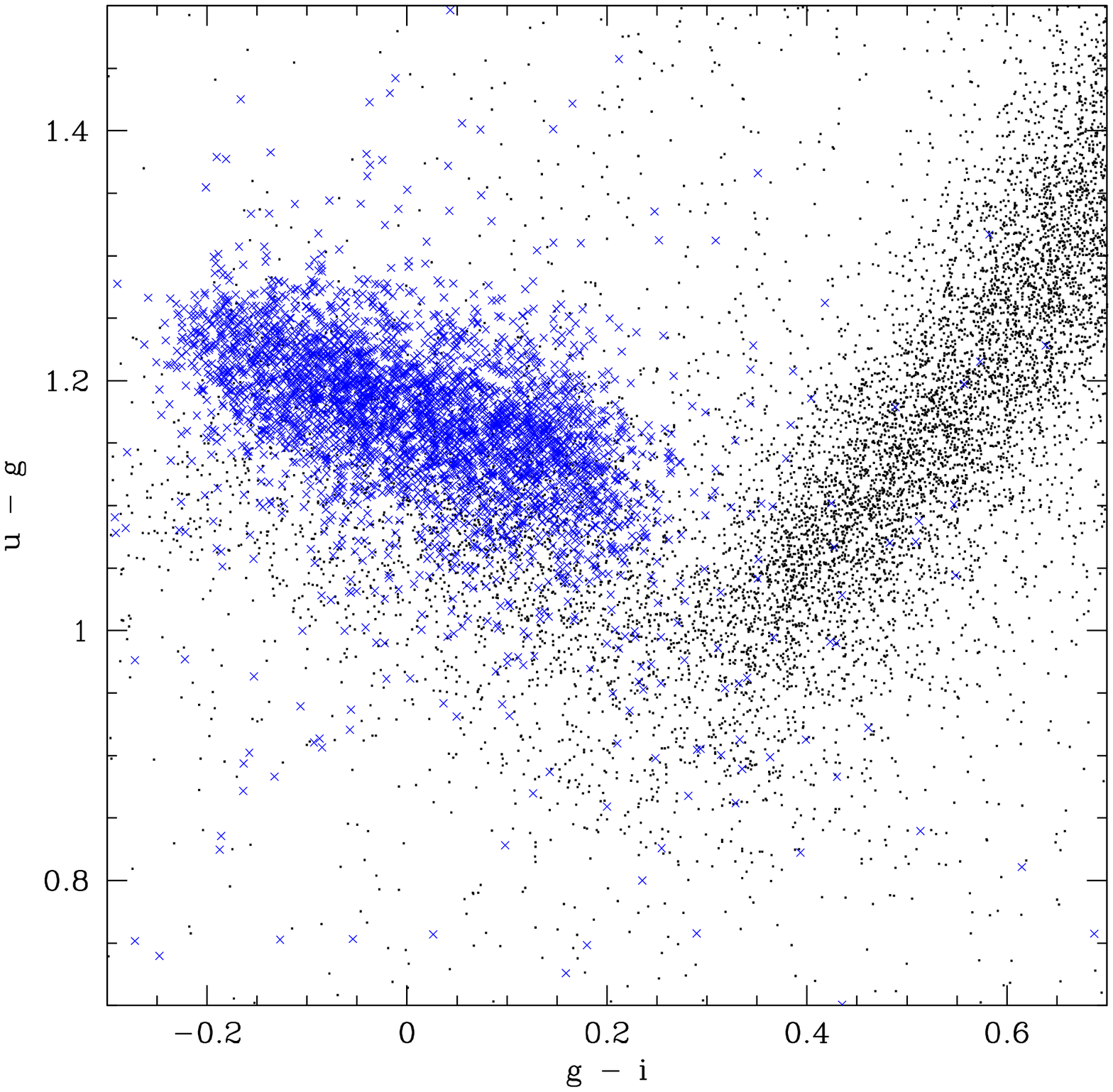}
\caption{\label{SDSS}
The colours of type-c RR Lyrae and eclipsing binaries from SDSS DR10 photometry. 
In the right panel presents the colour range of these eclipsing binaries. 
The left panel presents an expanded view of the horizontal branch region.
The symbols follow those in Figure \ref{Mtest}.
}
}
\end{figure}

In Figure \ref{SDSS}, we plot the SDSS $u-g$ and $g-i$ colours for contact binaries and RRc's. Once again the sources
are quite well separated since the contact binaries are mainly main-sequence stars, while RRc's are horizontal branch (HB) 
stars. There still are contact binaries with similar $g-i$ colours to the RRc's. However, on average, these sources are offset 
from RRc's by $\sim 0.1$ mags in their $u-g$ colours. Thus both WISE and SDSS provide a mean of separating the two
types of sources.

As a final test, we matched the RRc's and contact binaries with spectra from SDSS-DR10 (Ahn et al.~2013). 
We found $\sim 6000$ of the periodic variables had SDSS spectra. The SDSS analysis pipeline produces
calibrated values of abundances, surface gravity and radial velocity for all spectra with sufficient
signal-to-noise. As RRc's are halo giants they are expected to generally have low metallicities ($[Fe/H] < -1$) 
and surface gravities, in addition to a higher velocity dispersion than disk stars.

\begin{figure}[ht]{
\epsscale{0.7}
\plotone{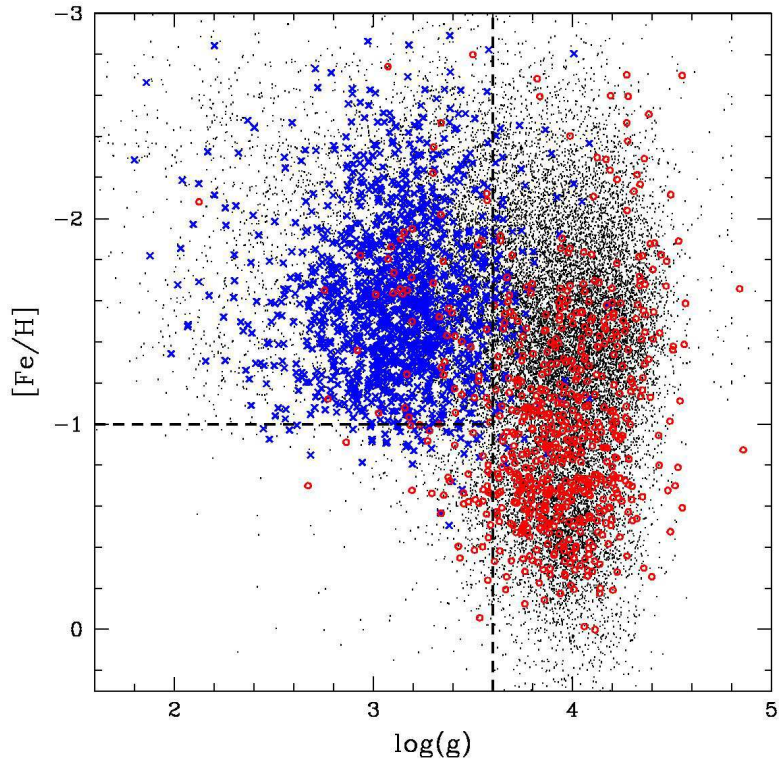}
\caption{\label{SDSSlgfe}
The distribution of surface gravity and metallicity for RRc's and contact binaries
based on SDSS DR10 spectra. The blue crosses present the values for RRc's.
The red circles plot the values for contact binaries and the black points
plot the values for ten thousand randomly selected A-type stars.
}
}
\end{figure}

In Figure \ref{SDSSlgfe}, we plot the surface gravity and metallicity measurements for the two types of variables along
with a sample of $10,000$ A-type stars (which includes both main sequence and HB stars). The low-metallicity,
low-surface gravity RRc's stars are well separated from the main-sequence contact binaries.

\begin{figure}[ht]{
\epsscale{0.7}
\plotone{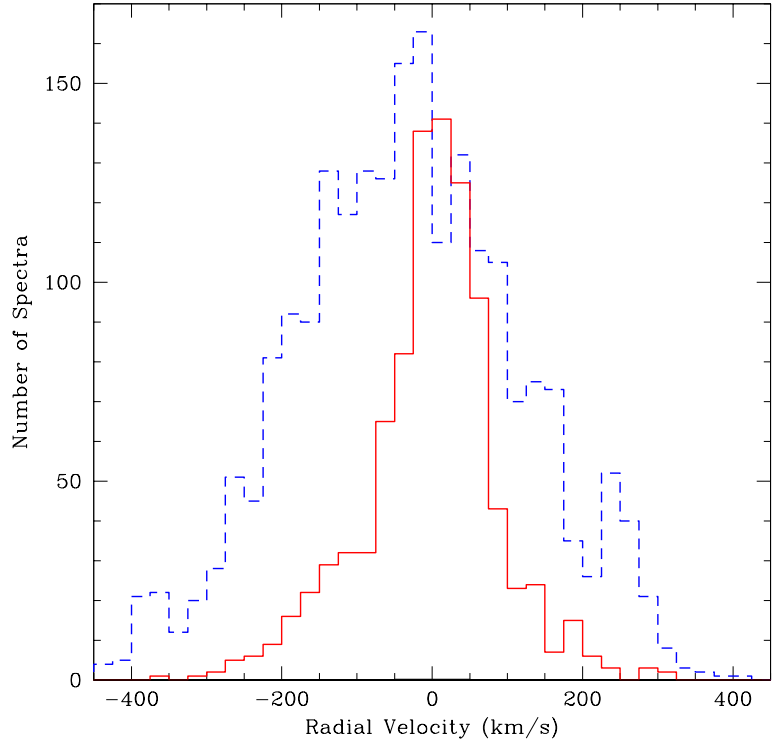}
\caption{\label{SDSSvel}
The distribution of radial velocities for RRc's and contact binaries from SDSS 
DR10 spectra. The solid shows the distribution for contact binaries while
the dashed line presents values for RRc's.
}
}
\end{figure}

In Figure \ref{SDSSvel}, we plot the Galactocentric radial velocities for the two groups. The velocity 
dispersion of the RRc's is clearly much larger than for the eclipsing binaries. 

We investigated all the sources that we had classified as RRc's and contact binaries based on light curve morphology,
which had colours or other information suggesting membership of the other class.  In cases where SDSS imaging was
available, we viewed the images and discovered that unusual colours or spectroscopic values were skewed by the presence
of blended sources. In almost all cases the original classification was not changed.

Overall, by investigating results from differences in light curve morphology (M-test), the period distribution, the
amplitude distribution, the optical and IR colours, log(g), [Fe/H] and velocities, we find that the number of 
RRc's that are likely to be misidentified contact binaries is only of order 1\%. The level of contamination is low due
to three factors. Firstly, most of the contact binaries in our data are brighter sources. The peak of the distribution
being at $V \sim 16$. As they are bright, their light curves are well sampled and they usually have sufficient signal to 
identify type by morphology. Secondly, as we noted above, only the long-period contact binaries can masquerade as RRc's 
and there are only a small fraction of long period contact binaries in our data. Lastly, the long-period contact binaries
that have blue colours similar to RRc's are the brightest contact binary systems. These sources have much higher S/N
than the bulk of the binaries which have short periods and moderately red colours.

\subsection{Eclipsing binaries}

Eclipsing binaries offer the opportunity of determining stellar parameters with a high degree of accuracy using
constraints on the geometry of the system (Southworth 2012 and refs therein). Among other things, eclipsing binaries can
provide a direct measurement of the radius of each star in the system if the period, inclination, and radial velocity of
each star is known.  Under the right circumstances eclipsing binaries can also be used as standard candles (e.g. 
Pietrzynski et al.~2013).  Eclipsing binaries include contact, semi-detached and detached systems.

\subsubsection{Contact binaries}

\begin{figure}[ht]{
\epsscale{0.7}
\plotone{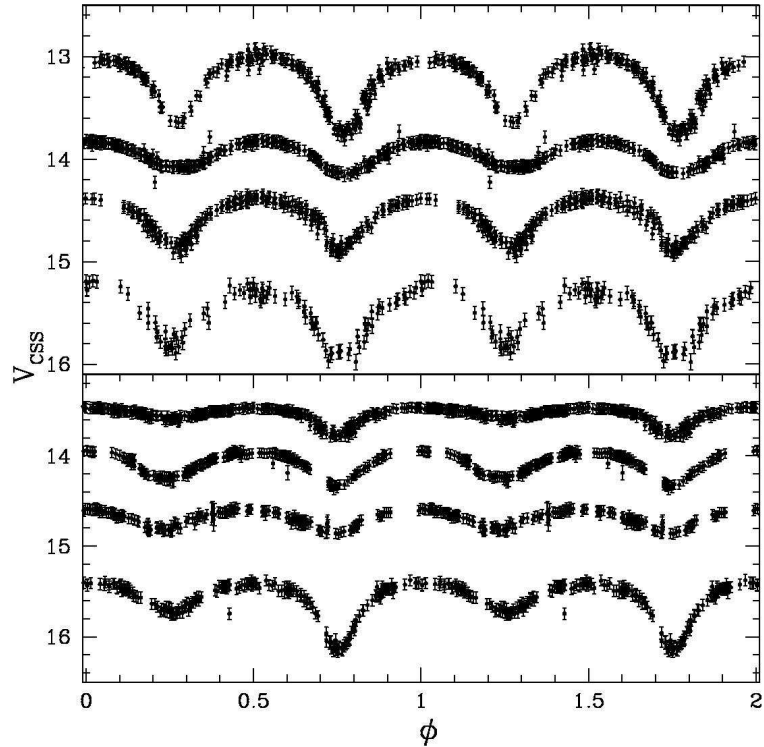}
\caption{\label{contact}
Examples of contact eclipsing binary light curves.
In the top panel we plot systems with similar primary
and secondary eclipse depths. In the lower panel we plot
systems with varying eclipse depths.
}
}
\end{figure}

Contact binary systems occur when both components of the binary fill their Roche lobes. 
Eclipsing contact binaries are referred to as W Ursae Majoris (W UMa's) stars, or EW's.
Since these systems are in contact mass flows from one star to the other and both stars usually
have similar temperatures and types. Slight differences in the eclipse depth are still possible 
and reflect remnant differences in the temperature of the component stars. In Figure \ref{contact}, 
we present an ensemble of different kinds of contact binary light curves. The top four are 
typical examples while the lower four exhibit effects of varying component temperature, degrees 
of contact, and inclination. In general contact binaries have previously been found to have a minimum 
period near 0.22 days (Rucinski et al.~1992, 1997). In our analysis we have detected a number of 
ultra-short-period eclisping binary systems below this value. These objects have been analysed and 
are present in Drake et al.~(2014).

\subsubsection{O'Connell Effect Binaries}

One of the poorly understood features of contact binary light curves are cases where the two maxima of the system have
different luminosities. Such cases are unexpected since the stars are side by side at the time of maximum. This asymmetry is
called the O'Connell effect (O'Connell 1951). Wilsey \& Beaky (2009) reviewed this problem and noted that there are three
possible causes: star spots, gas stream impacts, and circumstellar matter.  In the star spot model, chromospheric and
magnetic activity lead to the production of stars spots on the surface of at least one of the stars. In this model one
expects that the size of the star spots to evolve as they do with RS CVn binary systems.

\begin{figure}[ht]{
\epsscale{0.7}
\plotone{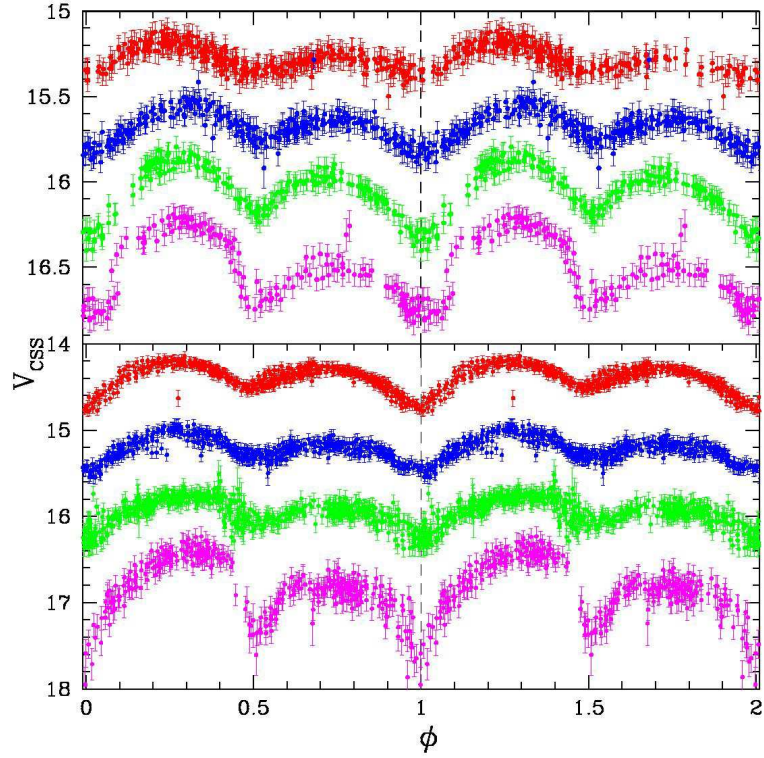}
\caption{\label{ocon}
Examples of contact eclipsing binary light curves presenting the O'Connell effect.
}
}
\end{figure}

In Figure \ref{ocon}, we plot the light curves of systems that exhibit the O'Connell effect.  Our data shows that there
is a significant diversity among these binaries. Furthermore, we see no evidence for changes in the maxima that are
expected as star spot numbers or sizes vary.  Since the CSDR1 data is taken over a baseline of thousands of days this 
suggests a cause for the O'Connell effect other than star spots. This is in agreement with the findings of Wilsey \& Beaky (2009).

\begin{figure}[ht]{
\epsscale{0.7}
\plotone{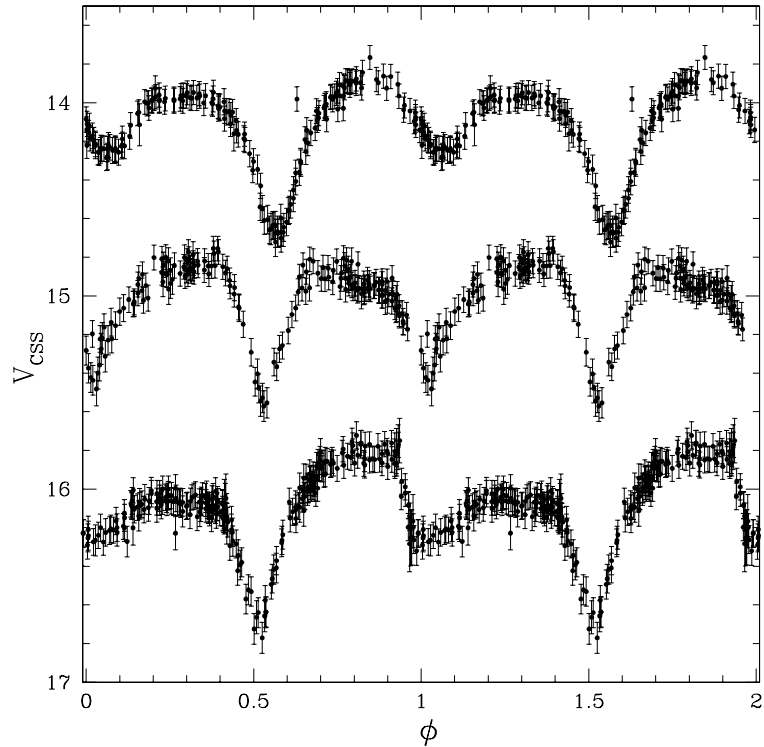}
\caption{\label{EclUn}
Examples of highly asymmetric contact binaries.
}
}
\end{figure}

In Figure \ref{EclUn}, we present the light curves of contact binaries exhibiting high levels of asymmetry. The bottom
light curve in this figure is very similar to that of V361 Lyr. Hilditch et al.~(1997) explain the light curve of V361 Lyr
as being due to the exchange of mass between two stars of significantly different mass. Among the 31 thousand
contact binary systems there are no more than a dozen of this kind, suggesting that the mass-transferring 
process must be very short-lived.

\subsubsection{Spotted Contact Binaries}

\begin{figure*}[ht]{
\epsscale{1.0}
\plottwo{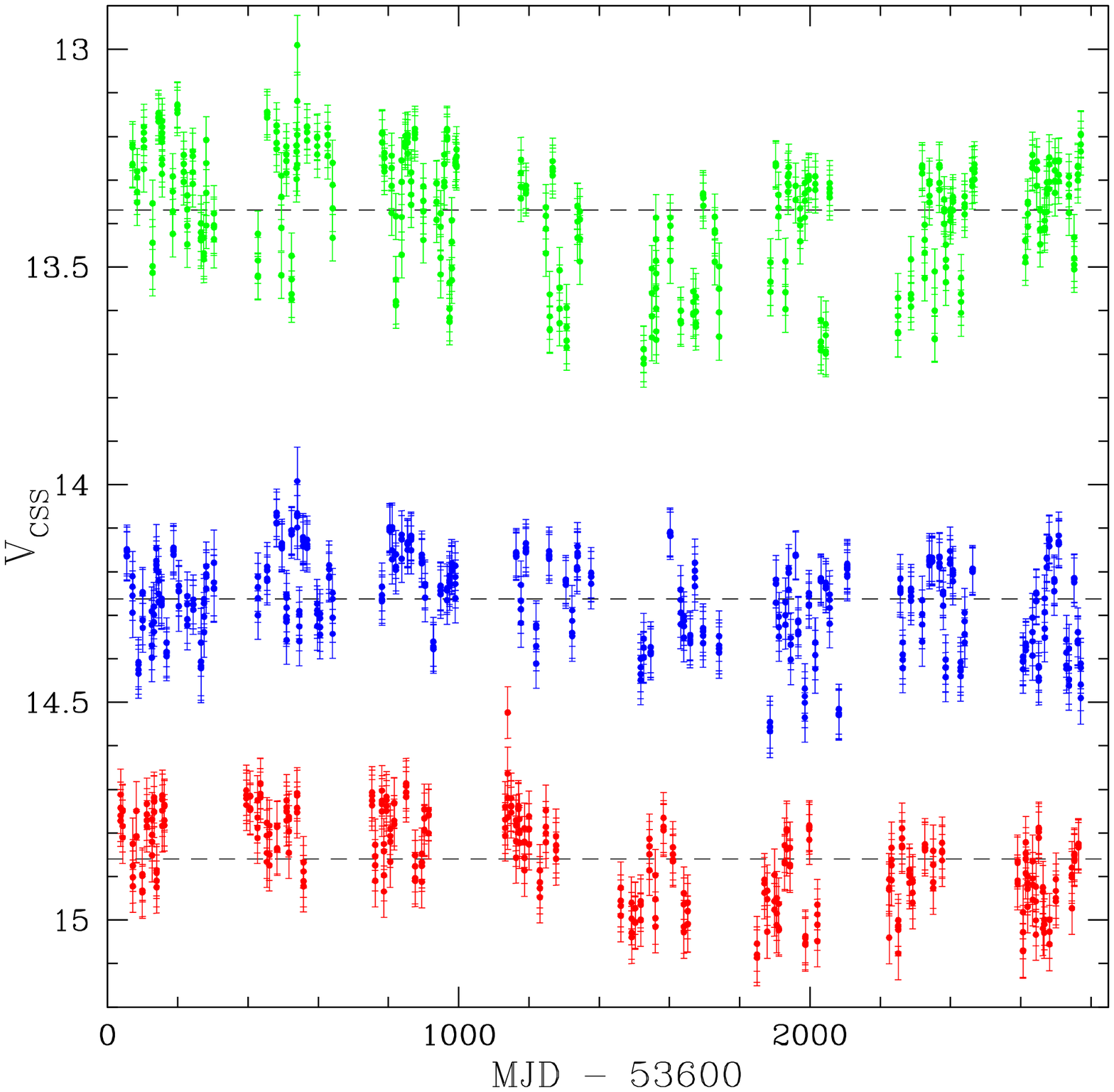}{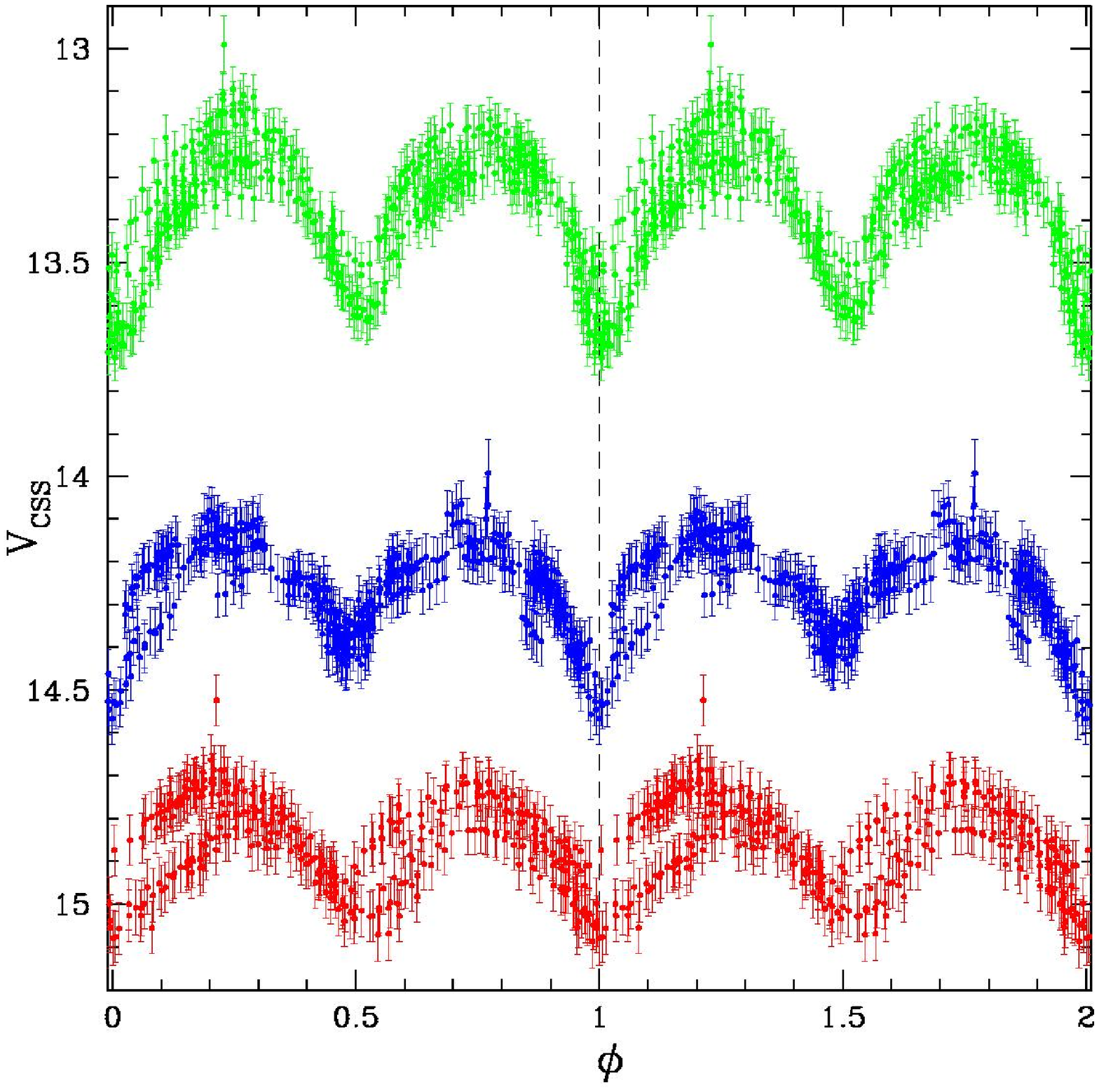}
\caption{\label{EclSpot}
Examples of spotted contact binaries.
Left panel: observed light curves of three 
spotted eclipsing binary systems. Right panel: 
Phased light curves of the same systems.
}
}
\end{figure*}

During our inspection of periodic variable light curves we noted the presence of many contact eclipsing binary systems
with varying mean brightness. In Figure \ref{EclSpot}, we present the observed and phased light curves of three of these 
systems. Large variations in average brightness are commonly seen in RS CVn systems where chromospheric activity causes 
varying levels of star spot coverage (e.g. Drake~2006). However, short-period RS CVn's ($P < 1$ day) are detached 
or semidetached binaries (Hall 1976) where the variation is due to spots or discrete eclipses. The systems observed 
here are clearly in contact. The time dependence of the light curves is strong evidence for the presence of star spots in 
these systems. However, the observed level of variation provides further evidence that the O'Connell effect systems noted 
above are due to a different effect.

\subsubsection{Semi-detached binaries}

During our inspection we separated sources with significant variations
in depth and V-shaped eclipses from the contact binaries. These sources
consist of semi-detached and detached eclipsing binaries.

Semi-detached eclipsing binaries, including $\beta$ Lyrae-type variables (EBs), 
consist of pairs of stars where one of the stars has a full Roche lobe
and the other has not. This enables the transfer of gas from the Roche 
lobe-filling star to the other.

\begin{figure}[ht]{
\epsscale{0.7}
\plotone{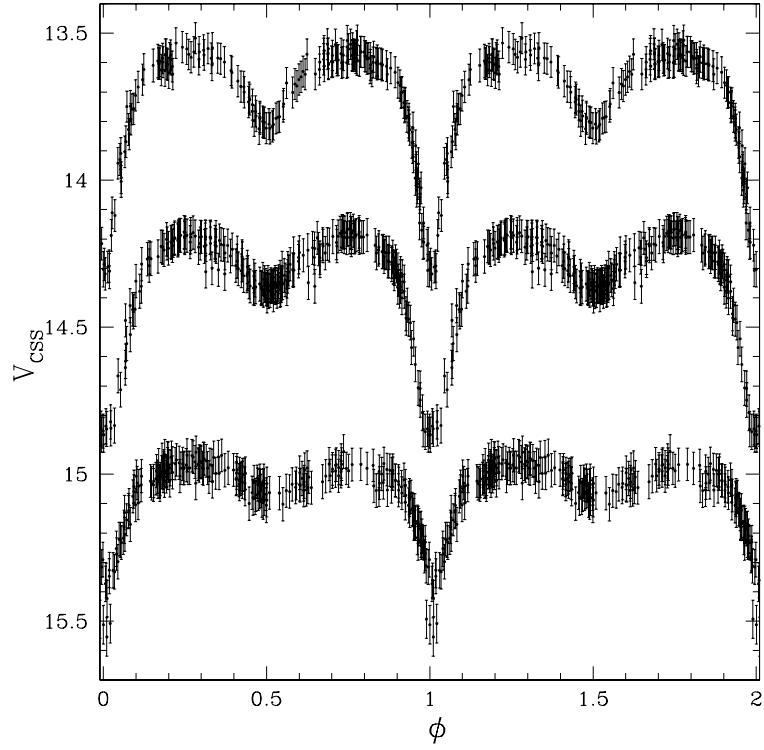}
\caption{\label{EB}
Examples of semi-detatched binary light curves.
}
}
\end{figure}

Semi-detached eclipsing variables can be distinguished by light curves that
continuously vary between eclipses due to ellipsoidal variations of the distorted
star. Unlike contact systems, the depth of the eclipses are unequal and more V-shaped. 
However, unlike detached binary systems it is not possible to distinguish the point at 
which an eclipse begins or ends. In Figure \ref{EB}, we present examples of these
objects.

After reviewing all of the detached candidates we separated the sample into semi-detached 
and detached binaries based on whether it was possible to determine the start or end 
time of the eclipses. Given the similarity of the light curves this process is
uncertain.

\subsubsection{Detached binaries}

\begin{figure}[ht]{
\epsscale{1.0}
\plottwo{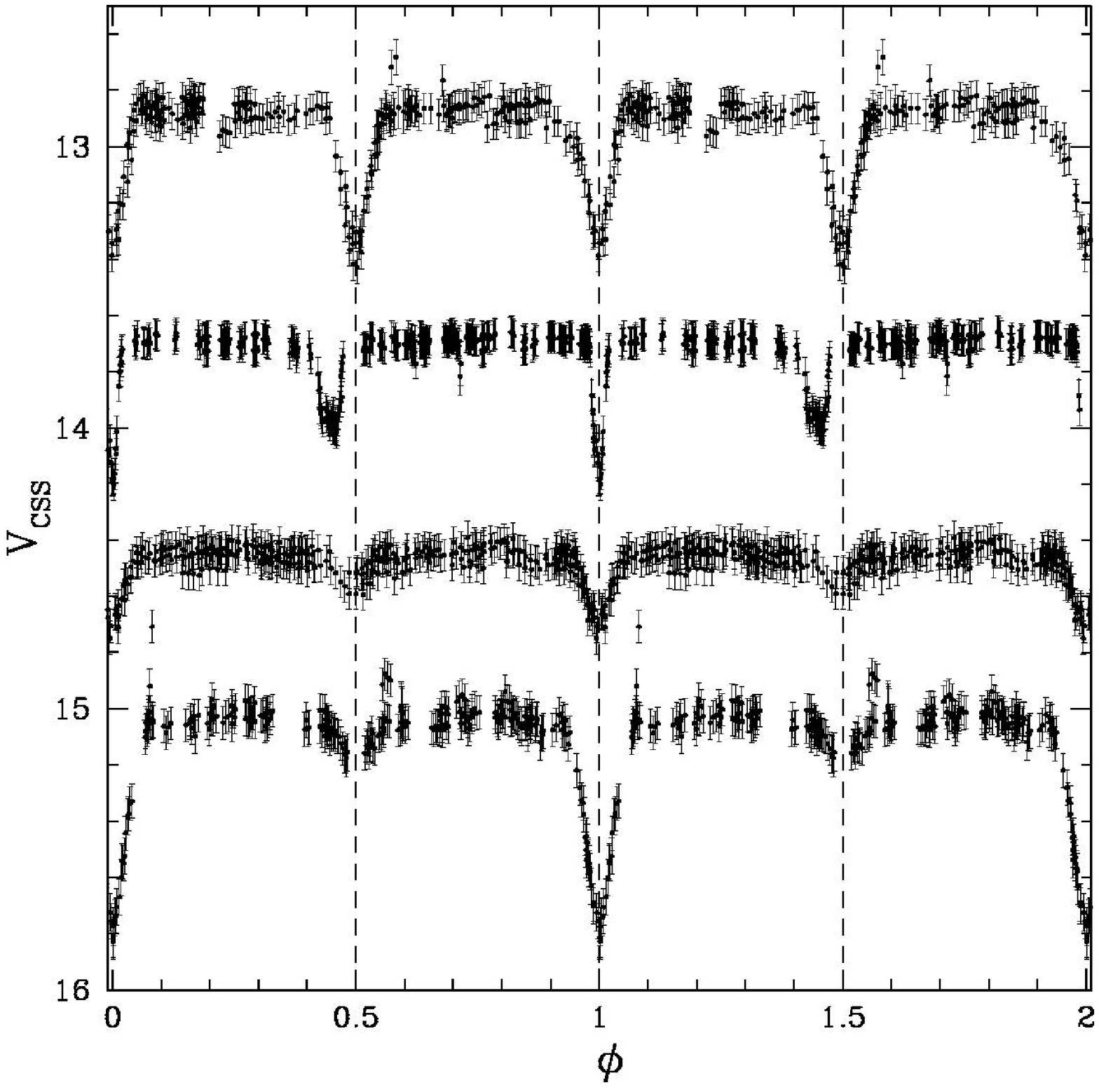}{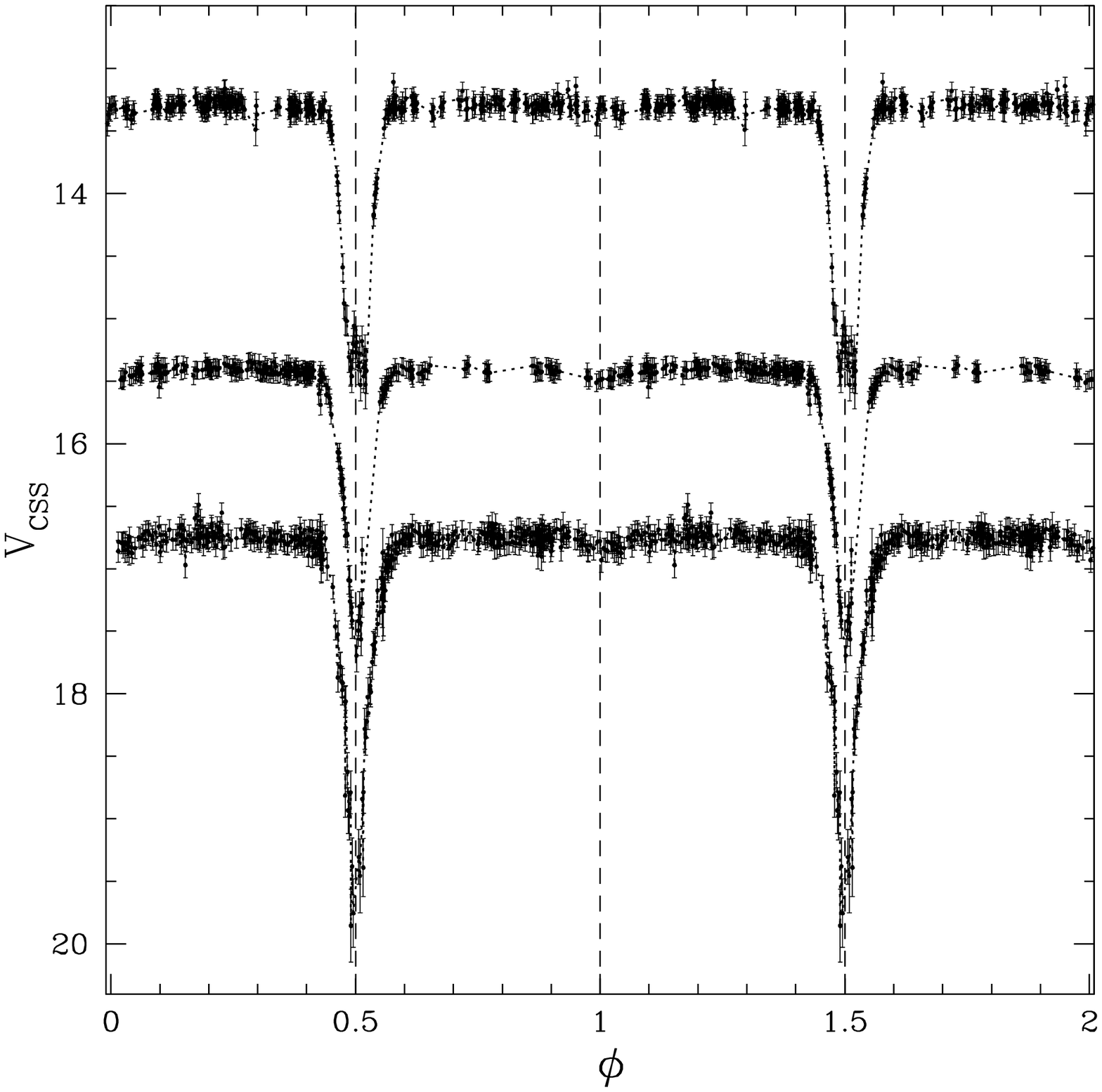}
\caption{\label{EA}
  Examples of detached eclipsing binary light curves.  In the left panel, we plot EAs where the secondary eclipse is
  clearly seen. The top most light curve is due to two stars of similar temperature. The next from the top shows a system
  where the secondary eclipse is earlier than the others due to an elliptical orbit (the phase difference between
  eclipses is not 0.5). In the right panel, we plot EAs
  where the components have a very large difference in the temperature, giving rise to high-amplitude eclipses.
}
}
\end{figure}

Detached eclipsing binaries, often noted as EAs (or Algol types), consist of two separated stars aligned closely along
our line-of-sight. Unlike contact binaries, these stars can have very different temperatures resulting in systems with
high degree of variability. EAs can also have highly elliptical orbits. In such cases, the primary and secondary
eclipses are not evenly spaced.  In Figure \ref{EA}, we present examples of EAs with eclipse depths ranging from 0.4
to 3 magnitudes. Binaries with eclipse depth greater than a magnitude result from objects of significantly different
temperatures. We denote these as deeply eclipsing systems.

\begin{figure}[ht]{
\epsscale{0.7}
\plotone{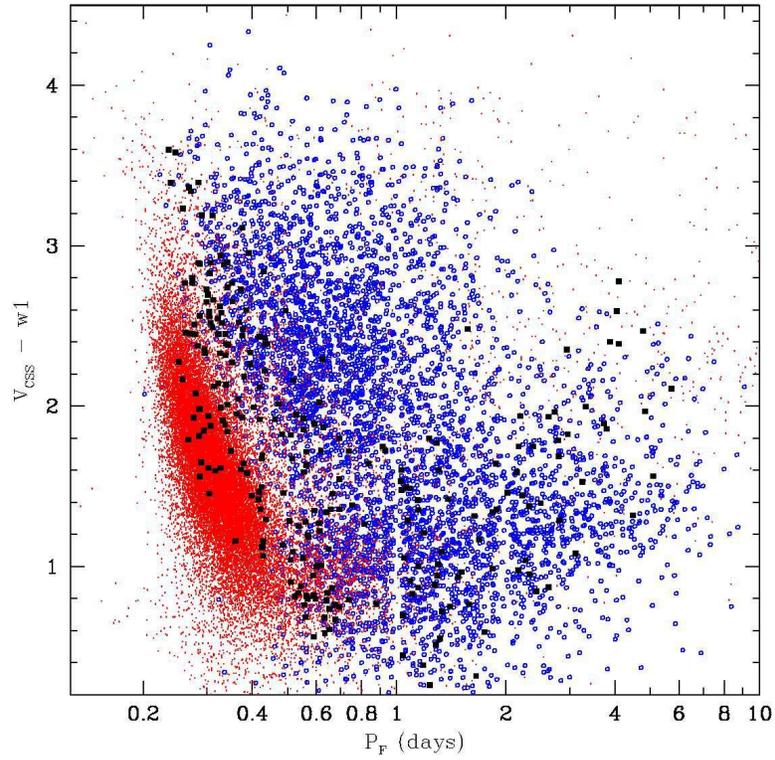}
\caption{\label{WISEecl}
The period-colour distribution of W UMa, $\beta$ Lyrae and Algol binaries 
The red points show the contact binaries. The black squares show the
$\beta$ Lyrae candidates and the blue circles show the detached binaries.
}
}
\end{figure}

In Figure \ref{WISEecl}, we plot the distribution of $V_{CSS} - w1$ colours as a function of period for EAs, EBs and
EWs.  As expected, the contact systems have the shortest periods for any given colour, while the increasingly separated
semi-contact and detached binaries have longer periods.

\subsection{Compact Eclipsing Binary Systems}

Compact binaries can often exhibit orbital periods below 0.2 days. Such binaries include systems with white dwarfs (WDs)
and subdwarfs (mainly sdB's and sdO's).  Post-common-envelope binaries (PCEBs) include WD-dM systems and are related to
interacting close binaries such as CVs.  These systems can aid our understanding of the complex common envelope
evolutionary phase in binary systems.  Subdwarf binaries (HW Vir stars) may also form through a common envelope phase.
Not all such system have to be eclipsing to be detected as binaries.

\begin{figure}[ht]{
\epsscale{0.7}
\plotone{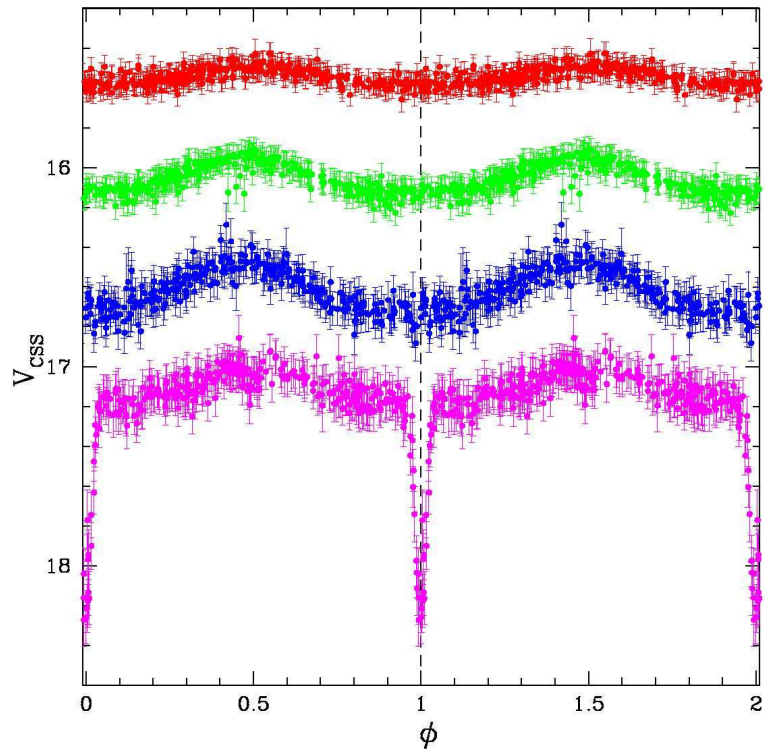}
\caption{\label{WD}
Examples of compact binary light curves.
The top light curve is that of sdB star 2MASS J23014582+1338374
while the other three light curves are of WD-dM binaries.
}
}
\end{figure}

In Figure \ref{WD}, we plot the light curves of four compact binary systems. In these light curves the modulation is due
to the distortion of the secondary star. The bottom light curve shows an example where the hot WD primary is eclipsed by
the distorted companion. The other lightcurves do do exhibit eclipses. Similar light curves are observed for gamma-ray
pulsars such as PSR J2339-0533 (Romani et al.~2011) and AY Sex (Wang et al.~2009; Tam et al.~2010).  Approximately 100
compact binaries were found in this work of which approximately half are new discoveries. Hence, this work constitutes a
significant addition. However, further work is required to identify the component stars in each system.

\subsection{RR Lyrae}

RR Lyrae stars are pulsational variables that can be used as standard candles (Catelan 2009, and references therein). In
Drake et al.~(2013a,b) we used CSS data to find RR Lyrae in the halo and thus determine the distance and location to the
Sagittarius tidal stream.  Because of the potential confusion between eclipsing binaries and RRc's, we used only RRab's
in our previous work.  As we have demonstrated, the level of contamination in our selection is no more than a few percent.
Thus RRc's can also be tracers of halo structure when they are well sampled.

\subsubsection{RRc's}

\begin{figure}[ht]{
\epsscale{0.8}
\plotone{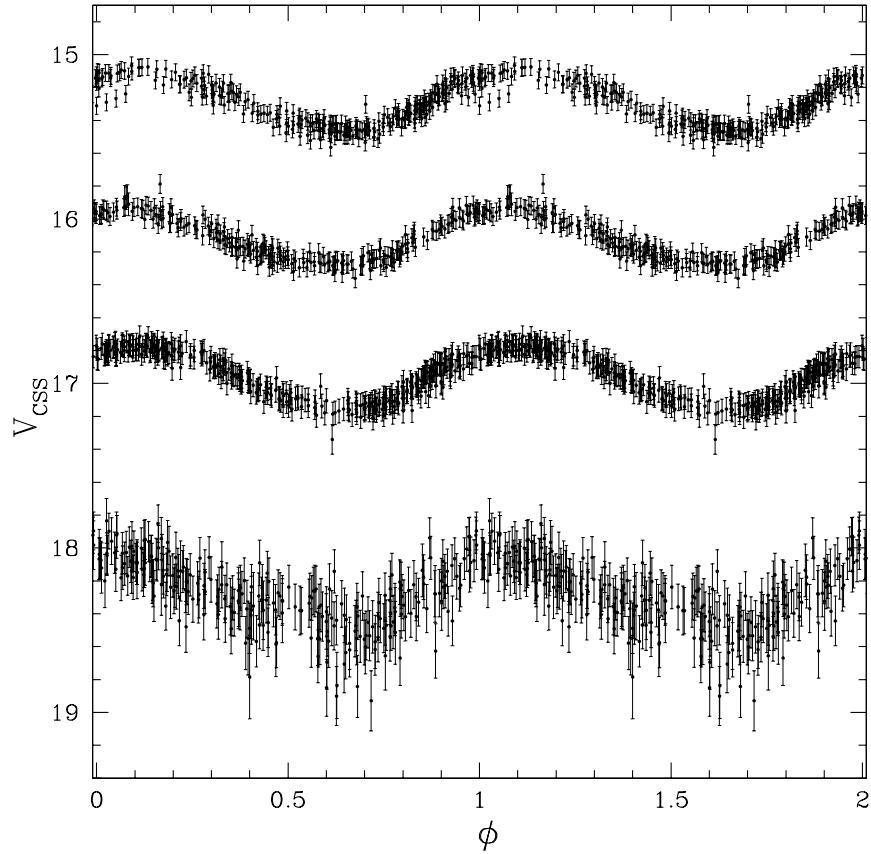}
\caption{\label{RRcLC}
Light curves of RRc's of varying average brightness.
}
}
\end{figure}

RRc's pulsate in the first radial overtone mode. They have bluer colours and very different light curves than RRab's,
which pulsate in the fundamental mode. The variation amplitudes of RRc's are approximately half those of short-period
RRab's.  This makes faint RRc's more difficult to detect than short-period RRab's of comparable brightness.
Nevertheless, as we have shown in Figure \ref{PerHist}, we were able to discover RRc's as faint as $V_{CSS}=19.5$. In
Figure \ref{RRcLC}, we present the light curves of four RRc's with a range of apparent brightnesses.

\begin{figure}[ht]{
\epsscale{0.9}
\plotone{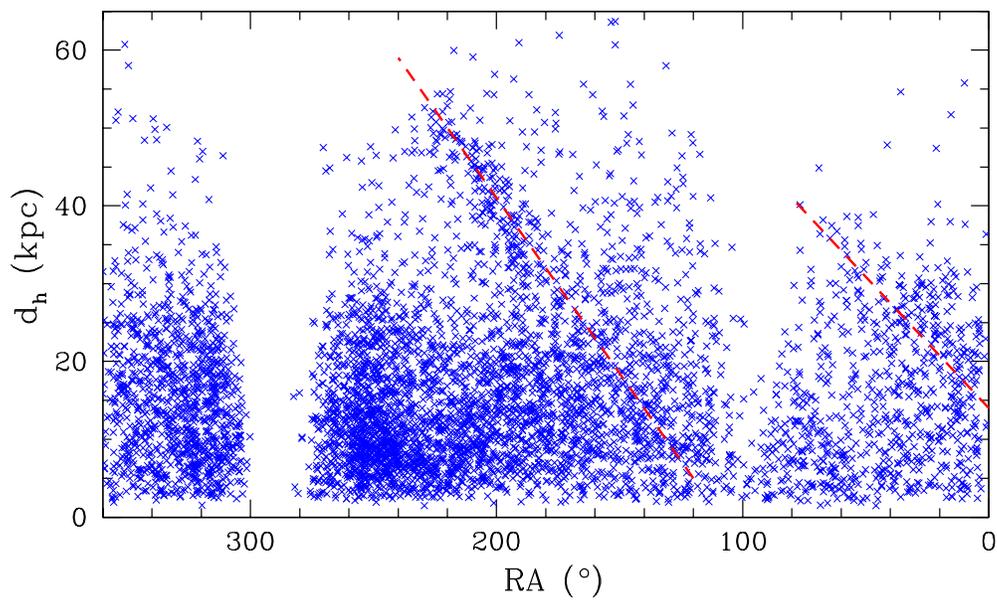}
\caption{\label{RRcDist}
The distribution of heliocentric distances for RRc's from CSDR1.
The dashed lines shows the location of the Sagittarius
tidal stream as given by Drake et al.~(2013a).
}
}
\end{figure}

Following our previous analysis we determine the distances to each of the RRc's assuming the same absolute magnitudes
for RRc's as RRab's, and using the calibration between absolute magnitude and metallicity from Catelan and Cort\'es
(2008). As RRc light curves are nearly sinusoidal, we have not corrected the average magnitudes from the fits to static values as 
is necessary for the asymmetric light curve shapes of RRab's. However, a slight correction may also be necessary for RRc's 
(Bono et al.~1995). In Figure \ref{RRcDist}, we plot the distances to the $\sim 5,500$
RRc's detected in this analysis. The presence of RR Lyrae associated with the Sagittarius tidal stream produces a strong
feature in the region $140\arcdeg < RA < 230 \arcdeg$ at distances beyond 30 kpc, as with the RRab's in Drake et al.~(2013a,b).

We found 2169 SDSS DR10 spectra matching 1136 of the RRc's in the catalog. This is a much larger fraction 
than RRab's with SDSS spectra (Drake et al.~2013a,b), as RRc's have greater overlap with the colours of 
blue HB (BHB) stars. It is the BHB stars that were the targets of the SDSS SEGUE-1 and SEGUE-2 projects (Yanny et al.~2009)
where they were used to determine distances based on single epochs of SDSS photometry.  However, since RRc's 
have very similar colours and spectra to BHB's, their presence within SDSS BHB catalogs limits the overall accuracy 
of distances based on BHB candidates (e.g.~Ruhland et al.~2011). Clean separation of BHB's and RRc's requires 
assessment of variability via multiple epochs of photometry or spectra.

To compare the distances and velocities of the RRc's with the RRab's from Drake et al.~(2013a,b), we selected 
the stars with $d_h > 30$ kpc that lie within $\sim 15$ degrees of the plane of the Sagittarius tidal stream as defined by 
Majewski et al.~(2003). We found 177 RRc spectra from 146 RRc's meeting these criteria. As RRc's have much
shorter periods and smaller pulsational velocities than RRab's (Liu 1991; Fernley \& Barnes 1997; Jeffery et al.~2007)
and the SDSS composite spectra are observed over a period of hours (Drake et al.~2013a), the radial velocities are 
smeared out over a range of pulsation phases. To account for this smearing, we artificially increase the measured 
uncertainties by $15 km/s$, since we assume that pulsation amplitudes are $\sim 30 km/s$.

\begin{figure}[ht]{
\epsscale{0.9}
\plotone{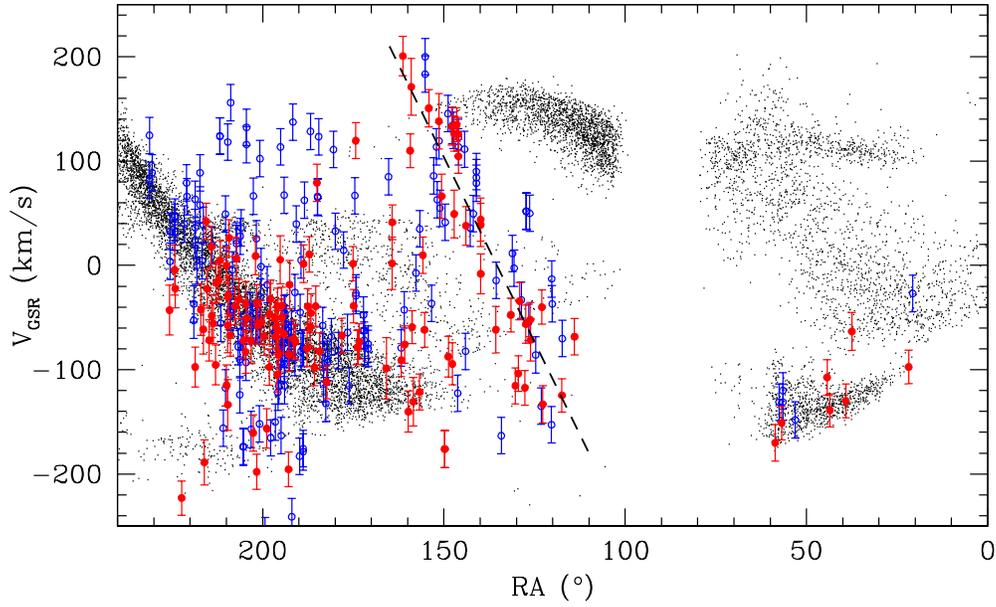}
\caption{\label{RRcSgr}
The distribution of  Galactocentric velocities for 146 RRc's and 130 RRab's
within $15\arcdeg$ of the plane of the Sagittarius stream at distances $d_h > 30$ kpc. 
The red dots are velocities of Drake et al.~(2013a) RRab's, while the blue circles
are RRc velocities. The small dots show the locations of simulated sources within 
the Sagittarius tidal stream based on the Law \& Majawski~(2010) model. The dashed line
presents the approximate location of a velocity feature within the data first noticed
by Drake et al.~(2013a,b) and recently confirmed by Belokurov et al.~(2014) using SDSS 
spectra of M giants.
}
}
\end{figure}

In Figure \ref{RRcSgr}, we plot the RRc Galactocentric radial velocities along with those of RRab from Drake et
al.~(2013a). The new data provides additional evidence for a halo structure with a velocity component within the 
range $110\arcdeg < RA < 160 \arcdeg$ as noted by Drake et al.~(2013a,b). This feature is not explained by the 
Law \& Majawski~(2010) model of the Sagittarius tidal stream. This was recently confirmed by 
Belokurov et al.~(2014) based on SDSS spectra of M-giants. The exact origin of this feature remains uncertain.
However, it may be associated with the distant Gemini tidal stream noted by Drake et al.~(2013b). 
Nevertheless, since the RRc's with SDSS spectra are half the distance of the most distant sources in 
the Gemini tidal stream, this suggests that this halo structure is dispersed over a large range of distances.

\subsubsection{RRab's}

\begin{figure}[ht]{
\epsscale{0.8}
\plotone{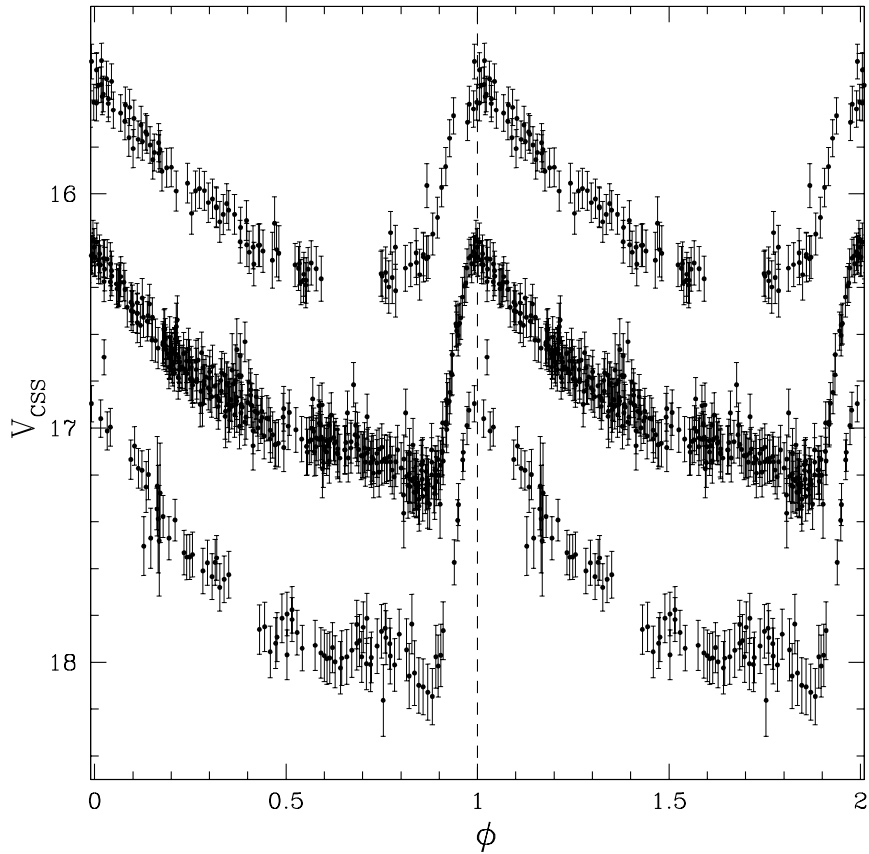}
\caption{\label{RRabLC}
Examples of light curves for three newly discovered RRab's.
}
}
\end{figure}

In Drake et al.~(2013a,b), we discovered $\sim 15,000$ RRab's in CSS data. In this analysis we examined sources
with a new $J_{WS}$ variability threshold, as well as objects that were outside the $0.34-1.5$ day period
range. Based on our detection efficiency simulations, we were $\sim 70\%$ complete for sources brighter than 
$V = 17$ in our original analysis. 2,000 more RRab's were given in Drake et al.~(2013b), and 
$\sim 2,400$ from this work. Combining the total number of RRab's we therefore expect to be 90\% complete 
for sources with $V < 17$. However, we expect the completeness to be much lower for RRc's because of their generally lower
variability amplitudes. In Figure \ref{RRabLC}, we present the light curves of three newly discovered RRab's.

\subsubsection{Anomalous Cepheids or long-period RRab stars}

\begin{figure}[ht]{
\epsscale{0.7}
\plotone{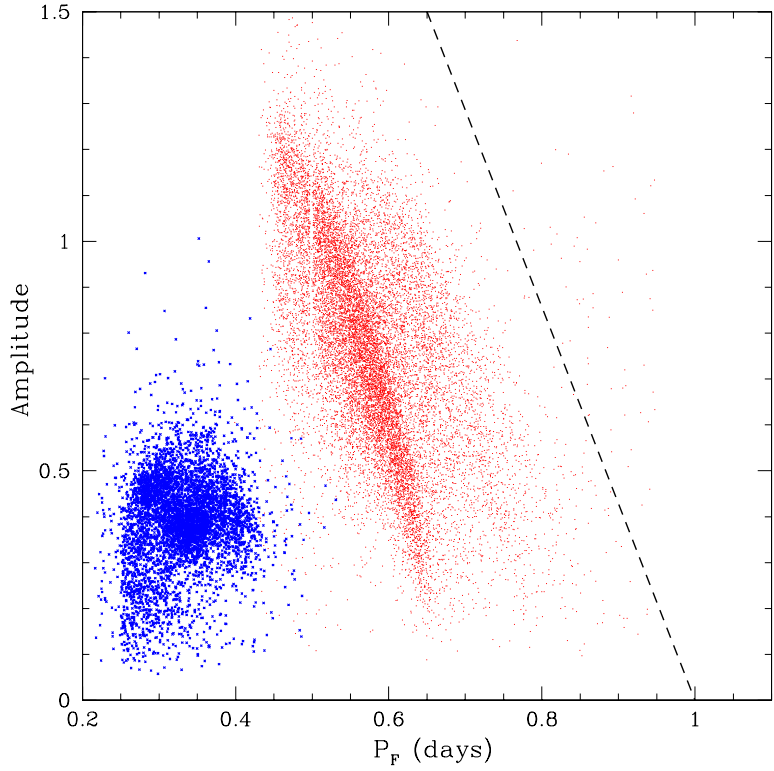}
\caption{\label{Per_Amp}
The period-amplitude diagram for RR Lyrae. The RRab's are given by red dots and the 
RRc's are given by blue crosses. The dashed line shows the division used to select 
high-amplitude, long-period sources.
}
}
\end{figure}

During our analysis of RR Lyrae we discovered many long-period sources with unexpectedly high-amplitudes. These objects
have light curves that resemble RRab's with much shorter periods, or fundamental-mode classical Cepheids. However,
classical Cepheids are due to a young population and are thus limited to the Galactic plane. In Figure \ref{Per_Amp}, we
plot the period-amplitude distribution of the RRc's and RRab's (including those from Drake et al.~2013a,b).  We see
that the RRab's are highly concentrated to periods $< 0.8$ days.

We selected RR Lyrae with amplitudes $A > 4.3 - 4.3 \times P_F$, for periods $P_F > 0.6$ days. Based on our examination
we found some sources in this region were caused by period alias of short-period RRab's. However, many of light curves 
are well enough sampled that they are clearly not aliases of either shorter or longer period variables.

We found high-amplitude sources with periods ranging from 0.77 days to 2.4 days.  Their light curves appear too similar
to be due to separate types of variables.  Only slight evolution in morphology was seen with increasing period. This
suggests that these sources are part of a single population.

Among the periodic variables in this group, a number were already known.  Some of these sources had previously 
been classified alternately as Cepheids, and RR Lyrae by different groups of authors. Matching these objects 
with SDSS, we found that the objects had the same colours as RRab's.
Given the Galactic latitude limits of CSS data ($|b| > 10 \arcdeg$), the sources are unlikely to be classical 
Cepheids and the light curve morphology is distinctly different from that of type II Cepheids.  
The light curves of the objects also resemble ACEP's, which have heretofore mainly been classified in dwarf 
spheroidal galaxies (Coppola et al.~2013). 
ACEP's have periods matching those of these objects.

\begin{figure}[ht]{
\epsscale{0.7}
\plotone{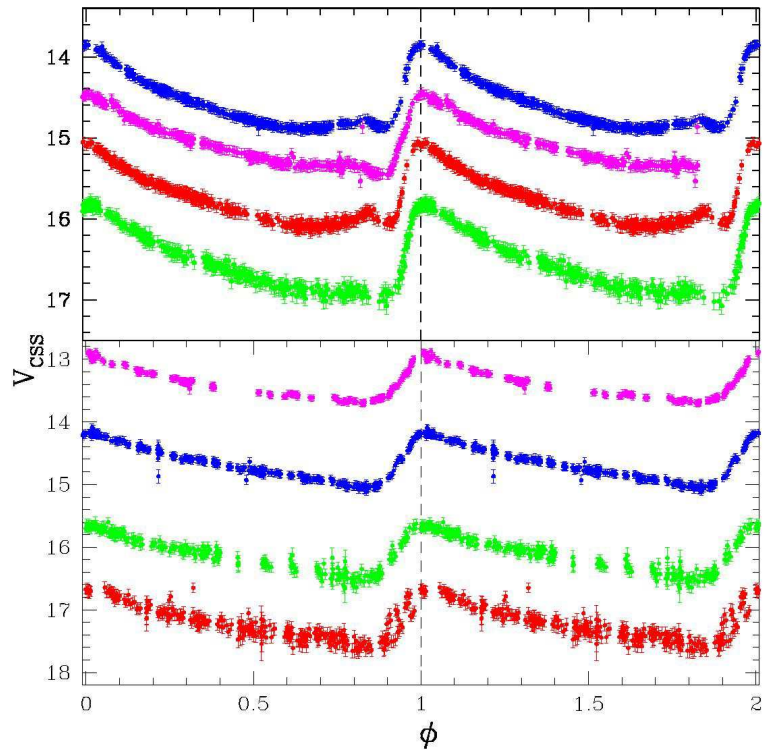}
\caption{\label{AC_RR}
Examples of anomalous Cepheid light curves. In the top panel we plot 
four objects with periods 0.77 to 1.1 days and in the lower panel we 
plot sources with periods from 1.5 to 2.1 days.
}
}
\end{figure}

In Figure \ref{AC_RR}, we plot the light curves of eight ACEP variable stars. 
After inspection, we find 61 new variables that fall in this class. Most of 
the objects are brighter than $V=16$ and they are distributed at Galactic 
latitudes ranging from 14 to $70 \arcdeg$, with average $43\arcdeg$.

\begin{figure}[ht]{
\epsscale{1.0}
\plottwo{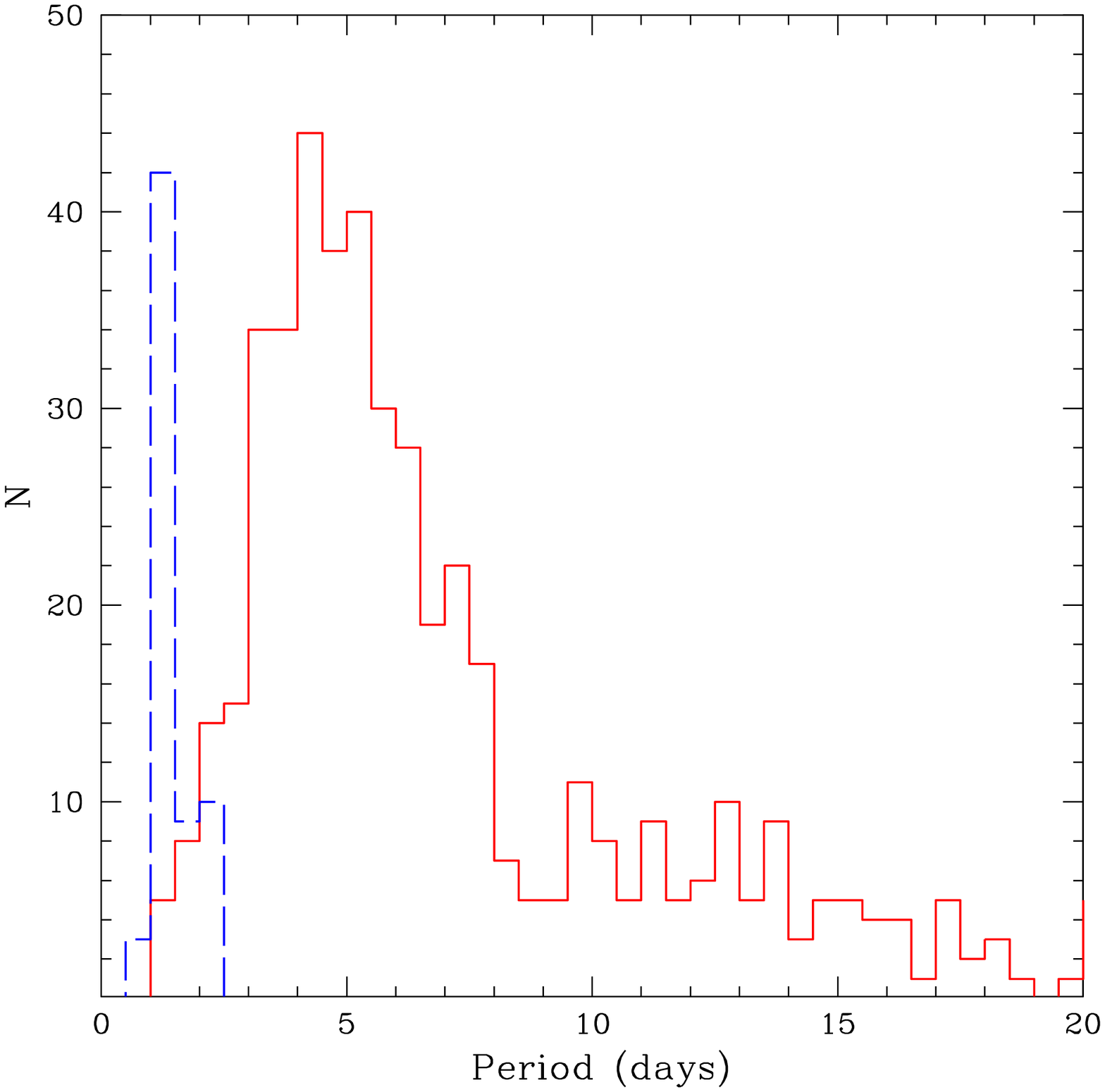}{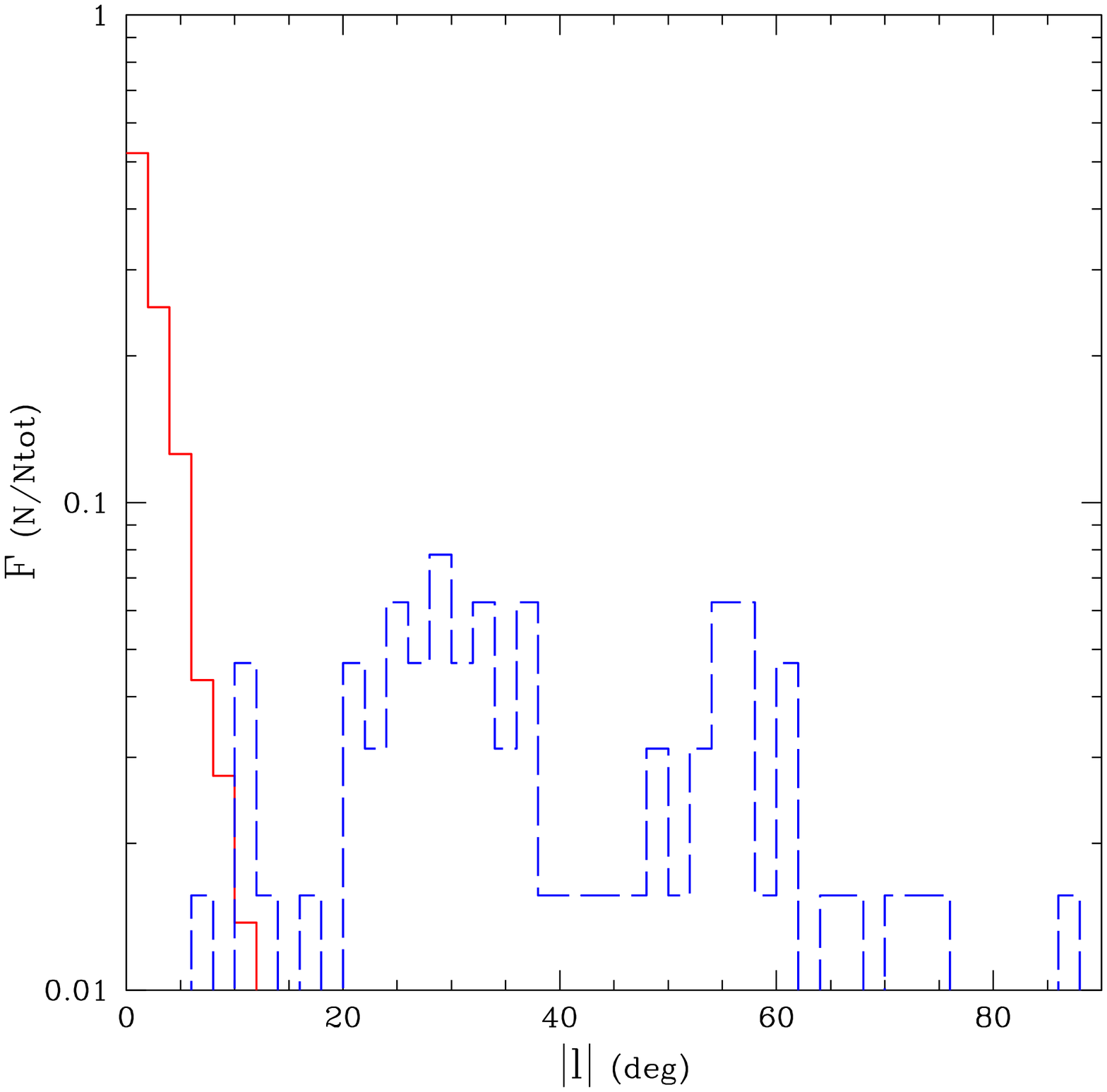}
\caption{\label{CepD}
The distribution of Cepheids. In the left panel, we plot the period distribution 
for the Anomalous Cepheid candidates (dashed blue line), and known 
classical Cepheids (solid red line). In the right panel we plot the Galactic 
latitude (l) distribution for the Anomalous Cepheid candidates (dashed blue line),
and known classical Cepheids (solid red line).
}
}
\end{figure}

In Figure \ref{CepD}, we plot the distribution of periods and Galactic latitudes of the ACEP candidates along with that
of 500 classical Cepheids from the Fernie et al.~(1995) catalog\footnote{http://www.astro.utoronto.ca/DDO/research/cepheids/}.  
These sources clearly have a different periods and spatial distribution than classical Cepheids.  Since we found no clear
association between the objects and Globular clusters, or other sources with known distances, the absolute magnitudes
of these sources remain uncertain.

ACEP's have been found at periods below 0.8 days in Carina (Dall\'Ora et al. 2003, Vivas \& Mateo 2013). Such stars could 
well be mistaken for RR Lyrae in our analysis since the distances to the sources are unknown.
However, ACEP's are much rarer than RR Lyrae so we do not expect they are present in very large numbers.

\subsubsection{RRd's and Blazkho RR Lyrae}

RR Lyrae are known to evolve across the HB between the red and blue ends.  As they do so, they cross the instability
strip becoming fundamental-mode pulsators (RRab's), on the red side and first-overtone pulsators (RRc's) on the 
blue side. This evolution is expect to take millions of years (e.g., Bono et al.~1997).
However, apart from fundamental and first overtone pulsators, RR Lyrae are also well known to exhibit multi-modal 
variations.  Type-d RR Lyrae (RRd's) oscillate in both the fundamental and first overtone modes simultaneously.
These two modes exhibit a period ratio of $\sim 0.74$ between the two components (see Catelan 2009, for a review).
The light curves of RRd's resemble poorly phased periodic variables. In contrast to RRd's, Blazkho RR Lyrae 
exhibit a modulation in amplitude and phase. Nevertheless, on long timescales this also makes them appear 
like variables with poorly determined periods.

\begin{figure}[ht]{
\epsscale{0.8}
\plotone{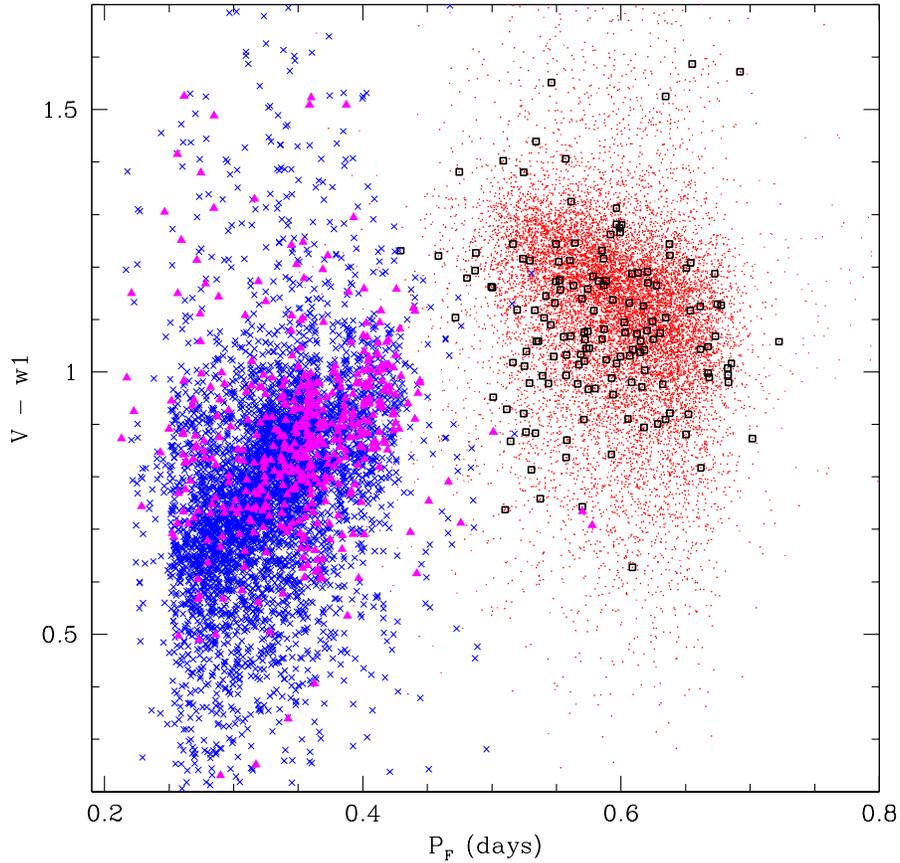}
\caption{\label{RRdBlaz}
The period-colour distribution of RR Lyraes.
Here we present the periods and colours for RRc's (blue crosses), RRd's (magenta triangles), RRab's (red dots)
and Blazkho (black squares) RR Lyrae.
}
}
\end{figure}

In Figure \ref{RRdBlaz}, we present the period-colour distribution of all RR Lyrae discovered in CSS data. The RRd's
have dominant single periods and colours similar to RRc's, while the Blazkhos have periods and colours similar to
RRab's. Because of the possible confusion of RRd's and Blazkhos with RR Lyrae having poorly determined periods, it is
likely that some of the RRd and Blazkho candidates presented here are misclassified.

\begin{figure}[ht]{
\epsscale{1.0}
\plottwo{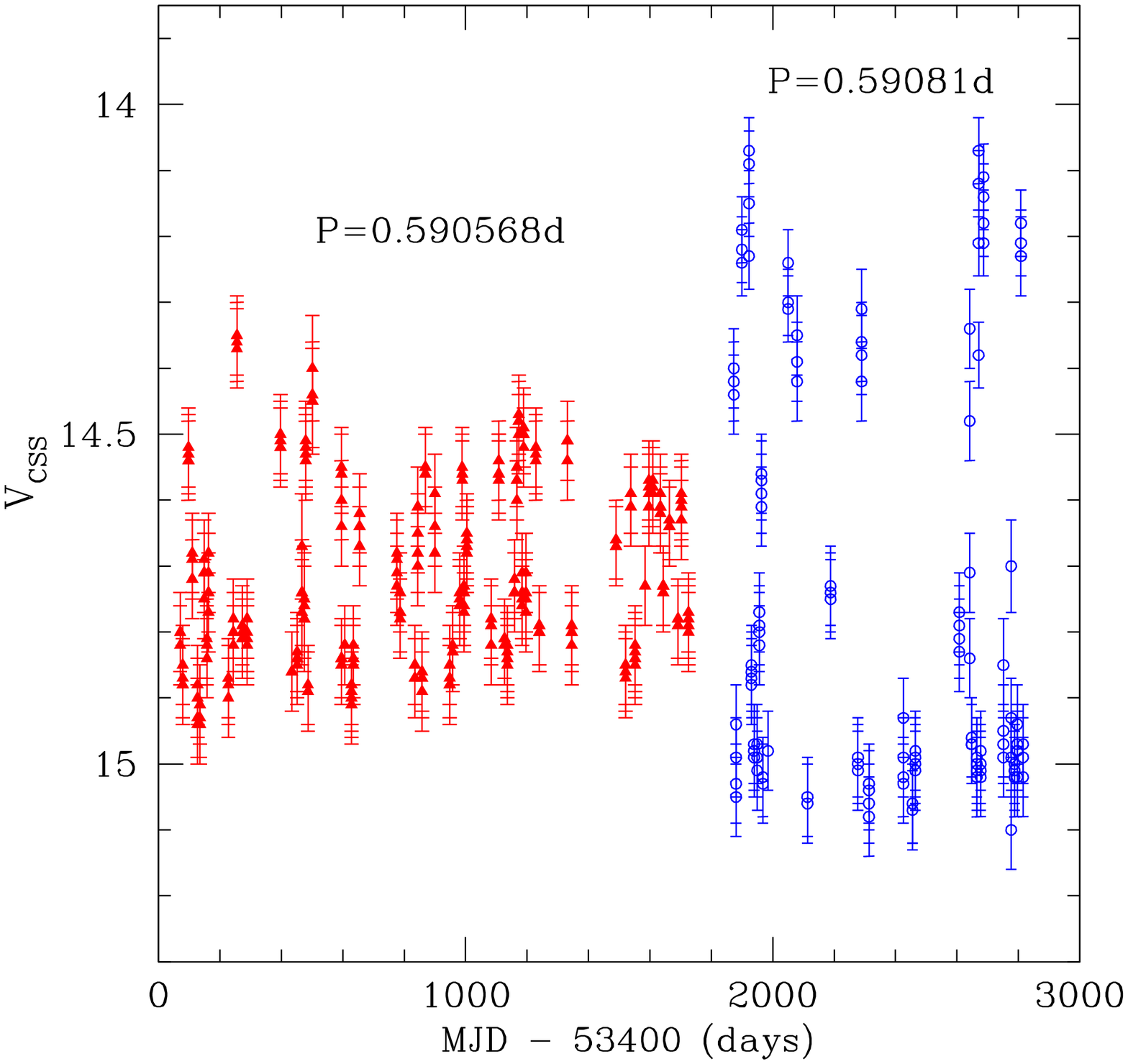}{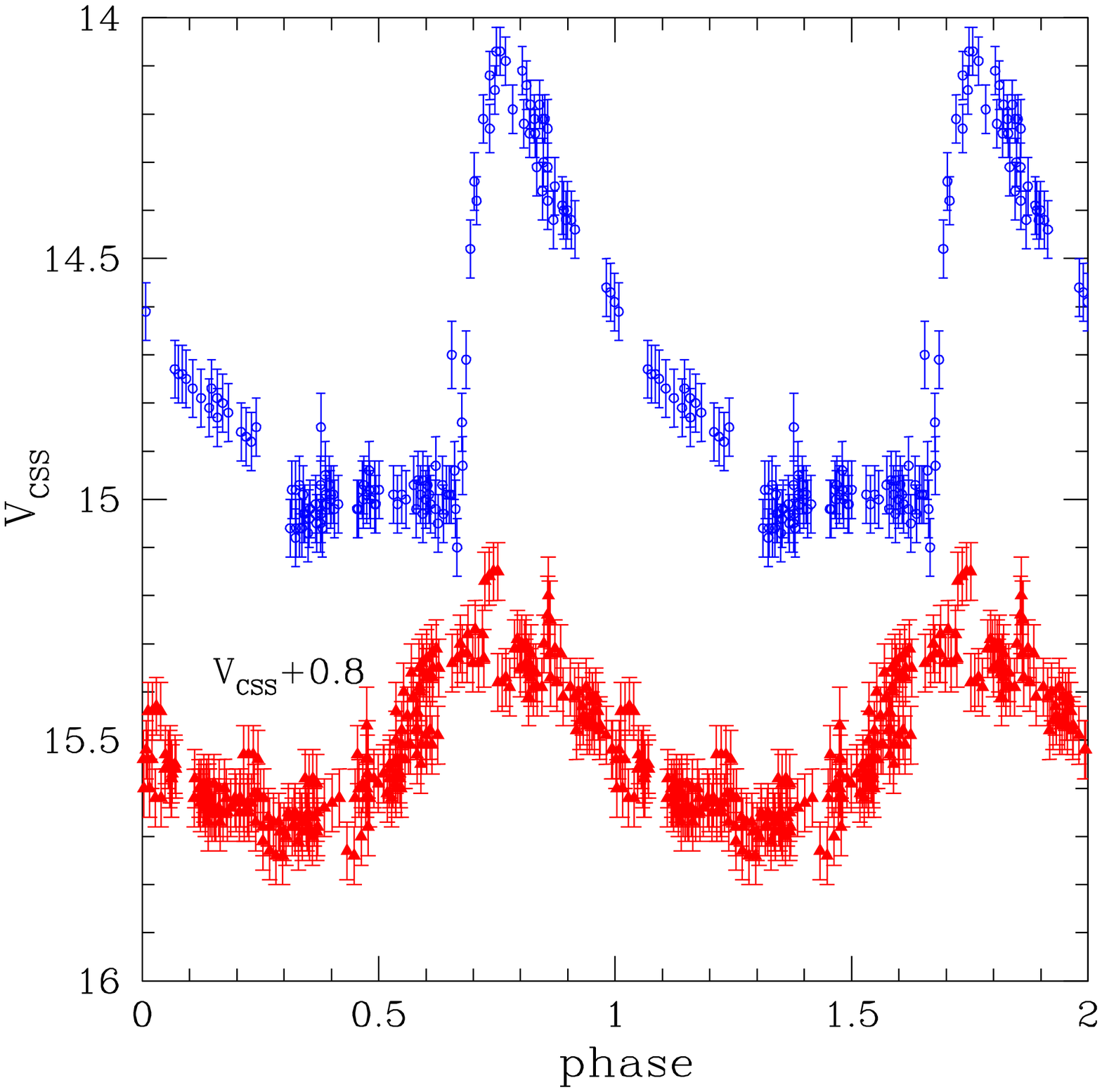}
\caption{\label{RRchange}
  The light curve of a mode-changing RR Lyrae, CSSJ172304.0+290810.  In the left panel, we present the light curve with the
  times when the system was observed in separate modes given in red and blue, respectively. In the right panel, we plot
  the phased light curves for the two separate pulsation modes based on the times given in the left panel.
}
}
\end{figure}

In addition to these sources we found six examples of RR Lyrae where the mode of pulsation appeared to change on a
timescale of months.  In Figure \ref{RRchange}, we plot an example of an RR Lyrae that underwent a sudden change from
what appears to be a first-overtone pulsator, to a fundamental-mode pulsator. The observed change in pulsation period 
between these two modes is only $\sim 21$ seconds. 

Additional RR Lyrae exhibiting such changes include V442 Her (Schmidt \& Lee 2000), 
V15 in NGC6121 (Clementini et al.~1994), and  V18 in M5 (Jurcsik et al.~2011). 
In the case of V18, the source was not covered continuously during the variation, so the timescale of the change is
poorly determined.  Both V442 Her and V18 have much shorter periods than CSSJ172304.0+290810 (0.48 days and 0.44 days, 
respectively). These values are consistent with RRd stars, whereas the period of CSSJ172304.0+290810 is
most consistent with an RRab.

Clementini et al.~(2004) also note that V21 in M68 changed from a double-mode RR Lyrae to a fundamental-mode
system and then subsequently became a double-mode object again.  They also note three additional RRd's in M3 (M3-V166,
M3-V200 and M3-V251) that switched their dominant pulsation modes within a year.  With a period of 0.59 days, 
CSSJ172304.0+290810 is at the limit of periods observed in double-mode RR Lyrae (Clementini et al.~2004).

Period change rates of $0.1 - 0.2\,{\rm d/Myr}$ have been observed for RR Lyrae (Le Borgne et al.~2007). Such rates are
an order of magnitude higher than predicted by stellar evolution models (e.g., Catelan 2009, and refs therein). However,
these have been found to be highly variable between RR Lyrae, even within individual globular clusters (Kunder et
al.~2011).  As noted by Kunder et al.~(2011), rapid changes have been attributed to mixing events (Sweigart \& Renzini
1979), magnetohydrodynamic events (Stothers 1980), and convection (Stothers 2010).

The abrupt period change for CSSJ172304.0+290810 appears to have occurred within a 
120-day window suggesting a rate of change of at least $2\,{\rm d/Myr}$. This 
is consistent with some more extreme period changes observed by Kunder et al.~(2011)
and Figuera Jaimes et al.~(2013).

\subsection{$\delta$ Scutis}

\begin{figure}[ht]{
\epsscale{0.8}
\plotone{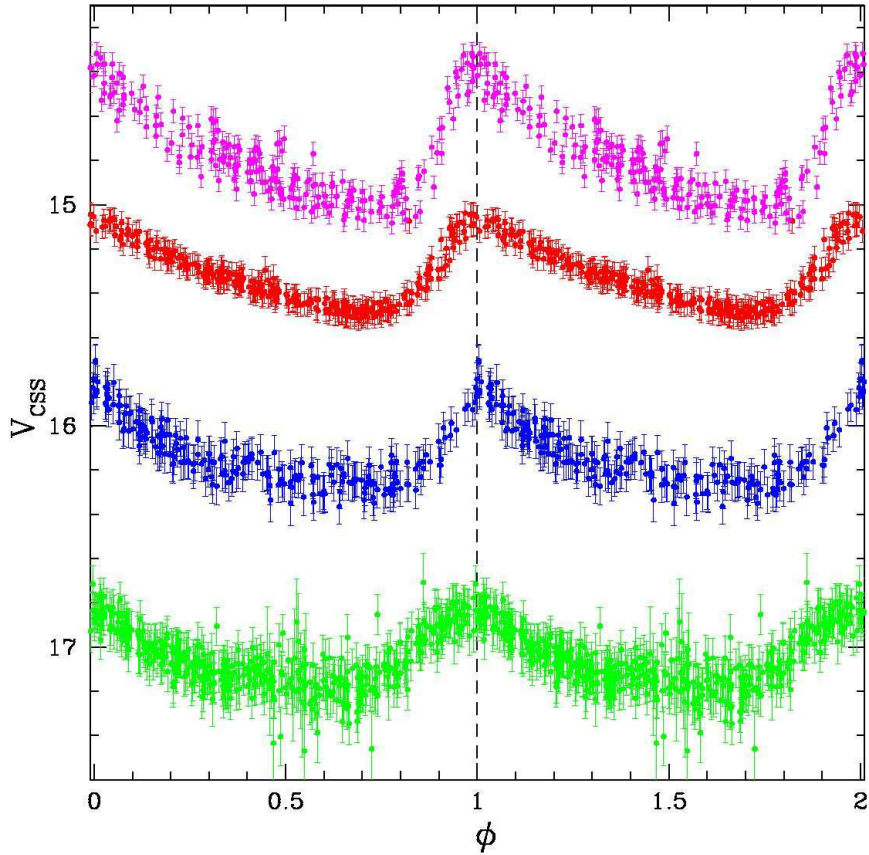}
\caption{\label{Delscu}
Examples of high amplitude $\delta$ Scuti star light curves.
}
}
\end{figure}

$\delta$ Scuti variables can exhibit brightness variations from 0.003 to 0.9 mags in $V$ and have periods of a few
hours. High-amplitude delta Scutis (HADS, AL Velorum stars) have amplitudes greater than 0.1 mags (Alcock et al 2000c),
while low-amplitude delta Scutis (LADS) have smaller amplitudes.  In Figure \ref{Delscu}, we plot four examples of the
HADS discovered. In our analyisis we are mainly sensitive to variations $> 0.1$ mags and periods of hours where a single
pulsation mode dominates, so it is likely that we did not detect all the $\delta$ Scutis within CSDR1 data.  Metal-poor
$\delta$ Scutis, called SX Phoenicis stars (SX Phe), are found within the halo and in globular clusters and naturally
have halo velocities and metalicities.

$\delta$ Scutis often exhibit multi-periodic behaviour. Most $\delta$-Scuti variables are main-sequence stars with blue
colours similar to those of RR Lyrae.  This can lead to confusion for colour-selected variables with insufficient
sampling to determine their periods (Sesar et al.~2010).
In this work, the presence of hundreds of observations and the short period (30 mins) between sets of four observations 
strongly limits the misidentification of short period sources as longer period ones. For example, HADS are likely to 
exhibit significant variation over the span of four observations, while RRc's (which have periods of many hours), are
not. The clear separatation between $\delta$ Scutis and RRc periods is shown by Palaversa et al.~(2013).

Further evidence against significant $\delta$ Scuti-RRc confusion in our data comes from Figure \ref{SDSSlgfe}. Here we
again note that objects classified as RRc's have low surface gravities and low metalities. The velocities shown in Figure
{SDSSvel}, also show that the RRc's form a halo population. Thus, the velocities, metallicities and surface gravities
rule out the presence of significant fraction of $\delta$ Scutis (within the area covered by SDSS).  On the other hand,
SX Phe stars have halo population properties like RR Lyrae. Yet, as with $\delta$-Scutis they are fainter, have 
higher surface gravities, and much shorter periods the RR Lyrae (Cohen \& Sarajedini 2012).

\subsection{Type II Cepheids}

Type II Cepheids are metal-poor Cepheids that are found in galaxy halos. These stars can be distinguished from classical
Cepheids by their amplitudes, light curves, spectral characteristics, and radial velocity curves. They are fainter than
classical Cepheids and are divided into three sub-classes that separated by increasing period and luminosity 
as defined by Wallerstein~(2002).

These sub-classes are BL Herculis variables (BL Her), with periods between 1 and 5 days, W Virginis variables (W Vir)
with periods of 5 to 20 days, and RV Tauri variables (RV Tau) with periods greater than 20 days.  As with classical
Cepheids, these variables can be useful for measuring distances since they obey a period-luminosity relationship
(e.g., McNamara 1995; Pritzl et al.~2003; Soszynski et al.~2008).

\subsubsection{BL Her}

\begin{figure}[ht]{
\epsscale{0.65}
\plotone{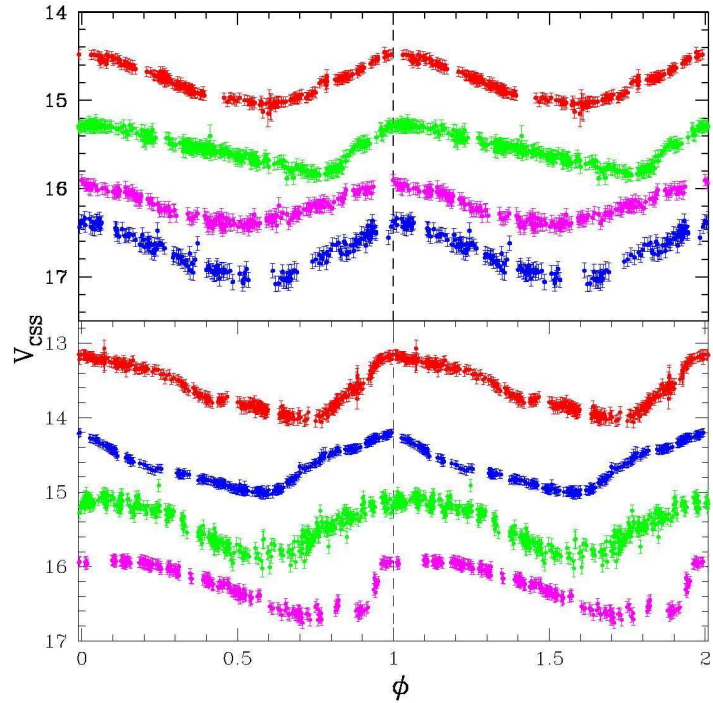}
\caption{\label{BLH}
Examples of BL Her type variable light curves.
In the top panel objects have periods from 0.89 to 1.04 days 
and in the lower panel 1.14 to 2.25 days.
}
}
\end{figure}

BL Her-type Cepheids usually show a bump on the descending side of their light curves
at short periods. This bump is seen on the ascending side at longer periods (Soszynski et al.~2008).
BL Her's have similar spectral types to RR Lyrae, but are slightly brighter.
In Figure \ref{BLH}, we plot the light curves of a few of the BL Her-type Cepheids.

\subsubsection{W Vir and RV Tau Cepheids}

\begin{figure}[ht]{
\epsscale{0.65}
\plotone{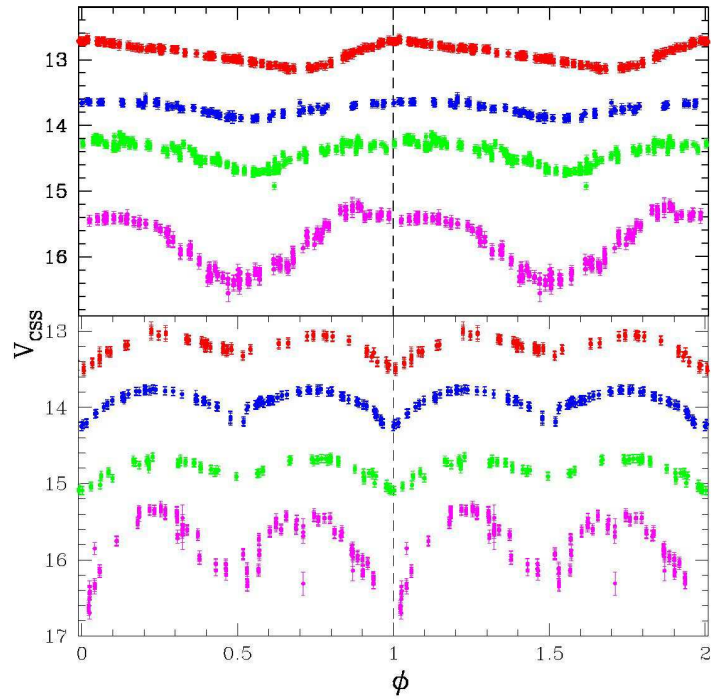}
\caption{\label{WVir}
Examples of W Vir and RV Tau type Cepheid light curves.
Top panel: W Vir type Cepheids with periods from 6.4
to 13.9 days,
Bottom panel: RV Tau-type Cepheids with periods from 22.3
to 56.9 days.
}
}
\end{figure}

W Virginis is the prototype for the population II Cepheids and has a period of 17 days.  Stars in the 
W Vir sub-type have periods longer than 5 days and do not exhibit the bumps of BL Her stars.
In contrast, RV Tau stars exhibit a secondary dip with distinctive alternating deep and shallow minima and 
periods longer than 20 days (Wallerstein~2002). In Figure \ref{WVir}, we plot examples of W Vir and RV Tau 
light curves within CSDR1 data.

\subsection{Rotational variables}

\begin{figure}[ht]{
\epsscale{0.65}
\plotone{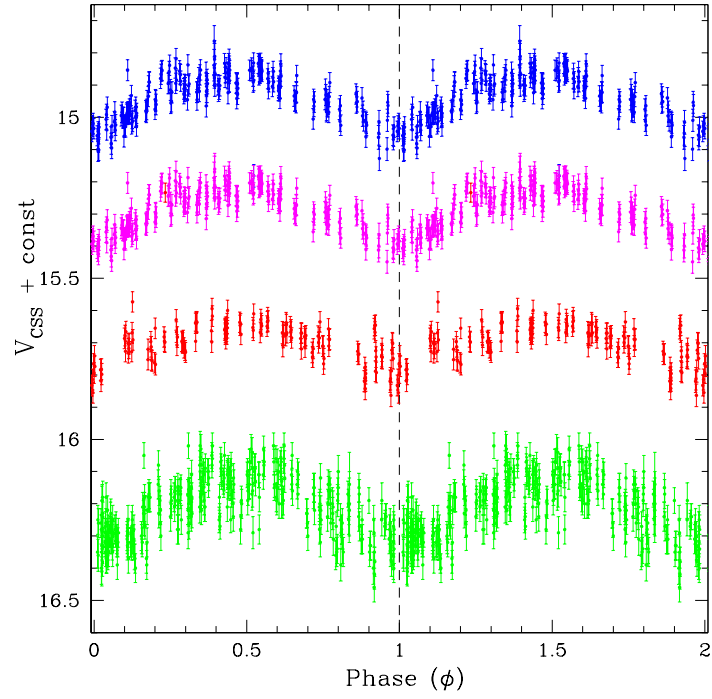}
\caption{\label{rot}
Examples of rotational variable light curves.
}
}
\end{figure}

RS Canum Venaticorum variables (RS CVn's) consist of spotted stars with periods from $< 1$ day for main-sequence stars,
to hundreds of days for giants (Drake 2006). The groups of spots on these systems can give rise to periodic variations
of $\sim 0.2$ mags. However, the numbers, sizes and locations of spots can change over time.  The chromospheric activity in
these stars is signaled by the presence of emission cores in the Ca II H and K resonance lines (Fekel et al.~1986).
Balmer, X-ray, and ultraviolet (UV) emission are also associated with their active chromospheres and transition regions
(Engvold et al.~1988, Rodriguez-Gil et al.~2011). Most of the RS CVn's discovered in this analysis have periods longer than
one day and moderately red colours consistent with the expected F or G-type stars. In Figure \ref{rot}, represent the
light curves of four stars rotational variable candidates.

\subsection{Long-Period Variables}

\begin{figure}[ht]{
\epsscale{0.65}
\plotone{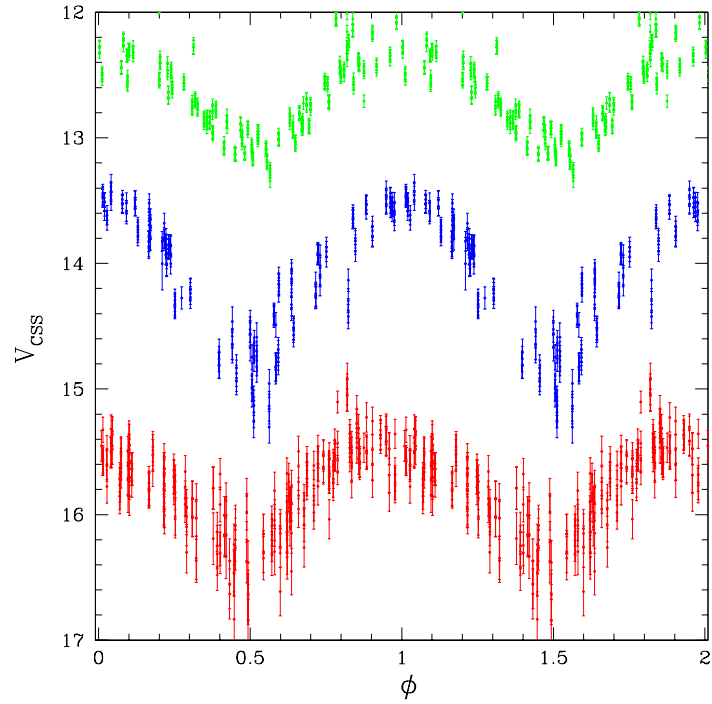}
\caption{\label{LPV}
Examples of long period variable light curves.
The brightest object among these saturates near 
maximum light.
}
}
\end{figure}

LPVs are cool pulsating giant stars with periods ranging from a few to 1,000 days. In Figure \ref{LPV}, we plot 
examples of LPV lightcurves. The amplitude of variation changes slightly between cycles so the the scatter in
the phased lightcurve is greater than the actual photometric uncertainty.
However, the brightest LPVs are saturated in catalina data.

In Figure \ref{Ait} (panel 2), we noted evidence for a spatial structure distribution of LPVs in the range $135\arcdeg <
\alpha < 250 \arcdeg$ from $\delta \sim 30\arcdeg$ to $\delta ~ -20 \arcdeg$. This feature mirrors the structure observed in
RRab's due to the tidal stream of the Sagittarius dwarf (Drake et al.~2013a). As this structure was originally discovered
by Majewski et al.~(2003) based on M-giants, it is of no surprise that it is seen among red giant variables.

To demonstrate the relationship between the variables and the Sagittarius stream,  we separated the LPVs by average
magnitude.  The brightest LPVs are foreground disk stars that are seen concentrated at low Galactic latitude, while the
faintest LPVs we detected come from nearby galaxies, such as M31. By selecting sources with $14.9 < \bar{V_{CSS}} < 15.9$ we 
find many halo LPVs. However, LPVs have a very broad range of absolute magnitude, and follow multiple families
of period-luminosity relations (e.g., Fraser et al.~2005).

\begin{figure}[ht]{
\epsscale{0.8}
\plotone{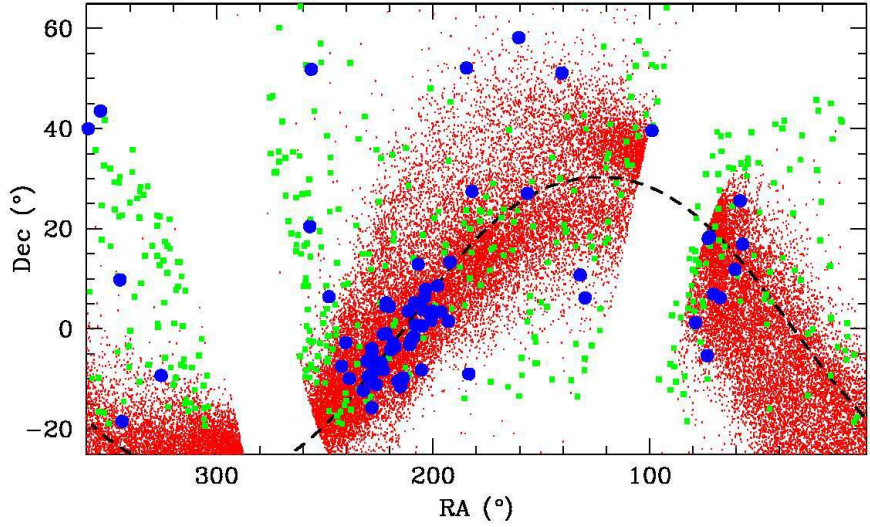}
\caption{\label{SgrLPV}
The distribution of LPVs compared to the Law \& Majewski (2010) model
of the Sagittarius stream.  The large blue dots show LPVs with 
$14.9 < V < 15.9$ and the green boxes show the locations of all
other LPVs. The dashed line shows the plane of the Sgr stream system
defined by Majewski et al.~(2013). The red points show the locations 
of simulated Sgr stream sources from  Law \& Majewski (2010).
}
}
\end{figure}

In Figure \ref{SgrLPV}, we plot the distribution of LPVs compared to the Sagittarius stream model of Law \& Majewski (2010). 
For the halo LPV sample we find a significant number coinciding with the Sgr stream in the region $180\arcdeg < RA < 245 \arcdeg$,
$-20\arcdeg < Dec < 15 \arcdeg$. Approximately 50 of the 80 LPVs in the range $14.9 < V < 15.9$ are found in this
region. An additional eight LPVs are found in the region $50\arcdeg < RA < 80\arcdeg$, $-10\arcdeg < Dec < 30 \arcdeg$ that also
overlaps with the Sgr stream.

\begin{figure}[ht]{
\epsscale{0.8}
\plotone{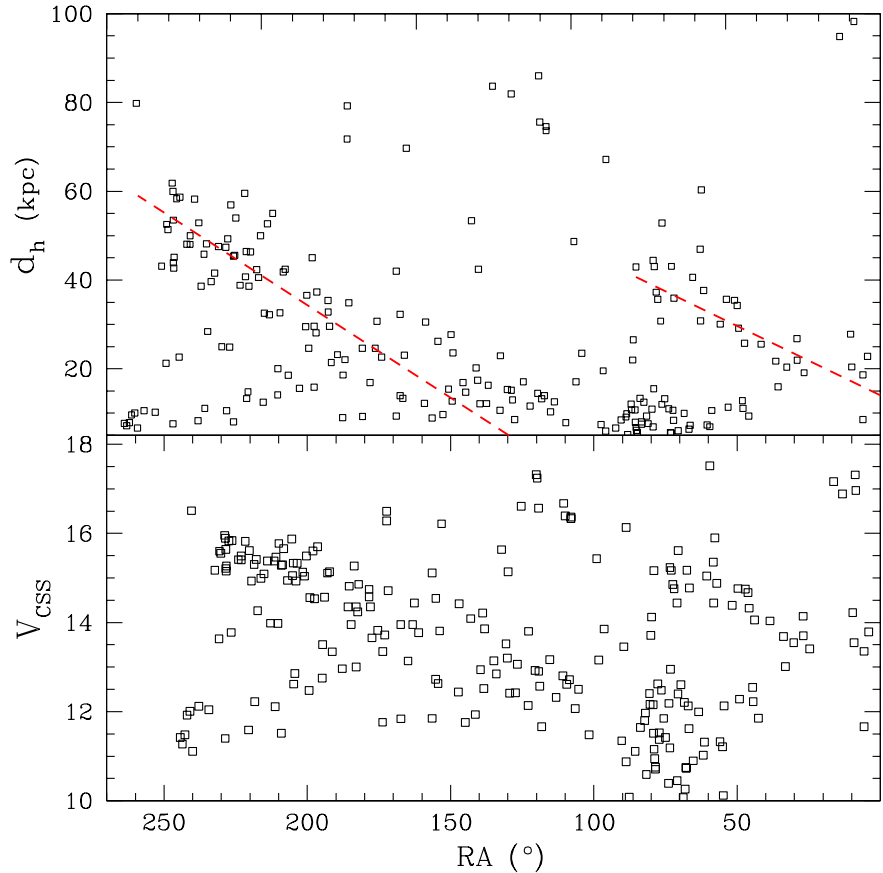}
\caption{\label{SgrLPVMag}
The spatial distribution of LPVs within $15\arcdeg$ of the plane 
of the Sagittarius tidal stream region. In the bottom panel we plot 
the average magnitudes for LPVs. In the top panel we plot distances
assuming the LPVs have $M_V = -3$. The dashed lines show the location 
of the Sagittarius streams as given by Drake et al.~(2013a).
}
}
\end{figure}

To further test the association of LPVs with the Sagittarius stream we selected LPVs within $15\arcdeg$ of the plane 
of the Sagittarius streams system defined by Majewski et al.~(2003). In Figure \ref{SgrLPVMag}, we plot the locations
of these objects.

Under the assumption that LPVs have representative luminosities of around $M_{V}=-3$ (Smak 1966) in CSDR1 data, we find
good agreement with the results based on RRab's and RRc's.  However, LPVs are known to occupy six separate
period-luminosity sequences (Fraser et al.~2005) and perhaps more (Mosser et al.~2013). As many of the LPV sequences
overlap in period range, to derive more accurate absolute magnitudes, hence distances, one must first determine the
sequence of the variable. A combination of light curve morphology and multi-wavelength observations may enable this
determination. Nevertheless, the figure does show that there is a strong trend in the average brightness of LPVs
which is consistent with membership of the Sagittarius tidal stream.

\subsection{Miscellaneous variable sources}\label{Misc}

\begin{figure}[ht]{
\epsscale{1.0}
\plottwo{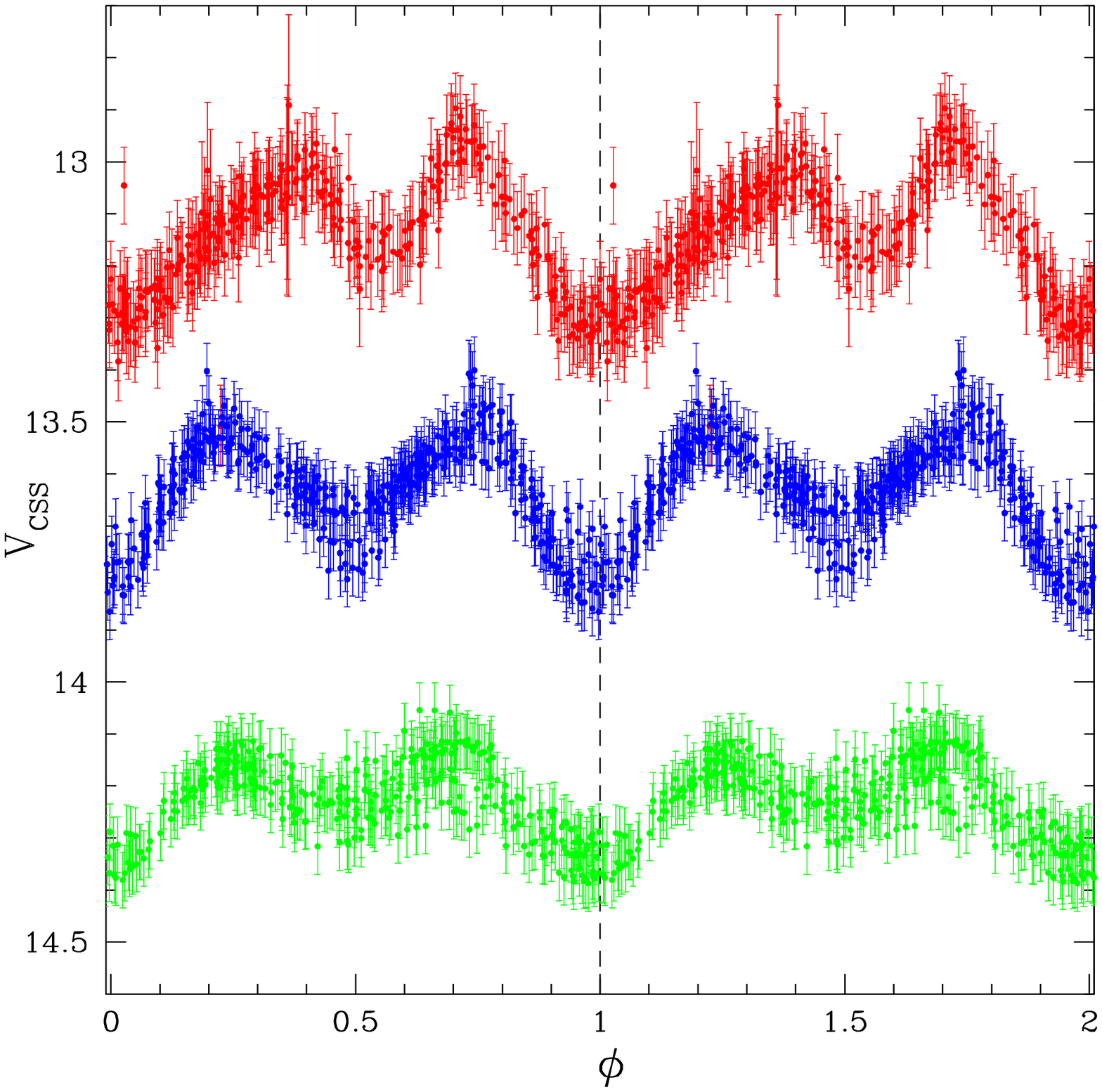}{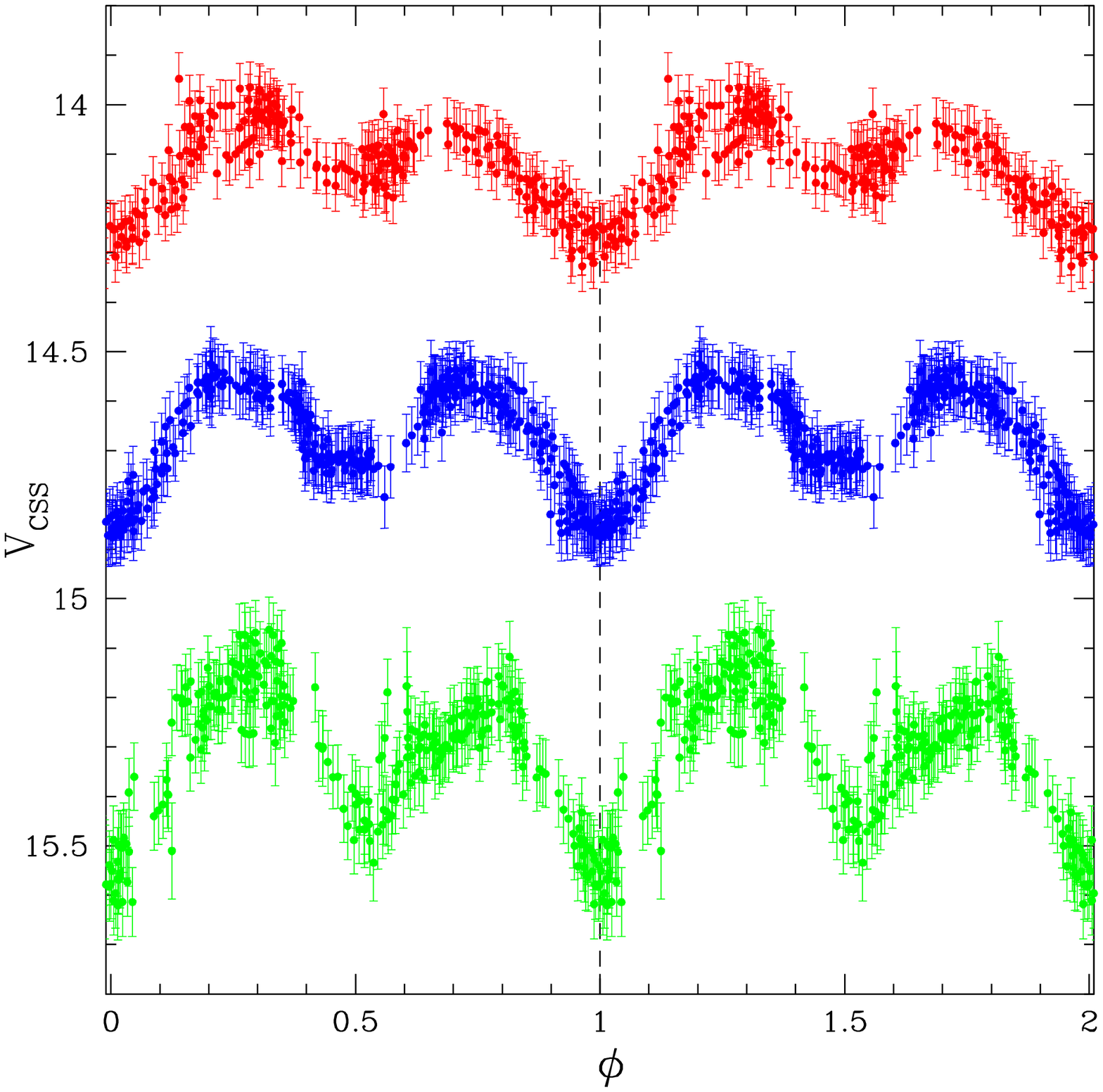}
\caption{\label{Zig}
Light curves of unclassified periodic variables.
The objects in the left panel have periods of 0.19 days (red), 0.25 days (blue) and 0.28 days (green).
The objects in the right panel have periods of 0.42 days (red), 0.45 days (blue) and 0.80 days (green). 
}
}
\end{figure}

More than 99\% of the periodic variables inspected clearly fall into to the types described above. However, some of the
periodic variables are difficult to classify.  In Figure \ref{Zig}, we plot the light curves of six periodic variables
with uncertain classifications. The top two light curves in the left panel are indicative of objects that exhibit zig-zag
shapes yet vary from one object to the next. The lower two light curves in this panel exhibit smoother curves that may
be indicative of over-contact systems sharing a common envelope.  The light curves in the right panel of Figure \ref{Zig}
exhibit some of the features of contact binaries presenting the O'Connell effect. However, the shapes of these ``Hump''
variables are much more erratic. This may be due to the presence of gas streams or hot spots on their surfaces, such as 
noted by Wilsey \& Beaky (2009). In total there are 68 objects in this group of which 25 are placed in the Hump
group.

During inspection of the periodic candidates in this work we discovered a number of non-periodic sources and periodic
sources where the light curve did not fit any of the existing classifications.  Most of the non-periodic variables are
simply stars and QSOs exhibiting irregular variability. However, some sources exhibiting outbursts were also discovered.
Among the aperiodic sources we serendipitously detected 51 supernovae, of which 42 are new discoveries.  Eighteen of
these SN occurred during the operation of CRTS (Drake et al.~2009) and other large transient surveys, such as PTF (Law et
al.~2009) and PanSTARRS-1 (Hodapp et al.~2004). The discovery of such a large fraction of new SN suggests that current
transient surveys miss many nearby, bright supernovae. Further details of the sources are given in the appendices.

\section{Completeness, Accuracy and Purity}

One possible way of measuring the utility of a periodic variable star catalog is based on its purity, accuracy and
completeness.  A natural expectation is that a periodic variable catalog should consist purely of stellar sources.
Additionally, the catalog sources should provide accurate periods, magnitudes, amplitudes, and classifications. One
might also expect a level of completeness that is at minimum equivalent to other recent catalogs.

\subsection{Purity}

To determine the purity of our periodic variables we matched our initial candidates with sources in SDSS DR10 photometry
and spectroscopic catalogs ($3\arcsec$ radius). Among the spectroscopic matches there were 83 different types of spectra
recorded in SDSS. These types can be broadly separated into stars, galaxies and QSOs. In Table \ref{TabSDSSS}, we
present the number of sources with SDSS photometry and spectra data among the 112,000 pre-inspection candidates and
among selected sources.  This table shows that a large number of the initial periodic candidates had spectra classifying 
them as galaxies and QSOs.  However, as expected, after the light curves were inspected only a few were selected
as periodic variables.

We reviewed the SDSS spectra for the 38 objects marked as galaxies and five sources marked as QSOs. 28 of the spectra
marked as galaxies were G and K stars that were incorrectly classified as galaxies near redshift zero. 
Seven of the matches were due to variables blended with neighbouring faint galaxies. Only three sources were actually
galaxies. These objects were removed from the catalog. All five of the objects classified as QSOs by SDSS were systems 
with strong emission lines. Two of the AGN among these appeared to exhibit possible periodic variability, yet all 
were removed.  The possible periodic variability of the two AGN may be due to an unresolved variable
superimposed on the galaxies.  

Based on the total number of SDSS spectra, the false classification rate suggests that $\sim 0.3\%$ (8/2299) of the objects 
in the catalog are likely to be galaxies or QSOs. Thus, from the whole catalog one might expect $\sim 160$ such objects.  
However, SDSS used colour selection to choose targets for spectroscopic follow-up.
Within SDSS-DR10 more than two million spectra of galaxies and QSOs have been taken, while only $\sim 270,\!000$ stars
(see: http://www.sdss3.org/dr10/). This suggests that the spectra comprise a much more complete sample of QSOs and
galaxies than stars, so the number of non-stellar sources may actually be smaller.

In contrast to the SDSS spectroscopy, photometry has been undertaken by SDSS across a large fraction of the sky 
irrespective of the source type. This imaging provides much greater depth and resolution than CSS data, as well 
as multi-band photometry. This makes it an ideal source for investigating even the faintest of the CSDR1 variables.
In order to further test the purity of our sources we matched the periodic candidates with object in the SDSS DR10
photometric catalog. 
Approximately $\sim$ 28,000 of the sources from our periodic variable catalog have SDSS photometry. 
Additionally, every object in SDSS is classified as either a star, a galaxy, or an object of unknown 
type. As with the SDSS spectra, there were many sources marked as galaxies among our initial candidates.
However, a moderate number ($\sim 3\%$) of the objects that we classifed periodic variables were classified 
by SDSS as galaxies.

Upon inspection of the SDSS images we found that the automated SDSS classifications were generally incorrect. 
For example, 77\% of the catalog sources marked as galaxies had $V_{CSS} < 15$. These objects were found to 
be stars that were saturated in SDSS photometry, rather than actual galaxies.
It seems likely that the saturation causes the sources to appear extended, and thus be wrongly classified 
as galaxies. Nevertheless, in order to account for differences in the saturation limit within the five SDSS 
photometry bands, and for variation in star colour, we examined the SDSS images of all 386 objects marked 
as galaxies with $V_{CSS} > 14$. Only ten of these sources were found to be galaxies. These objects
were removed from the catalog.

We also reviewed the SDSS images for some faint variables that were marked as galaxies by SDSS.
In contrast to the saturated sources, almost all of the sources fainter than $V_{CSS} = 15$ were found 
to be blends between close pairs or groups of stars, although, there were also a few cases where a periodic 
variable was blended with a galaxy.  
More than a dozen of the faint sources classified as galaxies based on SDSS photometry 
had stellar SDSS spectra. We found that in many cases the objects marked as galaxies had multiple epochs 
of SDSS photometry where sources were alternatively marked as stars or galaxies, depending on seeing.
Based on inspection of SDSS images and the photometric colour information, we find $<< 0.1\%$ of the periodic catalog 
sources are in fact galaxies. However, since QSOs are marked as stars in SDSS photometry, it is likely that there 
is some contamination due to QSOs (as suggested by the SDSS spectra above).  We have removed all the clear galaxies 
and QSOs from the catalog.

\subsection{Completeness and Accuracy}

\subsubsection{Known variables from VSX}

The International Variable Star Index (VSX, Watson et al.~2006) provides what is very likely the most complete catalog
of known variable stars.  We matched the original $154\!,000$ CSDR1 periodic variable candidates with sources in the VSX
dataset and found 6459 matches. These objects comes from 176 different classes of variable stars. This very large number
of classes is due to subclasses of variables as well as sources having either multiple or uncertain classifications
(e.g. class EB|EW, signifies membership of either $\beta$ Lyrae or W UMa types).  However, many of the variable stars
within the VSX catalog are not specifically periodic or they have non-periodic outbursts (e.g. T-Tauri stars, YSO's,
CV's, AM CVn's). Therefore we do not expect to recover all of the VSX variables.

The VSX variable classes having more than 100 matching variables of one type include: eclipsing binaries (3201), 
RR Lyrae (1006), $\delta$ Scutis (240), U Geminorum-type variables (214), Mira (206) and miscellaneous variables 
with unknown variable types (382). 
Of the original periodic candidates, 6030 were among the $112,\!000$ periodic variable candidates that were inspected.
Of these, 4861 made it into the final catalog. In Table \ref{TabVSX}, we present the number of known variables 
of each main type passing our initial and final selections. 

Unsurprisingly, many of the stars missing from the catalog are dwarf novae (U Gem type). These are not in
our catalog since their light curves are dominated by large outbursts that occur at times that are quasi-periodic 
at best. The light curves of these known CVs are available through the Catalina website.

Approximately $89\%$ of the strictly periodic variables (eclipsing binaries, RR Lyrae and $\delta$ Scutis) 
with VSX matches among our original candidates were recovered, and 93\% of the periodic candidates passing 
our periodic candidate selection are within our final catalog. 
As a comparison, we found only 1144 matches to the much smaller GCVS variable star catalog (Samus et al.~2002-2013, 
June 2013 edition). Since the GCVS variable catalog is included within the VSX set, we do not consider it further.

\subsubsection{Comparison with LINEAR}

While the VSX catalog provides a very complete set of known variables stars, it is inherently heterogeneous 
(being derived from hundreds of authors using hundreds of different instruments).  Therefore, to test the 
accuracy of our catalog, we instead decided to compare our results with large sets of variable stars 
found using a single instrument.

Palaversa et al.~(2013), undertook the analysis of 200,000 variable star candidates selected from the 25 million objects
observed by the LINEAR experiment. They detected 7196 periodic variables in the range $14.5 < m_{linear} < 17$ from a
$\sim 10,000$ square degree region of the sky overlapping SDSS. In contrast to this work, where all $112,\!000$
candidates were inspected by a single person (AJD), the LINEAR analysis consisted of combining the results from eight people
who each inspected $\sim 25,000$ different light curves. In order to mitigate the biases of combining the separate
classifications, Palaversa et al.~(2013) reviewed all of the final candidates.

The LINEAR classification scheme consists of 11 types. Among these classes the three least populated have a combined
total of only twelve variables.  In comparison with this work, the LINEAR classification scheme does not include any
ellipsoidal or rotational variables (e.g. spotted stars such as RS CVns) which account for $>3\%$ ($>1500$ objects) of
our periodic variables. It also does not include Blazkho's, RRd's, or PCEBs (WD \& SD binaries).  Such systems are
also not included within their ``Other'' class.

In order to first determine the completeness of our survey compared to Palaversa et al.~(2013), we matched the
coordinates of our candidate periodic variables with LINEAR variables within $3\arcsec$.  Due to differences 
in the sky coverage of the two surveys, 6750 of the LINEAR variable stars with periods were covered by CSDR1. 
We found that only 645 (9.6\%) of the LINEAR objects were not in our initial selection.
Of the missing LINEAR variable stars, 60 were found to have amplitude $< 0.1$ mags in CSDR1.  
Most of the other missing sources were in areas having few observations in the CSDR1 data.

In our final periodic variable catalog, we found that 6016 of the 6750 ($89\%$) LINEAR periodic variables 
were included. This result is in good agreement with the completeness based on the VSX catalog above. However, since 
VSX and LINEAR data is magnitude-limited, we expect that our completeness is much lower below $V_{CSS} = 17$.

In order to compare the classifications given by LINEAR, we matched each type of variable given by Palaversa et
al.~(2013) with those from our catalog (including type-ab and Blazkho RR Lyrae from Drake et al.~2013a,b). In Table
\ref{TabLIN}, we present a so called confusion matrix for the 6016 objects in common. The agreement between the two
classifications is very good, with 94\% of the objects having the same classification. The sources with different
classifications include a number of $\beta$ Lyrae, RRd's and Blazkho RR Lyrae for which the Palaversa et al.~(2013)
catalog has no classification. Ignoring the 171 objects in the missing classes, the agreement is excellent, at $\sim
97\%$.

Aside from the results above we note that the LINEAR catalog classifies 20 objects as EA's with single eclipses.  
Only five of these objects match sources in our catalog. Four of the systems were found to have two eclipses with a 
significant difference in eclipse depth.  The remaining EA appears to have two eclipses of similar depth.  Three of 
the EA's were found to have been phased at half their true periods, thus explaining the single eclipses. All of the 
eclipsing binaries in our catalog are given at periods that provide two eclipses, since it is very common for single 
eclipses to appear when light curves are phased to half their true periods.

\begin{figure}[ht]{
\epsscale{0.7}
\plotone{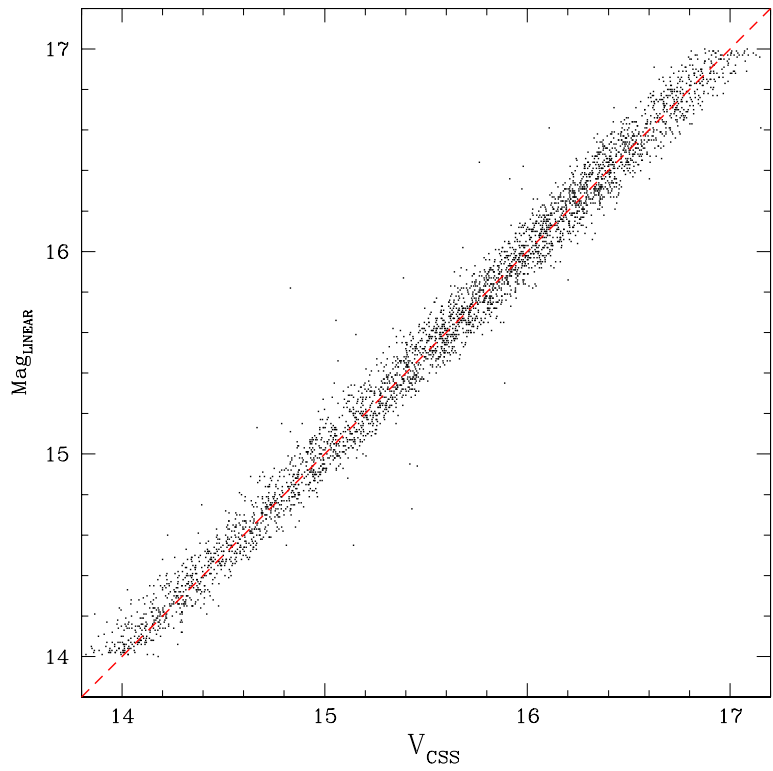}
\caption{\label{LinCal}
A comparison of median magnitudes for
variables found in LINEAR data with CSS
magnitudes.
}
}
\end{figure}

\begin{figure}[ht]{
\epsscale{0.7}
\plotone{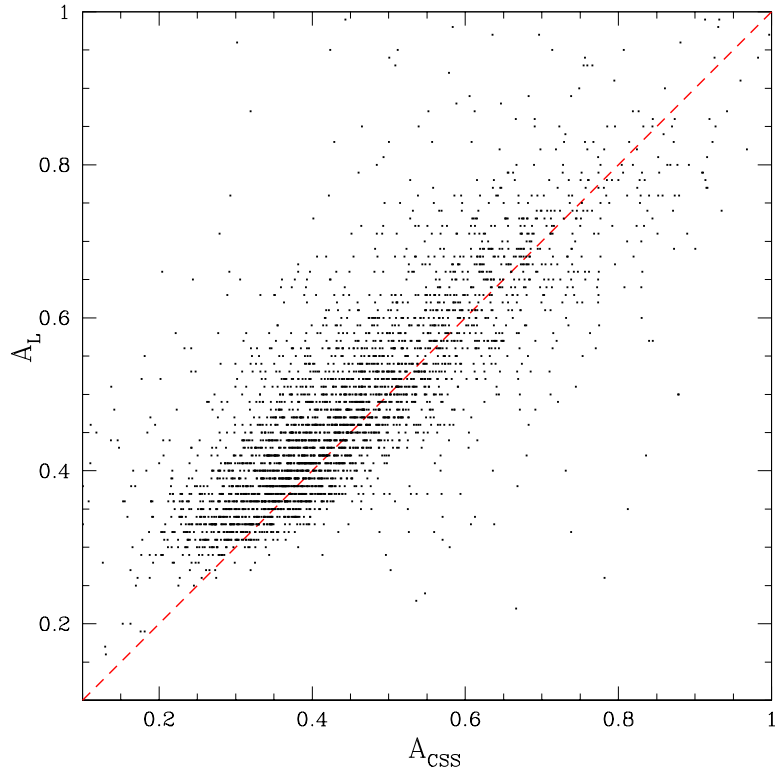}
\caption{\label{LinAmp}
The amplitudes of variables discovered in LINEAR
compared to the values from Fourier fits to CSS data.
}
}
\end{figure}

As the classes of variables are in excellent agreement, we can also compare the periods, amplitudes, and magnitudes of
the matching objects.  Both LINEAR and CSS observations are taken unfiltered. In Figure \ref{LinCal} we compare the
median LINEAR magnitudes with the average Fourier fit magnitudes from CSDR1. Once again we find excellent agreement
between the two systems over the magnitude range of the LINEAR variables.  In Figure \ref{LinAmp}, we compare the
variation amplitudes measured by LINEAR with those from our Fourier fits. Palaversa et al.~(2013) determined the amplitude as the
range of points between 5\% and 95\% in the magnitude distribution. In detached binaries systems, eclipses can last $<
10$\% of the orbital period, so we expect that the Palaversa et al.~(2013) scheme will often underestimate the actual
amplitude for such objects.  However, as these are only $4\%$ of the matching variables, the LINEAR and CSS amplitudes
are in good agreement.  For variables with the small amplitudes the LINEAR values are generally slightly larger. This is
likely to be due to cases where the uncertainty in the measurements is comparable with the amplitude of variability. At
the low-amplitude limit, the distribution of photometric measurements between 5\% and 95\% represents the photometric
scatter, rather than the amplitude of variability.

\begin{figure}[ht]{
\epsscale{0.7}
\plotone{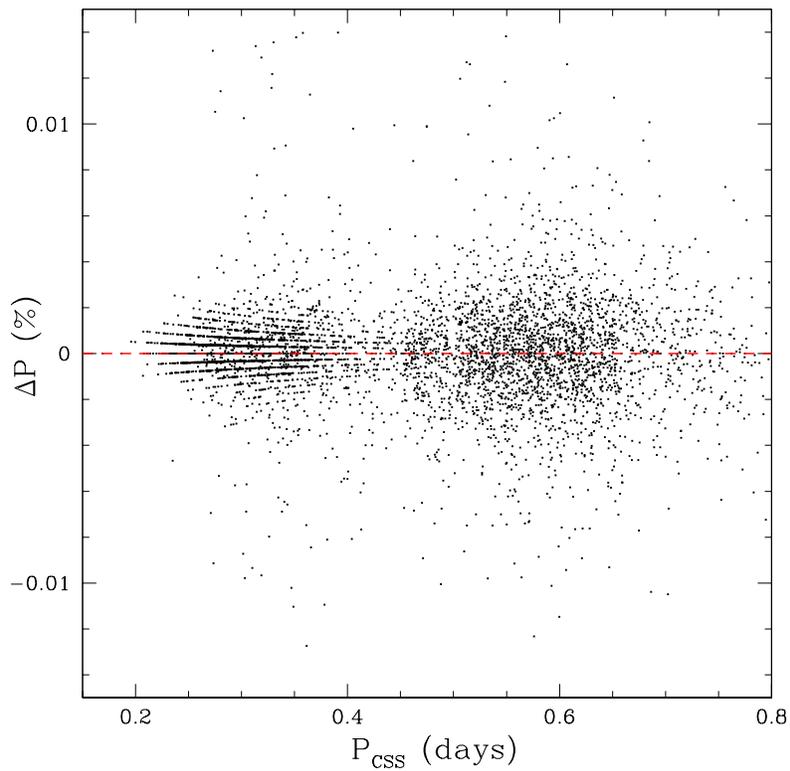}
\caption{\label{LinPer}
Period comparison. The percentage difference in period for sources present in both 
the LINEAR catalog (Palaversa et al.~2013) and this work, as function of the periods 
derived in this work. The features below 0.4 days are artificats due to rounding.
}
}
\end{figure}

In Figure \ref{LinPer}, we compare the periods determined by Palaversa et al.~(2013) with this work over the range where
$> 90\%$ of the variables reside.  The apparent features present below 0.4 days are merely due to the periods 
being rounded to the sixth decimal place.
Overall the differences in the periods are generally $< 0.002\%$. Once again this suggests
excellent agreement and independent confirmation of the variables. Combining the VSX and LINEAR catalogs we find that
overall 7856 objects were previously known periodic variables. The remaining $\sim 39,000$ are new discoveries.

\subsection{Comparisons with other variability surveys}

The largest existing sources of variable stars are microlensing surveys.
The four OGLE microlensing surveys (I-IV; Udalski et al.~1994) have together run for more 
than twenty years in search of microlensing events toward the Bulge and the Magellanic Clouds. 
As a by-product, the OGLE surveys have detected more than 200,000 periodic variables (e.g. Soszynski 
et al.~2009a,b, 2010a,b, 2011a,b; Graczyk et al.~2011; Pawlak et al.~2013). 
Most of the variable stars have been found within the Magellanic clouds.
Similarly, over a hundred thousand periodic variables have been discovered in data 
from the Macho microlensing project that was also taken toward the LMC and Bulge fields 
(e.g., Alcock et al.~1996, 1997, 1998, 1999, 2000b, 2001, 2002; Faccoli et al.~2007; Fraser et al.~2008).
Many of the variables detected by these surveys are common due to this overlap. Combined, 
these surveys cover $< 1000$ sq. deg on the sky.

Coverage of the Galactic bulge is now being expanded into the high extinction regions by the 
VVV survey. which also monitors the inner Galactic disk for variability in the near-IR
(Minniti et al.~2010; Catelan et al.~2011). 
This will ultimately cover 562 square degrees in $Z$, $Y$, $J$, $H$ and $K_s$ (Saito et al.~2012). 
Discoveries already include RR Lyrae, Cepheids, eclipsing binaries, LPVs and rotational 
binaries (Catelan et al.~2013). 

Over larger and less crowded regions of the sky, the very wide-field cameras of the ASAS survey (Pojmanski 1997) 
have discovered tens of thousands of variable stars down to $V \sim 14$ (e.g. Pojmanski 2000, 2002, 2003; 
Pojmanski \& Maciejewski 2004a,b; Pojmanski et al.~2005, Richards et al.~2012).
The NSVS survey has also covered a large fraction of the sky to similar depth ($8 < V < 15.5$, Wozniak et al.~2004)
using the Robotic Optical Transient Search Experiment (ROTSE-I, Akerlof et al.~2000). Searches of the ROTSE 
data by Kinemuchi et al.~(2006) and Hoffman et al.~(2008, 2009) have revealed many thousands of periodic variables.

Deeper searches for variability, in much smaller fields, have been carried out by QUEST (Vivas et al.~2004), LONEOS (Miceli et al.~2008), 
MOTESS-GNAT-1 (MG1, Krauss et al.~2007), {\em Kepler} (Matijevic et al.~2012) and SDSS Stripe-82 (Sesar et al.~2007, 
Bhatti et al.~2010, Becker et al.~2011).

In Table \ref{TabComp}, we compare the search parameters for surveys that have discovered large numbers of periodic variables
stars. While this table is not exhaustive, it does show that the CSDR1 is currently unequaled in its combination of sky area
and depth.  However, as microlensing surveys cover a similar number of stars within much smaller areas, and are observed 
at much higher cadence, it is not surprising that they continue to dominate the discovery statistics.

\section{Discussion and Conclusions}

In this paper we have presented an extensive search for periodic variable sources among 5.4 million candidate variables
selected from the CSDR1 data set.  When combined with variables from our prior analysis (Drake et al.~2013a,b), the
resulting catalog includes more Galactic eclipsing binaries and RR Lyrae than any existing survey. Comparison between
catalog sources and previously known periodic sources leads to the recovery of $\sim 90\%$ of known periodic objects in
the observated fields.  Of the 46668 periodic variables detected, 38812 (82\%) are new discoveries. Additionally, 396 of
these variables are detected twice  
due to overlap between fields. The high percentage of new discoveries is not unexpected since this analysis covers more
area to greater depth than any past search for periodic variables. Our variable star catalog does not include $\sim
4800$ sources that were flagged as potential periodic variables during the light curve inspection process.

Among the $\sim 31,000$ contact and ellipsoidal binaries we find almost 400 systems with periods below the 0.22 day
cutoff. These sources will be investigated in an upcoming paper (Drake et al. 2014 in prep.).  We also find many contact binaries that exhibit the
O'Connell effect.  The presence of stable light curves over $> 2000$ days suggests that this effect is not due to star
spots. The presence of long-timescale variations due to spots can be seen among other contact binaries, providing
further evidence that the O'Connell effect is not due to star spots.

We have investigated the possibility that contact binaries may have been misidentified as RRc's within our catalog.
Based on WISE and SDSS photometry as well as SDSS spectra, we find that the misidentification rate is at most a few
percent within our catalog. This result suggests that RRc's can be used as probes of halo
structure without significant contamination. Using the RRc's we confirm the presence of the Sagittarius tidal stream
structures. Additionally, we find strong evidence that a large fraction of the LPVs that we have discovered 
are part of the Sagittarius tidal stream.

We have found a small number of RR Lyrae that appear to undergo a rapid change in their pulsation mode.
Clementini et al.~(1994) called these objects switching-mode pulsators.  
The presence of sudden ($< 120$ days) changes may suggest abrupt changes to the stellar interior. 
Analysis of CSSJ172304.0+290810 suggests that some switching-mode pulsators are not Blazkho or RRd variables. 
However, more detailed analysis is required to fully test these possibilities.

In this analysis we have detected 85 PCEB candidates, of which 45 are new discoveries.  Parson et al.~(2013) recently
investigated PCEBs using Catalina data.  They selected candidate eclipsing systems based on the light curves of 835
spectroscopically confirmed WD-dM binaries. They found 29 new eclipsing systems bringing the total number of known systems
to 49. They also found 13 non-eclipsing ellipsoidal systems.  Unlike the Parsons et al.~(2013) analysis our selection is
not limited to sources with SDSS spectra. However, our ability to find such systems is strongly biased by the presence
of modulations within the light curve.  Only nine of the Parson et al.~(2013) systems are redetected, since PCEBs
without modulations, such as those discovered by Drake et al.~(2010), can not be detected. The current analysis, like 
that of Parsons et al.~(2013), is thus biased toward the detection of PCEB systems containing a luminous secondary.

We have detected a significant population of fundamental-mode pulsators with light curves that resemble both RRab's and 
classical Cepheids. These sources have amplitudes that are significantly larger than RRab's at their observed periods
and extend well beyond RR Lyrae (up to 2.5 days).  These sources have a significantly smaller range of periods
and a larger range of Galactic latitude than known classical Cepheids. They also do not exhibit the bumps seen among
short-period type-II Cepheids. We suggest that these objects are either ACEPs or RR Lyrae that have
recently evolved off the HB.

We are currently in the process of analyzing variable sources detected in data taken by the MLS and SSS telescopes
available in CSDR2. In this work we have parameterized Catalina light curves based on
statistical features such as periodicity, amplitude, variance, etc.  The current catalog will provide a training set for
selecting periodic variables within CSDR1 via Artificial Neural Networks, Random Forests and Support Vector Machines
following work performed to classify transient events (Mahabal et al.~2011, Djorgovski et al.~2012, Donalek et al.~2013).
In addition, we will apply Self-Organizing Maps and other machine learning techniques, similar to approaches
that recently been applied to large sets of variable stars (e.g., Dubath et al.~2011; Blomme et al.~2011; Richards
et al.~2012; Palaversa et al.~2013).

In this analysis we have exclusively searched for periodic variables in the sparse stellar fields that are observed away
from the Galactic disk. Extensive variability searches have also been undertaken in the Bulge and toward the Magellanic
Clouds as a byproduct of microlensing surveys. However, the dense stellar fields of the Galactic disk remain largely
unprobed for periodic variable stars.

In the near future the Gaia mission (Perryman et al.~2001) and the VVV survey (Minniti et al.~2010) will begin to
harvest of Galactic disk fields and are expected to find millions of periodic variables (Eyer et al.~2012; Catelan et
al.~2013). Although Gaia it only expected to reach stars to the same depth and number of epochs as CSDR1, it is expected
to have ultra-precise photometry. This will greatly increase the number of periodic variables discovered, even within
CSDR1 fields, since the fraction of variables increases as a power law with increasing photometric precision (Huber et
al.~2006, Howell 2008).  Likewise, while the LSST survey will have similar total numbers of observations (Ivezic et
al.~2008), it will reach far greater depths than any existing wide-field survey. LSST will thus find millions of faint
periodic variable sources.

\acknowledgements

CRTS and CSDR1 are supported by the U.S.~National Science Foundation under grant AST-1313422.
The CSS survey is funded by the National Aeronautics and Space Administration under Grant No. NNG05GF22G issued through the
Science Mission Directorate Near-Earth Objects Observations Program. J. L. P. acknowledges support from NASA through
Hubble Fellowship Grant HF-51261.01-A awarded by the STScI, which is operated by AURA, Inc.  for NASA, under contract
NAS 5-26555. Support for M.C. and G.T. is provided by the Ministry for the Economy, Development, and Tourism's Programa Iniciativa
Cient\'{i}fica Milenio through grant IC120009, awarded to Millennium Institute of Astrophysics (MAS); 
by Proyecto Basal PFB-06/2007; and by Proyecto FONDECYT Regular \#1110326 and \#1141141.
SDSS-III is managed by the Astrophysical Research Consortium for the Participating Institutions of the SDSS-III
Collaboration Funding for SDSS-III has been provided by the Alfred P. Sloan Foundation, the Partici:pating Institutions,
the National Science Foundation, and the U.S. Department of Energy Office of Science. The SDSS-III web site is
http://www.sdss3.org/.  This research has made use of the International Variable Star Index (VSX) database, operated at
AAVSO, Cambridge, Massachusetts, USA.

\begin{appendix}

During inspection of the periodic candidates in this work we discovered a number of non-periodic sources and periodic
sources where the light curve did not fit any of the existing classifications.  Most of the non-periodic variables are
simply stars and QSOs exhibiting irregular variability. However, some sources exhibiting outbursts were also discovered.

\vspace{0.3cm}\section{A: Supernovae}

Our review of periodic candidates led to the discovery of 51 variable sources exhibiting single, long, outbursts. Closer 
inspection showed that sources were supernovae associated with galaxies seen in SDSS and Catalina images. Nine of these
objects were previously discovered SN, while the rest were new discoveries.
Since supernovae are clearly not periodic, their discovery among our candidates is purely serendipitous. However,
this is not completely unexpected since Palaversa et al.~(2013) discovered a supernova during their inspection of 
periodic variable candidates from LINEAR data. In Table \ref{SN}, we present the parameters of the 42 newly 
discovered supernovae. Among all the supernovae discovered, two have SDSS spectra where some flux from the 
SN is seen. In one case this led to the host galaxy being classified by SDSS as a QSO.

To contrast these new detections with ongoing efforts to discover supernovae and other transients in the CSS data, we
note CRTS has so far discovered $> 1,\!000$ supernovae in seven years of CSS data. The fact that half of the new
supernovae occurred during the operation of CRTS, but were not detected, was found to be due to their association with
luminous host galaxies. In most cases, the increase in total brightness of the galaxy plus supernova is $< 1.5$ mags.
This is insufficient for automated detection by CRTS (Drake et al.~2009). In the near future, the CRTS-II project will
undertake image subtraction to find such supernovae. Our results suggest that the current detection rate should increase
significantly.

\begin{figure}[ht]{
\epsscale{1.0}
\plottwo{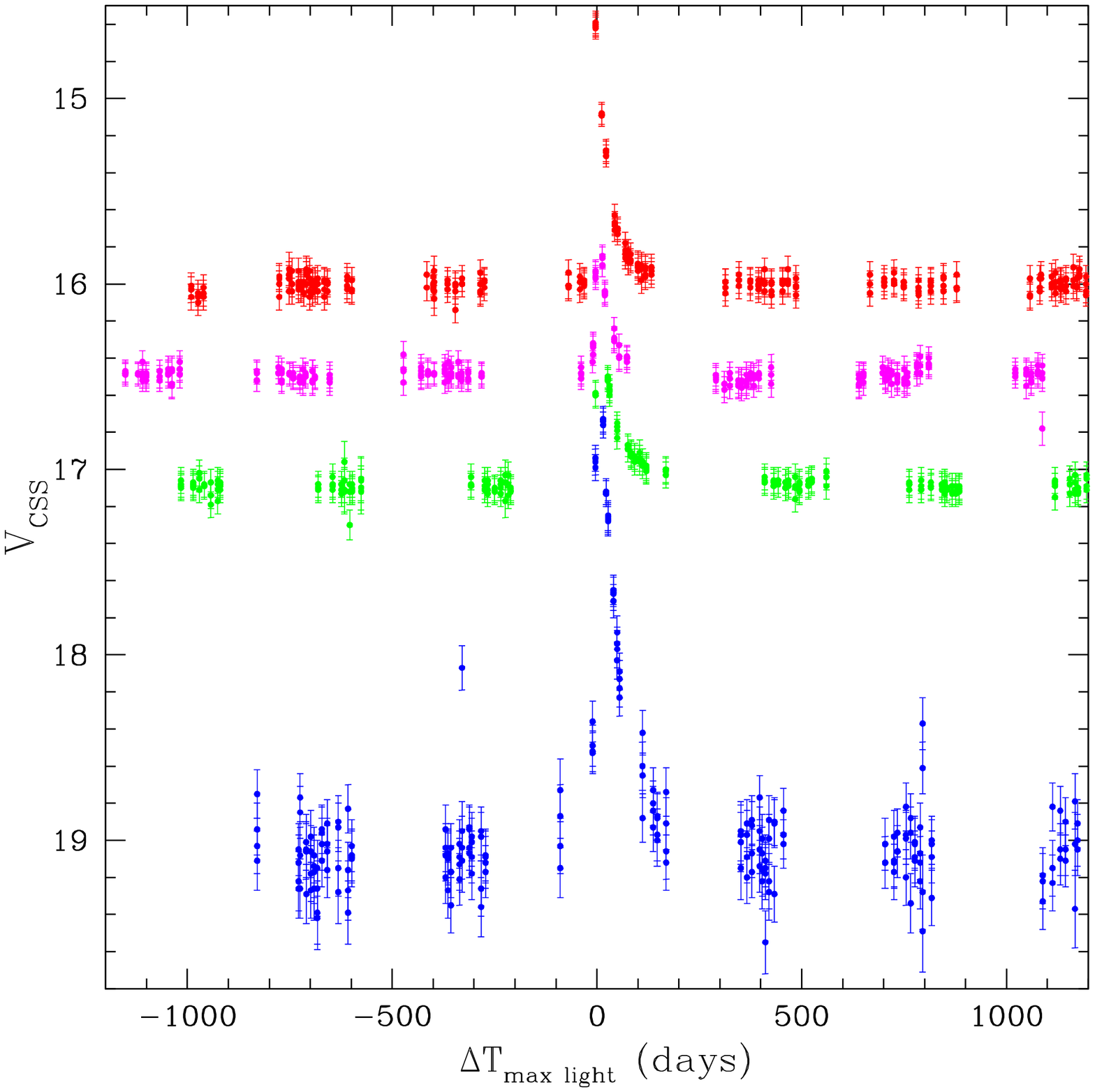}{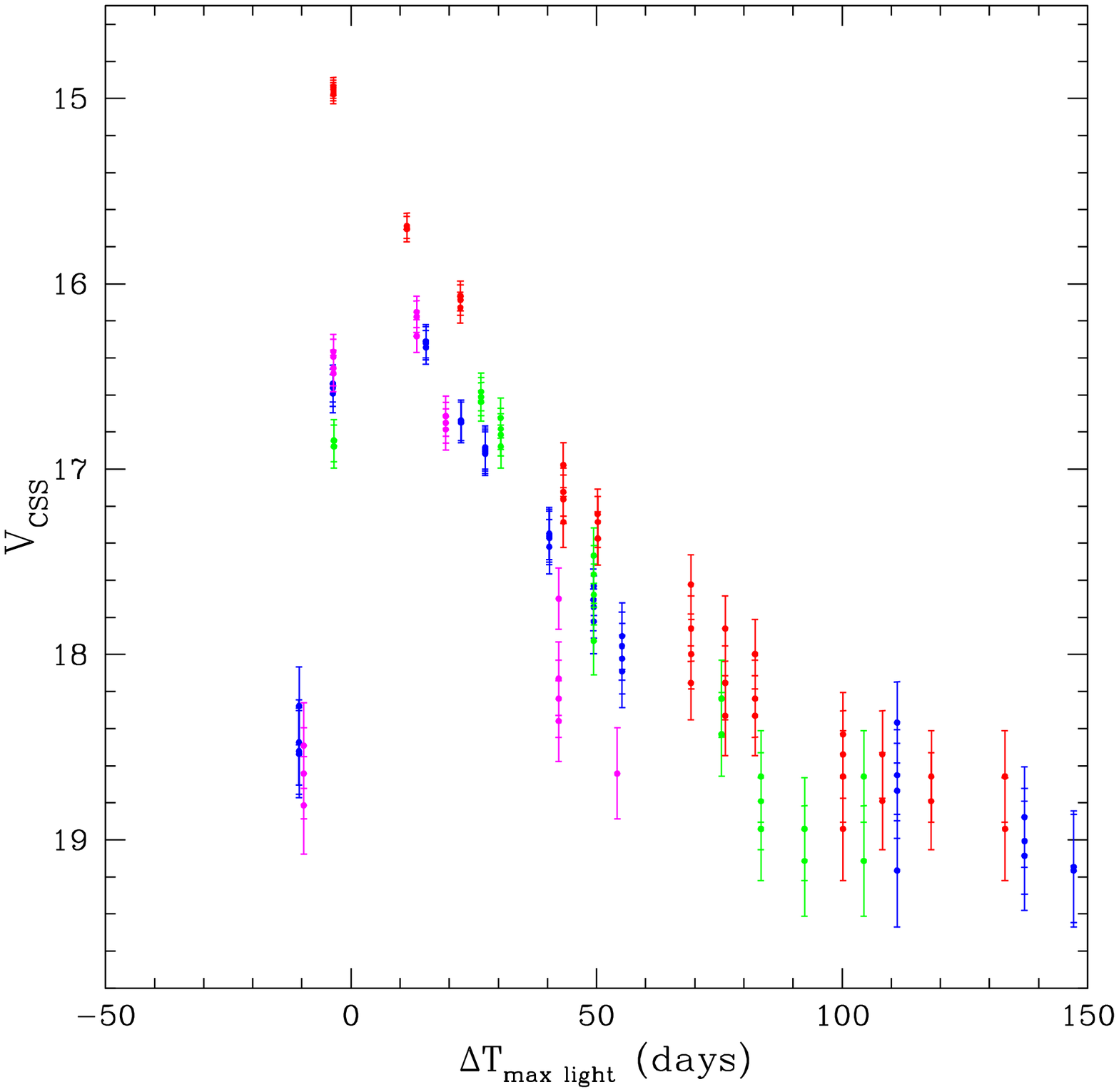}
\caption{\label{SNLC}
Light curves of serendipitously discovered supernovae.
Left: Observed SN plus host galaxy light curve.
Right: Light curve of the SN after subtracting the constant host galaxy flux.
}
}
\end{figure}

In Figure \ref{SNLC}, we present the light curves of four of the newly discovered supernovae.
We also show the apparent brightness of these events after the flux from the host galaxy 
has been subtracted.

\section{B: Cataclysmic Variables}

\begin{figure}[ht]{
\epsscale{0.8}
\plotone{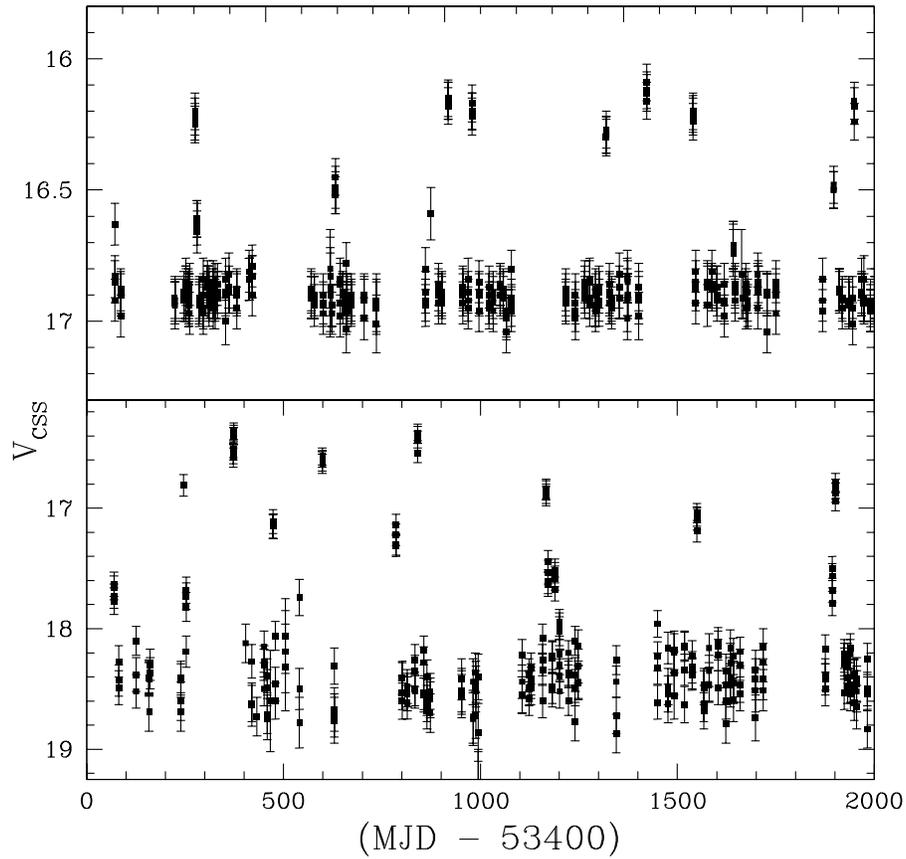}
\caption{\label{CVLC}
Light curves of two serendipitously discovered cataclysmic variables.
}
}
\end{figure}

In addition to supernovae, we discovered 18 new dwarf nova-type CVs. The light curves 
of these objects exhibit repeated outbursts. In Figure \ref{CVLC}, we plot two of the
sources and in Table \ref{CVtab}, we present the details of outbursting CVs discovered 
during the analysis. Drake et al.~(2013c) found that almost all CVs discovered by CRTS
have $u-g < 0.6$ and $g - r < 1$. Four of the 18 new CVs covered by SDSS photometry 
exhibit colour excess compared to Drake et al.~(2013c). This suggests CV systems
where the companion star dominates the observed flux.

\end{appendix}


\clearpage

\clearpage

\clearpage



\begin{table}
\caption{Periodic Variable Selection}
\label{Psel}
\begin{center}
\begin{tabular}{@{}lr}
\hline
Selection & Objects\\
\hline 
CSS Sources & 198 million\\
Variable Cand. &  5.4 million\\
LS $\eta < 10^{-5}$ & $\sim$154,000\\
$N_{blend} < 5\%$ & $\sim$144,000\\
$P \neq P_{alias}$ & $\sim$128,000 \\
$\chi^2_r < 5$ or $\eta < 10^{-9}$ & $\sim$112,000\\
Periodic & 46,668\\
\hline
\end{tabular}
\end{center}
Col. (1), selection used to select the sample of objects.
Col. (2), number of objects following the sub-selection.
These numbers do not include the $\sim 14,\!000$ known RRab's 
from Drake et al.~(2013a,b).
\end{table}

\begin{table}
\caption{Types of Periodic Variables}
\label{Vtype}
\begin{center}
\begin{tabular}{@{}lrrr}
\hline
Type & F [\%] & N & Class\\
\hline 
EW & 49.93 & 30743 & 1 \\
EA & 7.61 & 4683  & 2 \\
$\beta$ Lyrae &  0.45 & 279 & 3\\
RRab & 27.28 & 16797\tablenotemark{a} & 4\\
RRc &  8.88 & 5469 & 5\\
RRd &  0.82 & 502  & 6\\
Blazhko &  0.36 & 223\tablenotemark{a} & 7\\
RS CVn &  2.47 & 1522 & 8 \\
ACEP &  0.10 & 64 & 9\\
Cep-II &  0.20 & 124 & 10\\
HADS &  0.39 & 242 & 11 \\
LADS &  0.01 & 7 & 12 \\
LPV &  0.83 & 512 & 13\\
ELL &  0.23 & 143 & 14 \\
Hump &  0.04 & 25 & 15 \\
PCEB &  0.14 & 85 & 16\\
$\rm EA_{UP}$ &  0.25 & 155 & 17\\
\hline
\end{tabular}
\end{center}
The periodic variable classes are as noted in the text.
\tablenotetext{a}{Includes 14,362 known CSS RRab's and 149 known Blazkho types from Drake et al.~(2013a,b),
as well as 396 objects observed in multiple fields.
}
\end{table}

\begin{table*}
\caption{Periodic Variable Catalog}
\label{Full}
\begin{minipage}{186mm}
\begin{center}
\begin{tabular}{@{}lccclcl}
\hline 
CSS ID & RA & Dec.~(J2000) & $\rm \bar V_{CSS}$ & $\rm P_F$ & $\rm A_V$ & Class\\
\hline
CSS\_J000020.4+103118 & 00:00:20.41 & +10:31:18.9 &  14.62 & 1.491758 &  2.39 & 2\tablenotemark{d}\\ 
CSS\_J000031.5$-$084652 & 00:00:31.50 & $-$08:46:52.3 &  14.14 & 0.404185 &  0.12 & 1\\ 
CSS\_J000036.9+412805 & 00:00:36.94 & +41:28:05.7 &  17.39 & 0.274627 &  0.73 & 1\\ 
CSS\_J000037.5+390308 & 00:00:37.55 & +39:03:08.1 &  17.74 & 0.30691 &  0.23 & 1\tablenotemark{a}\\ 
CSS\_J000103.3+105724 & 00:01:03.37 & +10:57:24.4 &  15.25 & 1.5837582 &  0.11 & 8\\ 
CSS\_J000103.4+395744 & 00:01:03.46 & +39:57:44.5 &  15.51 & 1.9670131 &  0.14 & 1\\ 
CSS\_J000106.9+120610 & 00:01:06.96 & +12:06:10.3 &  15.85 & 0.297318 &  0.11 & 1\\ 
CSS\_J000110.8+400521 & 00:01:10.89 & +40:05:21.1 &  13.69 & 0.6606980 &  0.09 & 1\\ 
CSS\_J000131.5+324913 & 00:01:31.54 & +32:49:13.1 &  14.71 & 13.049549 &  0.17 & 8\tablenotemark{a}\\ 
CSS\_J000141.2+421108 & 00:01:41.28 & +42:11:08.2 &  18.45 & 0.308596 &  0.54 & 1\\ 
\hline
\end{tabular}
\end{center}
\end{minipage}\\
{\it The full table will be available online.}\\
Col (1), CSS ID;
Cols (2) \& (3), Right Ascension and Declination (J2000);
Col. (4), average magnitude from AFD;
Col. (5), period in days;
Col. (6), amplitude from AFD;
Col. (7), numerical class number based on Table \ref{Vtype}.
\tablenotetext{a}{Period is inexact based on light curve inspection.}
\tablenotetext{b}{An object exhibiting baseline variation due to spots.}
\tablenotetext{c}{Candidate ultra-short-period ellipsoidal binary.}
\tablenotetext{d}{Deeply eclipsing binary.}
\tablenotetext{e}{Light curve with unusual morphology.}
\tablenotetext{f}{Blended object.}
\end{table*}

\begin{table}
\caption{SDSS Classes of Periodic Candidates}
\label{TabSDSSS}
\begin{center}
\begin{tabular}{@{}lrrllcc}
\hline
Spectroscopic types & $\rm I$ & $\rm S$\\
\hline 
Total  & 5420 & 2299\\
Star   & 2918 & 2256 (2291\tablenotemark{a})\\
Galaxy & 1659 & 38 (3\tablenotemark{a})\\
QSO    &  843 & 5 (5\tablenotemark{a})\\
\hline
Photometric types & $\rm I$ & $\rm S$\\
\hline
Total   & 67076 & 27871\\
Star    & 58443 & 25639\\
Galaxy  & 5593  & 827 ($\sim 10$\tablenotemark{a})\\
Unknown & 3040  & 1405 \\
\hline
\end{tabular}
\tablenotetext{a}{Number based on inspection of SDSS data.}
\end{center}
Numbers and SDSS-DR10 object types from Catalina periodic variable 
candidates with SDSS photometry and spectra.
Col. (1) SDSS object type. Col. (2) Number of SDSS matches among 
the 112,000 inspected periodic variable candidates. Col. (3) Number of 
SDSS sources selected as periodic variables before inspection 
of SDSS images and spectra. 
\end{table}

\begin{table}
\caption{Matches with VSX Sources}
\label{TabVSX}
\begin{center}
\begin{tabular}{@{}llrr}
\hline
Var Type & $\rm O$ & $\rm I$ & $\rm S$\\
\hline
All     & 6454   &    6030  &  4861\\
EB      & 3201   &    3070  &  2834\\
RR Lyrae   & 1006   &     958  &   925\\
$\delta$ Scuti & 240    &     234  &   210\\
U Gem   & 214    &     119  &     6\\
Mira    & 206    &     187  &   162\\
Misc    & 382    &     328  &   217\\
\hline
\end{tabular}
\end{center}
Col. (1) Main types of variables with VSX matches.
Col. (2) Number of VSX variables of each type among the original 154 thousand
periodic candidates.
Col. (3) Number of VSX variables in the inspected set of 112 thousand candidates. 
Col. (4) Number of VSX variables selected for final periodic variable catalog.
We include both MISC and VAR VSX classes in our Misc category.
\end{table}

\begin{table*}
\caption{Catalina Variables with LINEAR matches}
\label{TabLIN}
\begin{minipage}{186mm}
\begin{center}
\begin{tabular}{@{}llllllllll}
\hline
Type & BLHer & ACEP & $\delta$-Scuti & EA & EB/EW & LPV & Misc & RRab & RRc\\
\hline
 Cep-II & \nodata  & \nodata & \nodata & \nodata & 2 & 1 & 1 & 1 & \nodata\\
 ACEP & 4  & \nodata & \nodata & \nodata & \nodata & \nodata & \nodata & 2 & \nodata\\
 HADS & \nodata  & \nodata & 14 & \nodata & \nodata & \nodata & \nodata & \nodata & \nodata\\
 LADS & \nodata  & \nodata & \nodata & \nodata & \nodata & \nodata & \nodata & \nodata & \nodata\\
 EA & \nodata  & \nodata & \nodata & 226 & 86 & \nodata & \nodata & \nodata & \nodata\\
 EW & \nodata  & \nodata & \nodata & 18 & 1943 & \nodata & 2 & \nodata & 14\\
 $\beta$ Lyrae & \nodata  & \nodata & \nodata & 1 & 27 & \nodata & \nodata & \nodata & \nodata\\
 ELL & \nodata  & \nodata & \nodata & \nodata & 3 & \nodata & \nodata & \nodata & \nodata\\
 LPV & \nodata  & \nodata & \nodata & \nodata & \nodata & 48 & \nodata & \nodata & \nodata\\
 RRab & \nodata  & 2 & \nodata & \nodata & 2 & \nodata & 6 & 2640 & 18\\
 RRc & \nodata  & \nodata & \nodata & \nodata & 9 & \nodata & 2 & 5 & 773\\
 Blazkho & 1  & \nodata & \nodata & \nodata & \nodata & \nodata & 1 & 38 & 3\\
 RRd & \nodata  & \nodata & \nodata & \nodata & 1 & \nodata & 5 & 6 & 96\\
 RS CVn & \nodata  & 1 & \nodata & \nodata & \nodata & 2 & 1 & 1 & \nodata\\
 PCEB & \nodata  & \nodata & \nodata & 2 & \nodata & \nodata & \nodata & \nodata & \nodata\\
 Hump & \nodata  & \nodata & \nodata & \nodata & 1 & \nodata & \nodata & \nodata & 1\\
 $\rm EA_{UP}$ & \nodata  & \nodata & \nodata & 4 & \nodata & \nodata & \nodata & \nodata & \nodata\\
 $\rm P_U$ & \nodata  & \nodata & \nodata & 1 & 1 & \nodata & \nodata & \nodata & \nodata\\
\hline
\end{tabular}
\end{center}
\end{minipage}\\
\medskip
The first row gives the types periodic variable from Palaversa et al.~(2013) based on LINEAR data,
and the first column gives the classes based on the current analysis.
\end{table*}

\begin{table}
\caption{Periodic Variables from Large Surveys}
\label{TabComp}
\begin{minipage}{186mm}
\begin{center}
\begin{tabular}{@{}lrrrrrrl}
\hline 
Survey  & Sources & $\rm N_{P}$ &  Area & Range & Span & Epochs & Filters \\
 & ($10^6$) &  &  (sq. deg.) & (mags) & (yrs) &  &  \\
\hline
OGLE I-IV      & 20-200  &   $>$200,000  &  100-630  &   12 $-$ 21    &  21   &  200-5000  & V,I\\
MACHO          &     60  & $\sim$120,000  &      100  &   13 $-$ 19.5  &  8   &  150-2000  & V,R\\
{\bf CSDR1}    &    200  & {\bf 61,000}\tablenotemark{a}  &   20,000  & 12.5 $-$ 19.5  &  7    &    70-350  & C\\
ASAS           &     15  &       12,000  &   30,000  &    8 $-$ 14    &  9    &       540  & V,I\\
LINEAR         &     25  &        7,000  &   10,000  &   14 $-$ 17    &  9    &       250  & C\\
NSVS           &     14  &        5,600  &   30,000  &    8 $-$ 15.5  &  1    &    100-500 & C\\ 
MG-1           &    2.1  &        5,200  &      300  &   13 $-$ 19    &  2    &        200 & R\\
{\em Kepler}         &   0.13  &        2,600  &      116  &    9 $-$ 16    &  3.5  &  $>$50,000 & C\\
LONEOS         &      1  &          840  &     1430  &   13 $-$ 18.5  &  2    &      28-50 & C\\
Quest          &    1.2  &          500  &      380  &  13.5 $-$ 19.7 &  2.3  &      15-40 & U,B,V,R,I\\
SDSS Stripe-82 &      1  &          500  &      290  &   15 $-$ 21    &  3    &         70 & u,g,r,i,z\\
\hline
\end{tabular}
\tablenotetext{a}{Including RRab's from Drake et al.~(2013a,b).}
\end{center}
\end{minipage}
\medskip
The number of periodic variables discovered by major projects based on references given within the text.
Col. (1) Names of survey where the data was taken.
Col. (2) Number of sources covered by survey. 
Col. (3) Number of known periodic variables. 
Col. (4) Total sky coverage.
Col. (5) Range of survey photometry magnitudes.
Col. (6) Total time span of the survey observations.
Col. (7) Number of image epochs taken per band.
Col. (8) Observation bands (C = clear/no filter).
Note: the data searched within the NSVS and LONEOS data is a small fraction 
of the total data taken by these surveys.
\end{table}

\begin{table}
\caption{Parameters of Newly Discovered Supernovae}
\label{SN}
\footnotesize
\begin{minipage}{186mm}
\begin{center}
\begin{tabular}{@{}lcccc}
\hline
CRTS ID & R.A. & Dec (J2000) & UT Date & $z$\\
\hline 
CSS\_J221632.6-020841 & 22:16:32.67 & -02:08:41.2 & 2005-09-24 & \nodata \\
CSS\_J021056.9+362733 & 02:10:56.92 & +36:27:33.9 & 2005-10-25 & \nodata \\
CSS\_J035255.3-112753 & 03:52:55.33 & -11:27:53.5 & 2005-11-02 & \nodata \\
CSS\_J235535.6+291220 & 23:55:35.62 & +29:12:20.4 & 2005-11-20 & \nodata \\
CSS\_J044104.9+093109 & 04:41:04.90 & +09:31:09.4 & 2005-11-25 & \nodata \\
CSS\_J092459.1+013839 & 09:24:59.17 & +01:38:39.2 & 2005-12-28 & \nodata \\
CSS\_J050141.1-015936 & 05:01:41.19 & -01:59:36.2 & 2005-12-31 & \nodata \\
CSS\_J162513.0+112756 & 16:25:13.06 & +11:27:56.0 & 2006-03-24 & \nodata \\
CSS\_J123716.5+264710 & 12:37:16.59 & +26:47:10.9 & 2006-04-26 & \nodata \\
CSS\_J023657.0+192258 & 02:36:57.06 & +19:22:58.0 & 2006-11-11 & \nodata \\
CSS\_J102255.7+163606 & 10:22:55.71 & +16:36:06.4 & 2006-12-14 & 0.0449\\
CSS\_J112851.2+390458 & 11:28:51.24 & +39:04:58.7 & 2007-02-27 & \nodata \\
CSS\_J141731.1+431843 & 14:17:31.16 & +43:18:43.0 & 2007-03-11 & \nodata \\
CSS\_J113517.2-152048 & 11:35:17.23 & -15:20:48.8 & 2007-03-30 & \nodata \\
CSS\_J144959.8+084819 & 14:49:59.89 & +08:48:19.6 & 2007-04-11 & \nodata \\
CSS\_J130404.0+443720 & 13:04:04.08 & +44:37:20.3 & 2007-06-16 & \nodata \\
CSS\_J013303.2-082303 & 01:33:03.28 & -08:23:03.6 & 2007-09-06 & \nodata \\
CSS\_J014920.9+313213 & 01:49:20.94 & +31:32:13.6 & 2007-09-12 & \nodata \\
CSS\_J090316.2+373746 & 09:03:16.25 & +37:37:46.4 & 2007-11-20 & \nodata \\
CSS\_J083533.4+015103 & 08:35:33.48 & +01:51:03.2 & 2007-12-17 & \nodata \\
CSS\_J152459.7+045422 & 15:24:59.73 & +04:54:22.7 & 2008-01-09 & 0.04373\\
CSS\_J161542.7+332401 & 16:15:42.77 & +33:24:01.0 & 2008-02-12 & 0.02969\\
CSS\_J151555.7+165902 & 15:15:55.75 & +16:59:02.4 & 2008-04-15 & \nodata \\
CSS\_J221502.0+151854 & 22:15:02.00 & +15:18:54.2 & 2008-09-22 & \nodata \\
CSS\_J100004.1+060544 & 10:00:04.18 & +06:05:44.7 & 2008-12-06 & \nodata \\
CSS\_J085423.4+512330 & 08:54:23.41 & +51:23:30.1 & 2009-02-24 & \nodata \\
CSS\_J141043.2+115704 & 14:10:43.22 & +11:57:04.5 & 2009-03-18 & \nodata \\
CSS\_J224534.9+263535 & 22:45:34.94 & +26:35:35.8 & 2009-07-31 & \nodata \\
CSS\_J011524.5-100823 & 01:15:24.54 & -10:08:23.5 & 2009-09-16 & \nodata \\
CSS\_J020949.5+294021 & 02:09:49.56 & +29:40:21.7 & 2009-09-18 & \nodata \\
CSS\_J073643.3+473127 & 07:36:43.32 & +47:31:27.5 & 2009-11-14 & \nodata \\
CSS\_J114817.6+292102 & 11:48:17.61 & +29:21:02.3 & 2009-11-24 & 0.0476\\
CSS\_J020439.7+130902 & 02:04:39.76 & +13:09:02.9 & 2009-12-07 & \nodata \\
CSS\_J103418.6+225056 & 10:34:18.65 & +22:50:56.2 & 2009-12-16 & \nodata \\
CSS\_J133147.6+072652 & 13:31:47.69 & +07:26:52.8 & 2010-03-15 & 0.0209\\
CSS\_J145206.1+175523 & 14:52:06.12 & +17:55:23.3 & 2010-03-19 & \nodata \\
CSS\_J151658.3+052245 & 15:16:58.32 & +05:22:45.3 & 2010-05-06 & \nodata \\
CSS\_J081617.0+014822 & 08:16:17.04 & +01:48:22.5 & 2010-10-19 & \nodata \\
CSS\_J090424.8+210542 & 09:04:24.89 & +21:05:42.6 & 2010-12-01 & \nodata \\
CSS\_J101727.6+155844 & 10:17:27.62 & +15:58:44.5 & 2011-01-13 & \nodata \\
CSS\_J082143.2+034733 & 08:21:43.25 & +03:47:33.1 & 2011-01-13 & 0.02975\\
CSS\_J132915.1+154719 & 13:29:15.19 & +15:47:19.2 & 2011-03-05 & \nodata \\
\hline
\end{tabular}
\end{center}
\end{minipage}
Col. (1) Catalina ID. Cols. (2) \& (3) Right Ascension and Declination. Col. (4) Date of maximum light.
Col (5) Host galaxy redshift (if known).
\end{table}

\begin{table}
\caption{Parameters of Newly Discovered CVs}
\label{CVtab}
\begin{minipage}{186mm}
\begin{center}
\begin{tabular}{@{}lcccccccc}
\hline
CRTS ID & R.A. & Dec (J2000) & $\bar{V}$  & $u$ & $g$ & $r$ & $i$ & $z$\\
\hline 
CSS\_J003304.0+380105 & 00:33:04.09 & +38:01:05.8 &   17.8 &   20.5 &   20.5 &   20.6 &   20.5 &   19.8\\
CSS\_J004024.1+334131 & 00:40:24.18 & +33:41:31.9 &   18.3 &   20.4 &   19.0 &   18.4 &   18.1 &   18.0\tablenotemark{a}\\
CSS\_J031245.8+304246 & 03:12:45.82 & +30:42:46.9 &   17.2 & \nodata & \nodata & \nodata & \nodata & \nodata \\
CSS\_J033104.4+172540 & 03:31:04.41 & +17:25:40.7 &   17.9 &   19.3 &   19.4 &   19.0 &   18.6 &   18.3\\
CSS\_J035318.1-034846 & 03:53:18.19 & -03:48:46.8 &   17.2 & \nodata & \nodata & \nodata & \nodata & \nodata \\
CSS\_J035806.6+340343 & 03:58:06.69 & +34:03:43.8 &   16.7 & \nodata & \nodata & \nodata & \nodata & \nodata \\
CSS\_J035905.8+175034 & 03:59:05.89 & +17:50:34.7 &   17.4 &   16.9 &   16.6 &   16.7 &   16.8 &   16.8\\
CSS\_J041138.6+232219 & 04:11:38.60 & +23:22:19.9 &   17.8 & \nodata & \nodata & \nodata & \nodata & \nodata \\
CSS\_J042933.6+312927 & 04:29:33.65 & +31:29:27.7 &   18.3 & \nodata & \nodata & \nodata & \nodata & \nodata \\
CSS\_J043019.9+095318 & 04:30:19.98 & +09:53:18.2 &   18.2 &   18.6 &   18.8 &   18.5 &   18.2 &   18.0\\
CSS\_J050236.5+134916 & 05:02:36.58 & +13:49:16.9 &   17.4 &   18.4 &   17.6 &   16.2 &   15.1 &   14.5\tablenotemark{a}\\
CSS\_J050253.1+171041 & 05:02:53.10 & +17:10:41.5 &   17.9 & \nodata & \nodata & \nodata & \nodata & \nodata \\
CSS\_J075311.6+352631 & 07:53:11.61 & +35:26:31.4 &   16.9 &   16.9 &   16.4 &   16.2 &   16.0 &   16.0\\
CSS\_J082925.3-001351 & 08:29:25.37 & -00:13:51.0 &   16.9 & \nodata & \nodata & \nodata & \nodata & \nodata \\
CSS\_J101233.9+074644 & 10:12:33.92 & +07:46:44.5 &   16.9 &   15.6 &   15.4 &   15.8 &   15.9 &   16.1\\
CSS\_J134504.3+004253 & 13:45:04.32 & +00:42:53.1 &   14.4 & \nodata & \nodata & \nodata & \nodata & \nodata \\
CSS\_J152351.2+083606 & 15:23:51.22 & +08:36:06.9 &   18.7 &   14.7 &   14.7 &   15.4 &   14.4 &   14.0\\
CSS\_J164825.6+142411 & 16:48:25.62 & +14:24:11.3 &   18.0 &   21.9 &   19.8 &   18.6 &   18.1 &   17.9\tablenotemark{a}\\
CSS\_J172951.6+220807 & 17:29:51.64 & +22:08:07.2 &   17.1 & \nodata & \nodata & \nodata & \nodata & \nodata \\
CSS\_J180222.4+455245 & 18:02:22.42 & +45:52:45.3 &   17.4 &   19.5 &   19.4 &   19.3 &   19.2 &   18.9\\
CSS\_J205518.8-162640 & 20:55:18.83 & -16:26:40.6 &   17.7 &   17.1 &   17.0 &   17.2 &   17.3 &   17.3\\
CSS\_J210732.6-095659 & 21:07:32.61 & -09:56:59.1 &   17.4 &   16.9 &   16.3 &   16.5 &   16.6 &   16.7\\
CSS\_J212625.0+201946 & 21:26:25.03 & +20:19:46.4 &   16.4 &   16.5 &   16.4 &   16.6 &   16.8 &   16.9\\
CSS\_J213319.4+190155 & 21:33:19.47 & +19:01:55.0 &   16.8 &   19.0 &   17.7 &   17.0 &   16.7 &   16.5\tablenotemark{a}\\
CSS\_J213438.4-055319 & 21:34:38.44 & -05:53:19.2 &   18.1 &   18.3 &   18.2 &   18.3 &   18.3 &   18.3\\
CSS\_J220031.2+033430 & 22:00:31.23 & +03:34:30.4 &   18.0 &   18.3 &   18.5 &   18.3 &   18.0 &   17.9\\
CSS\_J220321.3+144606 & 22:03:21.32 & +14:46:06.9 &   17.5 &   17.5 &   17.9 &   17.7 &   17.3 &   17.6\\
\hline
\end{tabular}
\tablenotetext{a}{System exhibiting colour excess}
\end{center}
\end{minipage}
Col. (1) Catalina ID.
Cols. (2) \& (3) Right Ascension and Declination
Col. (4) Average CSS magnitude.
Cols. (5) to (9). The SDSS-DR10 photometric magnitudes.
\end{table}

\end{document}